%% file: AWW-OpenFlavor-EtaShift.tex
\documentclass[twocolumn,epjc3]{svjour3}  

\journalname{EPJA}

\RequirePackage{graphicx}
\RequirePackage{flushend}
\RequirePackage[numbers,sort&compress]{natbib}

\usepackage{microtype}
\usepackage{upgreek}
\usepackage{amsmath,amssymb}
\usepackage{slashed}
\usepackage{booktabs}
\usepackage{nicefrac}
\usepackage{graphicx}
\usepackage{braket}
\usepackage{xtab}
\usepackage[super]{nth}

\usepackage[labelformat=empty]{subcaption}

\usepackage{tikz}
\usetikzlibrary{shapes}

\usepackage[
    pdftitle={AWW-OpenFlavor-EtaShift},
    pdfauthor={Dr. Thomas Hilger}
    bookmarks=true,
    pdfborder={0 0 0},
    bookmarksopen=true,
    unicode,
    hyperfootnotes=true,
    colorlinks=false,
    pagebackref=false,
    citecolor=blue,
    linkcolor=blue,
    urlcolor=blue,
    breaklinks,
    pdfpagelabels,
    ]{hyperref}
\usepackage[numbered]{bookmark}
\usepackage[all]{hypcap}

\usepackage[
    toc,
    acronym,
    nomain,
    style=long,
    nonumberlist,
    nowarn,
    nomain,
    nosuper,
    ]{glossaries}

\makeglossaries
\glsdisablehyper
\input{glossaries}

\newcommand{\ld}{: \!}
\newcommand{\rd}{\! :}

\newcommand{\Dp}[2]{\mathrm{d}^{#1\,} \! #2}

\AtBeginDocument{
\newboolean{draft}
\setboolean{draft}{true}
\newboolean{todo}
\setboolean{todo}{true}
\newcommand{\comment}[2]{\ifthenelse{\boolean{#1}}{\textcolor{blue}{\emph{\small #2}}}{}}
}

\begin{document}

\title{Aspects of open-flavour mesons in a comprehensive DSBSE study\thanksref{t1}
}

\institute{Institute of Physics, University of Graz, NAWI Graz, A-8010 Graz, Austria \label{addr1}
           \and
           ECT*, Villa Tambosi, 38123 Villazzano (Trento), Italy \label{addr2}
           \and
           Institute of High Energy Physics, Austrian Academy of Sciences, A-1050 Vienna, Austria \label{addr3}
}

\author{T. Hilger\thanksref{e1,addr1,addr3}
        \and
        M. G\'{o}mez-Rocha\thanksref{e2,addr2}
        \and
        A. Krassnigg\thanksref{e3,addr1}
        \and
        W. Lucha\thanksref{e4,addr3}
        }

\thankstext{t1}{This work was supported by the Austrian Science Fund (FWF) under project no.\ P25121-N27.}
\thankstext{e1}{e-mail: thomas.hilger@uni-graz.at}
\thankstext{e2}{e-mail: gomezr@ectstar.eu}
\thankstext{e3}{e-mail: andreas.krassnigg@uni-graz.at}
\thankstext{e4}{e-mail: Wolfgang.Lucha@oeaw.ac.at}


\date{Received: \today / Accepted: date}

\maketitle

\begin{abstract}
Open-flavour meson studies are the necessary completion to any comprehensive investigation of quarkonia. 
We extend recent studies of quarkonia in the Dyson-Schwinger-Bethe-Salpeter-equation approach to explore their results for all possible flavour combinations.
Within the inherent limitations of the setup, we present the most comprehensive results for meson masses and leptonic decay constants currently available and put them in perspective with respect to experiment and other approaches.
\end{abstract}

\maketitle

\section{Introduction}

In \gls{QCD}, mesons and bary\-ons are archetypical in different ways: baryons are dominantly abundant concerning hadronic matter, and every-day life even offers ample motivation for their study.
Mesons, on the other hand, are easily produced in particle-physics experiments due to the lack of a conserved meson-quantum number, and restrictions apply only via mass and other quantum numbers, such as flavour or electric charge.

With respect to the strong interaction, specifically, mesons are more immediate objects of study for theory due to their more basic setup in a quark-antiquark bilinear fashion.
Given the character of bound states and strong resonances within \gls{QCD}, a nonperturbative treatment is imperative to learn about the inner workings of such states in terms of quarks and gluons.
More specifically even, the open-flavour-meson case is of particular interest, since a number of issues are prominent: the appearance of different quark masses inside the state, the loss of $\mathcal{C}$-symmetry, and a variety of meson-mass ranges to challenge experiment and theory alike.
The sector of open-flavour mesons composed of a light quark, with mass below the \gls{QCD} scale, and a heavy quark, with mass above the \gls{QCD} scale, bridges the regime of chiral dynamics and heavy-quark symmetries.
It has even been suggested that both aspects amplify each other \cite{Hilger:2008jg,Hilger:2011cq}.
Therefore, open-flavour mesons are attractive probes to study genuine non-perturbative effects of the strong interaction.

From the theory point of view, one needs a thorough understanding of nonperturbative \gls{QCD}, which is sought via different approaches such as the traditional quark model in relativistic forms \cite{Eichten:1979ms,Godfrey:1985xj,Roberts:2007ni,Melde:2008yr,GomezRocha:2012zd,Gomez-Rocha:2014aoa,Li:2015zda,Segovia:2016xqb,Vary:2016emi}, effective field theories \cite{Berwein:2015vca,Pelaez:2015zoa,Terschlusen:2016cfw,Parganlija:2016yxq}, \glspl{QSR} \cite{Shifman:1978bx,Nielsen:2009uh,Narison:2014wqa,Ho:2016owu}, lattice-regularized \gls{QCD} \cite{Dudek:2010wm,Bouchard:2014eea,Lang:2015hza,Hutauruk:2016sug,Lang:2015ljt}, and covariant approaches to \gls{QCD} bound states at various levels of sophistication and aspects pertaining thereto \cite{Eichmann:2016yit,Blaschke:2000gd,Krassnigg:2003wy,Alkofer:2005ug,Eichmann:2007nn,Eichmann:2009qa,Blank:2010bz,Carbonell:2013kwa,Sauli:2014uxa,Biernat:2014xaa,Blum:2014gna,Biernat:2015xya,Bedolla:2015mpa,Raya:2015gva,Binosi:2016rxz,Serna:2016ifh,Maas:2016tci}.

Experimentally, the opportunities are as remarkable as the challenges and dedicated programs are on their way to advance our knowledge in this particular field, e.\,g., \cite{Bettoni:2014vka,Geesaman:2015fha,Olive:2016xmw}.

The \gls{DSBSE} approach treats light and chiral quarks on an equal footing with heavy quarks, and, therefore, constitutes a naturally unified access to both regimes.
Likewise, momenta corresponding to the perturbative and non-perturbative regimes are accessible equally well.
Our background in this manifestly covariant approach has enabled many interesting studies in theoretical hadron physics over the past years, e.\,g., \cite{Maris:1997tm,Maris:2000sk,Burden:2002ps,Maris:2005tt,Maris:2006ea,Bhagwat:2007rj,Eichmann:2008ef,Blank:2010sn,Mader:2011zf,Popovici:2014pha,Sanchis-Alepuz:2014sca,Eichmann:2015cra}, but neglected a comprehensive study of open-flavour mesons so far.
Our present study is the first step towards filling this gap in that we present a complete set of available calculated results that complement comprehensive analogous studies of quarkonia performed recently \cite{Krassnigg:2009zh,Blank:2011ha,Hilger:2014nma,Hilger:2015hka,Hilger:2015ora}.
Here, we present quarkonium results together with and as a basis for the open-flavour results, which makes this paper the first study to comprehensively report all flavour combinations found in a covariant model setup like ours.

Particular aspects of interest for us are the appearance and correspondence of so-called quasi-exotic quark-bilinear meson states as recently discussed in \cite{Hilger:2016efh,Hilger:2016drj}. 
Furthermore and in particular, we present our values for masses, leptonic decay constants, and in-hadron condensates due to the relevance of the topic for in-medium \gls{QCD} \cite{Friman:2011zz,Rapp:2011zz}, and discuss the results with regard to comparison to experimental data as well as estimating systematic uncertainties in our numbers.
Finally, we add a few conceptional and technical musings and improvements as they appeared in the context of our investigation.

\section{Setup}

\subsection{Quark DSE}
\label{sct:dse}

Herein, as is suitable for a comprehensive study, we employ the rainbow-truncated Euclidean \gls{DSE} for the non-perturbative quark propagator, which reads
\begin{subequations}\label{eq:dse}
\begin{align}
    &S(p)^{-1} = Z_2 \, \left({\rm i}\gamma\cdot p + Z_4 \, m_q\right) + \Sigma(p) \,,
    \\\label{eq:dse_sigma}
    &\Sigma(p) = {Z_2}^2 C_\mathrm{F} \, \int^\Lambda_q\!\! \mathcal{G}\!\left((p-q)^2\right) \, D_{\mu\nu}^\mathrm{f}(p-q) \,\gamma_\mu \,S(q)\, \gamma_\nu \,,
\end{align}
\end{subequations}
where $C_\mathrm{F} = {(N_\mathrm{c}^2-1)}/{2N_\mathrm{c}}$ and $N_\mathrm{c} = 3$.
$\Sigma(p)$ is the quark self-energy or mass-shift operator and $D_{\mu\nu}^\mathrm{f}(l) = \left(\delta_{\mu\nu} - {l_\mu l_\nu}/{l^2} \right)$ is the transversal-projector part of the free gluon propagator in Landau gauge.
The effective interaction $l^2 \mathcal{G}(l^2)$ is introduced in place of the combination of gluon propagator and quark-gluon-vertex dressing functions and is intended to suitably imitate the combined effects of omitted quark-gluon vertex terms and the true gluon propagator as well as other dressing functions.
$\int^\Lambda_q = \int^\Lambda \frac{\Dp{4}{q}}{(2\uppi)^4}$ is a translationally invariant Pauli-Villars regularized integration measure \cite{Holl:2003dq} with regularisation scale $\Lambda = 200\,\mathrm{GeV}$ \cite{Maris:1997tm}.
$Z_2$ and $Z_4$ are quark-wave-function and quark-mass renormalisation constants.
The current-quark mass is denoted by $m_q$ and its value given at the renormalisation point $\mu = 19\,\mathrm{GeV}$. 
Further details of the renormalisation procedure and a novel algorithm in the chiral limit are described in \ref{sec:renorm}.

With the decompositions
\begin{equation}\label{eq:dressing}
\begin{split}
    S(p)^{-1} &= {\rm i}\gamma\cdot p \; A(p^2)+B(p^2) \\
    &= Z^{-1}(p^2)\left({\rm i}\gamma\cdot p+M(p^2)\right)
    \\
    &= \left[ -{\rm i}\gamma\cdot p \; \sigma_\mathrm{V}(p^2) + \sigma_\mathrm{S}(p^2) \right]^{-1} \,,
\end{split}
\end{equation}
Eq.\,\eqref{eq:dse} defines a system of inhomogeneous, non-linear, singular, coupled Fredholm integral equations of the second kind for the propagator dressing functions $A$ and $B$ \cite{Hilger:2016drj}.
Depending upon details of the specific interaction, e.\,g.\ strength, multiple solutions of this equation can exist \cite{Williams:2007ey,Wang:2012me}.
In particular, different solutions are possible with regard to a possible dynamical breaking of chiral symmetry in QCD \cite{Hilger:2015zva}.
Most commonly, two different solution strategies are employed: fixed-point iteration and optimisation algorithms \cite{Bloch:1995dd,Krassnigg:2008gd,Dorkin:2013rsa}. 
Typical solutions of the gap equation are plotted, e.\,g., in Fig.\ 4 of \cite{Krassnigg:2016hml}, where usually the emphasis is on \gls{DCSB}, which is a clearly visible feature in the dressing functions independent of the current-quark mass.

\subsection{Effective Interaction}\label{sec:effint}

The effective interaction $\mathcal{G}$ in Eq.~(\ref{eq:dse}) has a long and diverse history, see, e.\,g., the corresponding discussion in \cite{Hilger:2014nma}. 
In essence, one can use the perturbative running coupling of QCD as a basis for this function \cite{Maris:1997tm,Maris:1999nt}.
However, there are two aspects to take care of: the infrared behaviour of the interaction as well as the intermediate-momentum strength and its effect on \gls{DCSB}, which is desired in order to work with a realistic setting. While the details of the effective interaction in the infrared are largely irrelevant for spectroscopy \cite{Blank:2010pa}, the intermediate-momentum regime is, in fact, a dominant factor. 
The correct \gls{UV} behaviour is important in high-energy processes, but \emph{de facto} optional for spectroscopy as well.

Thus, a few very similar effective interactions are currently used in spectroscopic \gls{DSBSE} studies, two of which are used herein.
The simpler \gls{AWW} parametrisation \cite{Alkofer:2002bp} of the interaction reads
\begin{equation}\label{eq:AWW}
    \mathcal{G}_\mathrm{AWW}(q^2) = 4 \uppi^2 D \frac{q^2}{\omega^6} {\rm e}^{-\frac{q^2}{\omega^2}} \, .
\end{equation}
It is \gls{UV} finite, all renormalisation constants are equal to one, and the limit $\Lambda \to \infty$ can be taken initially.
It also simplifies the computational effort somewhat; one can even perform one of the angular integrations exactly \cite{Alkofer:2002bp}.

The correct perturbative limit is ensured by adding the one-loop \gls{UV} term \cite{Maris:1997tm}
\begin{equation}\label{eq:UVmodel}
    \mathcal{G}_\mathrm{UV}(q^2)=\frac{4\uppi^2\;\gamma_\mathrm{m} \;\mathcal{F}(q^2) }{\frac12 \ln
\left[\tau\!+\!\left(1\!+\nicefrac{q^2}{\Lambda_\mathrm{QCD}^2}\right)^2\right]}
\end{equation}
with
\begin{equation}
{\cal F}(q^2)= \frac{1 - \mathrm{e}^{-\nicefrac{q^2}{4 m_\mathrm{t}^2}}}{q^2}\;,
\end{equation} 
where $m_\mathrm{t}=0.5$~GeV,
$\tau={\rm e}^2-1$, $N_\mathrm{f}=4$, $\Lambda_\mathrm{QCD}^{N_\mathrm{f}=4}=
0.234\,{\rm GeV}$, and $\gamma_\mathrm{m}=12/(33-2N_\mathrm{f})$, which is unchanged from Ref.~\cite{Maris:1999nt}.

Using two-loop expressions \cite{Jain:1993qh} is, in fact, unnecessary, since the intermediate-momentum-range model enhancement engulfs the effects beyond the one-loop formula; in addition, the two-loop form's singularity structure is more complicated and does not benefit the numerical procedures.
The \gls{MT} interaction model is parametrized as \cite{Maris:1999nt}
\begin{equation}\label{eq:MT}
    \mathcal{G}_\mathrm{MT}(q^2) = \mathcal{G}_\mathrm{AWW}(q^2) + \mathcal{G}_\mathrm{UV}(q^2)\;.
\end{equation}

We use these two model interactions side by side to capture prominent cases of model dependences as well as to determine possible effects of the \gls{UV} part, as described below.

\subsection{Meson BSE}
\label{sct:bse}

The quark-bilinear setup of the \gls{BSE} contains the quark-antiquark scattering kernel, which is an unknown in QCD. 
Thus, in an effective study, guidance on or restrictions of the kernel's form are welcome.
In fact, the relations used to provide such guidance are not only welcome and convenient, they are necessary constraints in order to respect the fundamental symmetries of the underlying theory and correctly implement them in the model computations. 
The most prominent example for light-meson physics is the would-be chiral symmetry of QCD, in particular including \gls{DCSB}. 
As a result of requiring to correctly implement these, the kernels of \gls{DS} and \gls{BS} equations are related via the \gls{AVWTI} \cite{Munczek:1994zz}.
This can be shown to hold for the \gls{RL} truncation \cite{Eichmann:2008ae} in the \gls{DSBSE} approach, which is the first term in a systematic truncation scheme explored in some detail for a very simple model interaction \cite{Bender:2002as,Bhagwat:2004hn,Holl:2004qn,Matevosyan:2006bk,Matevosyan:2007cx,Gomez-Rocha:2014vsa,Gomez-Rocha:2015qga,Jinno:2015sea,Gomez-Rocha:2016cji} or more complicated situations, e.\,g., \cite{Fischer:2009jm,Chang:2009zb,Williams:2015cvx}.

The \gls{RL} truncated Euclidean homogeneous meson \gls{BSE}  reads
\begin{multline}\label{eq:BSE}
    \Gamma(p;P) = -C_\mathrm{F}\, {Z_2}^2\!\int^\Lambda_q\!\!\mathcal{G}((p-q)^2)\; D_{\mu\nu}^\mathrm{f}(p-q)
        \;\chi(q;P) \,.
\end{multline}
$\Gamma(p;P)$ is the \gls{BSA} and 
\begin{equation}\label{eq:bsa-bsw}
\chi(q;P) \equiv S_1(q_+) \, \Gamma(q;P) \, S_2(q_-)
\end{equation}
is the \gls{BSW}, with the dressed-quark propagators $S_1$ and $S_2$ and their arguments being 
$q_+ = q +\eta P$, $q_- = q-(1-\eta)P$.

The solution strategy makes use of a covariant basis and a subsequent numerical treatment. 
We follow the methods and details given in Refs.\ \cite{Dorkin:2010ut,Blank:2010bp,UweHilger:2012uua,Dorkin:2013rsa,Hilger:2015ora}.
In particular, the \gls{BS} amplitude is expanded in a finite set of covariants $T_n(p,P,\gamma)$ \cite{Krassnigg:2010mh,Fischer:2014xha}, which specify the quantum numbers $J$ and $\mathcal{P}$, according to
\begin{equation}\label{eq:bsadecomp}
    \Gamma(p;P) = \sum_{n} \Gamma_{n}\!\!\left(p^2,\cos\sphericalangle(P,p);P^2\right) \, T_n(p,P,\gamma) \; ,
\end{equation}
with $\Gamma_{n}\!\!\left(p^2,\cos\sphericalangle(P,p);P^2\right)$ being the partial amplitudes, and the dependence on $\gamma$ refers to the dependence on constructions involving the four-vector of Dirac $\gamma$ matrices.
Equation \eqref{eq:BSE} is then projected via appropriate traces onto the covariants $T_n(p,P,\gamma)$, yielding $4$ ($J=0$) or $8$ ($J\geq 1$) coupled equations for the partial amplitudes $\Gamma_{n}\!\!\left(p^2,\cos\sphericalangle(P,p);P^2\right)$.
The respective traces have been generated and implemented via a unified and automatable framework, which likewise prevents implementation errors in the rather long expressions for higher $J$.
We solve Eq.~\eqref{eq:BSE} by evaluating the determinant via lower-upper factorization and by determining the eigenvalue spectrum for the largest $10$ eigenvalues and eigenvectors via the implicitly restarted Arnoldi method.
Canonical normalisation, leptonic decay constants, residual and (generalized) \gls{GMOR} relation are evaluated according to \cite{Maris:1999nt}; cf.\ \ref{sec:spice}.

We would like to point out that we have employed advanced technniques for the solution of the \gls{DSE} (see \ref{sec:DSEcontour}) as well as a careful and thorough definition of consistency and plausibility checks at intermediate and, in particular, at the level of \gls{BSE} on-shell results. 
This allows for the design and development of a unified and rigorous numerical \gls{DSBSE} framework to automatically evaluate meson properties for virtually any set of quantum numbers $J^{\mathcal{P}(\mathcal{C})}$.
In addition, this can be achieved with minimal need for manual interference and minimal setup expenses, and can be run on off-the-shelf desktop computers.

Note that we do not expand the angular dependence of the $\Gamma_{n}$ in, e.\,g., Chebyshev polynomials \cite{Krassnigg:2003dr}, but retain the full angular dependence in our calculation \cite{Maris:1997tm,Maris:1999nt,Bhagwat:2006pu}. 
This difference is more of a conceptual kind than a numerical one, since in a heavy-light meson a high number of Chebyshev polynomials is expected to be needed in order to properly capture the angular structure of the state \cite{Alkofer:2002bp}. 
As a result, we expect the number of necessary Chebyshev moments to be similar to the number of necessary angular integration points.
This numerical complication, however, simplifies the systematisation and ensures the full Poincar\'e covariance of the approach.
Indeed, it turns out that the $\eta$-dependence of the bound-state pole in $P^2$ is below the numerical accuracy, i.\,e.\ of the same order as the numerical tolerance of the $P^2$ determination itself.
This is the key to utilizing the $\eta$-shift method in order to exclude quark-propagator poles from the \gls{BSE} integration domain and to increase the applicable pole threshold.

\subsection{Spice Ingredients}\label{sec:spice}

According to \cite{Blank:2011qk}, charge-conjugation symmetry of an equal-flavour meson,
$[C \Gamma(-p;P) C^{-1}]^\mathrm{T} = c \;\Gamma(p;P)$, with the charge-conjugation matrix $C = \mathrm{i} \gamma_4\gamma_2$, translates for the \gls{BS} partial amplitudes to
\begin{equation} \label{eq:BSAparity}
    \Gamma_{n}\!\!\left(p^2,\cos\sphericalangle(P,p);P^2\right) = \bar \xi_n \Gamma_{n}\!\!\left(p^2,- \cos\sphericalangle(P,p);P^2\right) \, ,
\end{equation}
where
$c = \bar \xi_n \xi_n~\forall n$ (no summation over $n$ here) must hold, and the covariants transform under charge conjugation as $[ C T_n(-p,P,\gamma) C^{-1}]^\mathrm{T} = \xi_n T_n(p,P,\gamma)$.
After projecting onto a suitable set of covariants and discretization of the integrals, the meson \gls{BSE}, Eq.~\eqref{eq:BSE}, assumes the form
\begin{multline} \label{eq:BSEprojected}
    \Gamma_{n}\!\!\left(p^2,t_p;P^2\right)
        \\ = \sum_{m, q^2,t_q} K_{nm} (p^2, t_p, q^2, t_q; P^2) \Gamma_{m}\!\!\left(q^2,t_q;P^2\right) \, ,
\end{multline}
with $t_k \equiv \cos\sphericalangle(P,k) = P\cdot k / \sqrt{P^2 k^2}$ determining the hyper-angle and the $6$-dimensional \gls{BS} matrix $K$.
Equation \eqref{eq:BSEprojected} contains all $\mathcal{C}$-parity states, but doesn't imply that they are coupled to each other. These can be identified a posteriori by virtue of Eq.~\eqref{eq:BSAparity}.
However, by solving instead
\begin{multline} \label{eq:BSEprojectedC}
    \Gamma_{n}\!\!\left(p^2,t_p;P^2\right)
        \\ = \frac12\sum_{m, q^2,t_q}
        \Bigg[ K_{nm} (p^2, t_p, q^2, t_q; P^2)
        \\+ \left(\bar \xi_n\right)^{-1} K_{nm} (p^2, -t_p, q^2, t_q; P^2) \Bigg]
        \Gamma_{m}\!\!\left(q^2,t_q;P^2\right) \, ,
\end{multline}
the solutions are restricted a priori to states with certain $\mathcal{C}$-parity.
Technically, Eq.~\eqref{eq:BSEprojectedC} is simpler to implement than restricting expansions of the \gls{BSA}'s angular dependence to particular subsets of, e.\,g., Chebyshev polynomials as this amounts to a generic low-level array operation.

Once the meson \gls{BSE} and the quark \gls{DSE} have been solved consistently and the bound-state mass $M$ has been found, one can obtain the leptonic decay constant of a pseudoscalar meson by projection onto the axialvector current \cite{Maris:1997hd}:
\begin{equation}\label{eq:f0-}
    f_{0^-} = N_\mathrm{c} {Z_2} \left. \int^\Lambda_q \frac{\mathrm{Tr}[\gamma_5 \gamma\!\cdot\! P \; \chi^{0^-}\!\!(q;P) ]}{P^2} \right|_{P^2 = - M_{0^-}^2}
    \,,
\end{equation}
with the pseudoscalar \gls{BSW}
$\chi^{0^-}\!\!(q;P)$
and $\Gamma^{0^-}\!\!(q;P)$ the corresponding \gls{BSA}, the solution of Eq.~(\ref{eq:BSE}).
Comparing residues of a flavour-nonsinglet pseudoscalar pole in the dressed pseudoscalar and axialvector vertices as they appear in the \gls{AVWTI}, one arrives at the generalized \gls{GMOR} relation \cite{Maris:1997tm,Maris:1997hd,Holl:2004fr}, which reads
\begin{equation} \label{eq:GMOR}
    f_{0^-}\;M_{0^-}^2 = (m_q+m_{\bar{q}^\prime})\;r_{0^-} \,.
\end{equation}
The residue $r_{0^-}$ of the pseudoscalar's mass pole in the pseudoscalar vertex is given by \cite{Maris:1997tm}
\begin{equation}\label{eq:r0-}
    \mathrm{i}\;r_{0^-} = N_\mathrm{c} {Z_4} \left. \int^\Lambda_q \frac{\mathrm{Tr} [\gamma_5 \; \chi^{0^-}\!\!(q;P) ]}{\sqrt{2}} \right|_{P^2 = - M_{0^-}^2} \,.
\end{equation}
Equation \eqref{eq:GMOR} is valid for all current-quark masses and flavour-nonsinglet pseudoscalar states, including excitations as well as exotic states, where it is trivially satisfied by $f_{0^{--}}=r_{0^{--}}=0$.
The latter is a consequence of the antisymmetry of the corresponding integrands in Eqs.~(\ref{eq:f0-}) and (\ref{eq:r0-}) and is detailed further in \ref{sec:f0--}.
Thus, it provides a natural definition of the chiral condensate which is valid beyond the chiral limit \cite{Holl:2004fr,Roberts:2012sv}:
\begin{equation} \label{eq:qq}
    \langle \ld \bar q q \rd \rangle \equiv - f_{0^-} \: r_{0^-} = -\frac{f_{0^-}^2 M_{0^-}^2}{m_q+m_{\bar{q}^\prime}}\,.
\end{equation}
In the scalar-meson case, one has an analogous flavour-nonsinglet relation \cite{Qin:2011xq}, which generalises Eq.~\eqref{eq:GMOR} to
\begin{equation}\label{eq:genericGMOR}
f_{0^\mathcal{P}}\; M_{0^\mathcal{P}}^2=(m_q- \mathcal{P} \; m_{\bar{q}})\; r_{0^\mathcal{P}}\;,
\end{equation}
implying that all flavour-nonsinglet scalar quarkonia, conventional or exotic, have a vanishing leptonic decay constant \cite{Maris:2000ig,Bhagwat:2006py}.
It is noteworthy that Eq.~\eqref{eq:genericGMOR} is fulfilled to very high numerical accuracy ($10^{-3}$) on the mass shell.
In analogy to the two cases for $f$ defined above, the decay constants of the vector and axialvector mesons are obtained by projecting the \gls{BSA} onto vector and axialvector currents,
\begin{subequations}
\begin{align}
    f_{0^+} &= N_\mathrm{c} {Z_2} \left. \int^\Lambda_q \frac{\mathrm{Tr}[\gamma\!\cdot\! P \; \chi^{0^+}\!\!(q;P) ]}{P^2} \right|_{P^2 = - M_{0^+}^2}
    \,,
    \\
    f_{1^-} &= N_\mathrm{c} {Z_2} \left. \int^\Lambda_q \frac{\mathrm{Tr}[\gamma_\mu \chi^{1^-}_\mu\!\!(q;P) ]}{3 P^2} \right|_{P^2 = - M_{1^-}^2}
    \,,
    \\
    f_{1^+} &= N_\mathrm{c} {Z_2} \left. \int^\Lambda_q \frac{\mathrm{Tr}[\gamma_5\; \gamma_\mu \chi^{1^+}_\mu\!\!(q;P) ]}{3 P^2} \right|_{P^2 = - M_{1^+}^2}
    \,,
\end{align}
\end{subequations}
and \cite{Bhagwat:2006py}
\begin{equation}
    \mathrm{i}\, r_{0^+} = N_\mathrm{c} {Z_4} \left. \int^\Lambda_q \frac{\mathrm{Tr}[\chi^{0^+}\!\!(q;P) ]}{P^2} \right|_{P^2 = - M_{0^+}^2}
    \,.
\end{equation}

\subsection{In-Hadron Condensate}\label{sec:InHadronCond}

The pseudoscalar in-hadron chiral condensate defined in Eq.~\eqref{eq:qq} arises in the course of the derivation of the generalized \gls{GMOR} relation, Eq.~\eqref{eq:GMOR} \cite{Cloet:2013jya,Horn:2016rip}.
The latter one is satisfied by all quark bilinear pseudoscalar mesons which are subject to the \gls{AVWTI}, irrespective of quark content or quark-mass configuration.

In the chiral limit, comparison to the traditional \gls{GMOR} reveals that the thus defined in-hadron chiral condensate reduces to its current-algebra, or \gls{QSR}, counterpart, cf.\ Eq.~\eqref{eq:qq} \cite{Brodsky:2010xf,Chang:2011mu,Brodsky:2012ku}.
An appealing advantage of the in-hadron chiral condensate compared to other definitions in the \gls{DSBSE} context \cite{Williams:2006vva,Williams:2007ef,Williams:2007ey,Fischer:2014ata} is that it is well-defined, i.\,e., it exists for all pseudoscalar quark-bilinear meson states and is unique in the following sense.

The most common alternative definitions used beyond the chiral limit are solely based on the quark propagator itself and necessitate the subtraction of a scaled quark propagator with the correct \gls{UV} limit, in order to cancel an inherent and characteristic quadratic divergence of the momentum integral over the scalar projection of the quark propagator beyond the chiral limit.
Employing the correct scaling, any propagator with the correct \gls{UV} limit can be used, such as quark propagators for different quark masses or \gls{WW}-phase solutions to the \gls{DSE}.

However, due to the non-linearity of the quark \gls{DSE}, none of these linear constructions represent a solution of the quark \gls{DSE}.
Moreover, in the spirit of such a linear construction any arbitrary function with the correct \gls{UV} limit would, strictly speaking, suffice.
In principle, subtracting the \gls{WW}-phase propagator of the same quark flavour is the least ambiguous and therefore most preferable approach among the subtraction schemes \cite{Williams:2006vva,Williams:2007ef,Williams:2007ey,Fischer:2014ata}, because it is sufficient to render the di-quark condensate finite without the need for rescaling and without quark-flavour ambiguity. 
Nonetheless, most commonly an \gls{NG}-phase propagator in the strange-quark-mass regime is used.
The apparent caveat with the \gls{WW}-phase subtraction is that at high densities and temperatures the quark \gls{DSE} has no \gls{WW}-phase solution in some model interactions.
In general, each subtraction choice results in a different numerical value for the condensates, which is the governing ambiguity.
As the chiral condensate is of paramount importance in predictions of the \gls{DCSB} phase transition at large temperatures and densities \cite{Fischer:2014ata}, a definition which is free of such ambiguities is very tempting.

Furthermore, the notion of in-hadron condensates provides the opportunity to overcome the problem of a cosmological constant which is by orders of magnitude too large when the vacuum energy is induced by the \gls{QCD} chiral condensate \cite{Brodsky:2008xu,Brodsky:2009zd}.
With condensates being non-zero only inside hadrons, a zero chiral condensate outside of the hadron, strictly speaking, does not necessitate a restoration of the chiral symmetry; see the discussion and a collection of references in \cite{Hilger:2015zva}.

On a different note, the definition \eqref{eq:qq} indicates that in-hadron chiral condensates may keep remembrance of their hadronic origins, since they are different for different hadrons (flavour content, $\mathcal{C}$-parity, or excitation), which is in contrast to the usual \gls{QSR} assumption.
Within the latter, the \gls{OPE} aims at a neat separation of long- and short-range physics by absorbing inherent divergences of local operator products into \gls{QCD} condensates while encoding the perturbative part of the interaction in Wilson coefficients \cite{Shifman:1978bx}.
In this spirit, the  employed \gls{QCD} condensates are universal constants which quantify the complicated dynamical structure of the non-perturbative \gls{QCD} vacuum, and they do not vary for different hadrons.
While a space dependence might still be conceivable, these condensates are independent of the system under consideration, in particular, of excitation or charge-conjugation parity $\mathcal{C}$.
It is not yet clear how the picture of in-hadron condensates and \gls{QSR} condensates can be brought to a mutual agreement, nor if they indeed disagree.

Another issue concerns the heavy-quark mass expansion \cite{Shifman:1978by,Buchheim:2015yyc,Buchheim:2014uda} which allows to expand a charm- or bottom-quark condensate into a power series in the inverse heavy-quark mass with gluon condensates as coefficients.
For example, the charm-quark condensate is an order of magnitude smaller than the chiral condensate.
It often suffices to regard heavy quarks as static and to neglect heavy-quark condensates in heavy-quarkonium \gls{QSR} evaluations, which leaves the integrated spectral properties of states dominated by gluon condensates and the perturbative contribution \cite{Morita:2007pt,Hilger:2010zb}.

An interesting sector in between the chiral and heavy-quark regimes is the sector of open-flavour mesons such as the $D$-meson \cite{Hilger:2009kn,Hilger:2010zb,Hilger:2012db,Hilger:2011cq}.
For these states an infrared-finite \gls{OPE} requires the introduction of non-normal ordered condensates, which results in a cancellation of heavy-quark-mass expansion and renormalization contributions \cite{Zschocke:2011aa,UweHilger:2012uua}.

Future investigations should shed light on these issues in the scope of the notion of in-hadron condensates.
Finally, having a unique in-hadron condensate definition at hand for the chiral condensate immediately raises the question for analogous definitions for other condensates, e.\,g., four-quark condensates \cite{Buchheim:2015xka,Buchheim:2014rpa}, which can likewise be order parameters of \gls{DCSB} and which are of paramount importance for light-quark mesons and baryons and their properties under restoration of \gls{DCSB} \cite{Thomas:2006nk,Thomas:2007gx,Thomas:2007es,Hilger:2010cn}, or generalisations.
In this work, we present in-hadron condensates for all $0^{-(\pm)}$-states that can be analysed at the moment within the \gls{DSBSE} approach.
It is the first investigation covering all flavour combinations, charge parities, and excitiations which can be found below the pole threshold.


\subsection{Open-Flavour States in Quark-Mass-Independent Interaction Models}
\label{sct:independent}

The generalisation of any \gls{RL} setup as described above is straight-forward as long as the effective interaction does not change with the quark mass.
Simply put, the effective interaction in the \gls{BSE} kernel is connected to the dressed \gls{QGV}, which appears on one end of the effective dressed-gluon interaction only and thus, if the effective interaction is different for each quark propagator flavour, there is no clear correspondence here. 
We discuss such a situation below in \ref{sec:intdependent} in more detail. 
In the following, however, we assume the same interaction parameters for all quark flavours and masses as it was always the case before in the literature in studies of open-flavour states \cite{Jain:1993qh,Maris:1997tm,Maris:1999nt,Maris:2000sk,Jarecke:2002xd,Alkofer:2002bp,Horvatic:2007qs,Qin:2011xq,Bashir:2012fs,Fischer:2014xha,Fischer:2014cfa,Rojas:2014aka,Serna:2016kdb,El-Bennich:2016qmb,Hilger:2016efh,Hilger:2016drj}.

In a standard numerical setup, where one aims at a solution of the homogeneous \gls{BSE}, it is necessary to analyse the analytical structure of the dressed quark propagators and, in particular, their dressing functions. 
This can be done as demonstrated in \cite{Dorkin:2013rsa} by Cauchy's argument principle or by utilising a Newton-Krylov root finding method.
In the \gls{NG} phase, the quark propagator has a tower of complex conjugated poles off the real axis \cite{Dorkin:2013rsa}.
In such a case, standard numerical treatments are limited in bound-state mass by the appearance of the non-analytic structure closest to the sampling domain in the complex $p^2$ plane.
In \ref{sec:DSEcontour}, we detail our algorithm to uniquely locate this closest quark-propagator singularity in the complex $p^2$ plane and show how to maximally expand the parabolic contour on which the quark propagator is evaluated.

The sampling area necessary in the complex $q_{\pm}^2$ plane dictated by the kinematics in the \gls{BSE} is parabolic and its dimension is given by the maximal bound state's mass $M_{\mathrm{max}}$ squared in combination with the momentum-partitioning parameter $\eta$. 
While $\eta=0.5$ is ideal for quarkonia, the situation for unequal-mass constituents can be assessed by defining the maximal bound-state parameter $\chi_{\mathrm{max}}(m_{\bar{q}^\prime},m_q) = \left( \eta M \right)_\mathrm{max}$.

\begin{figure}
\includegraphics[width=.48\textwidth]{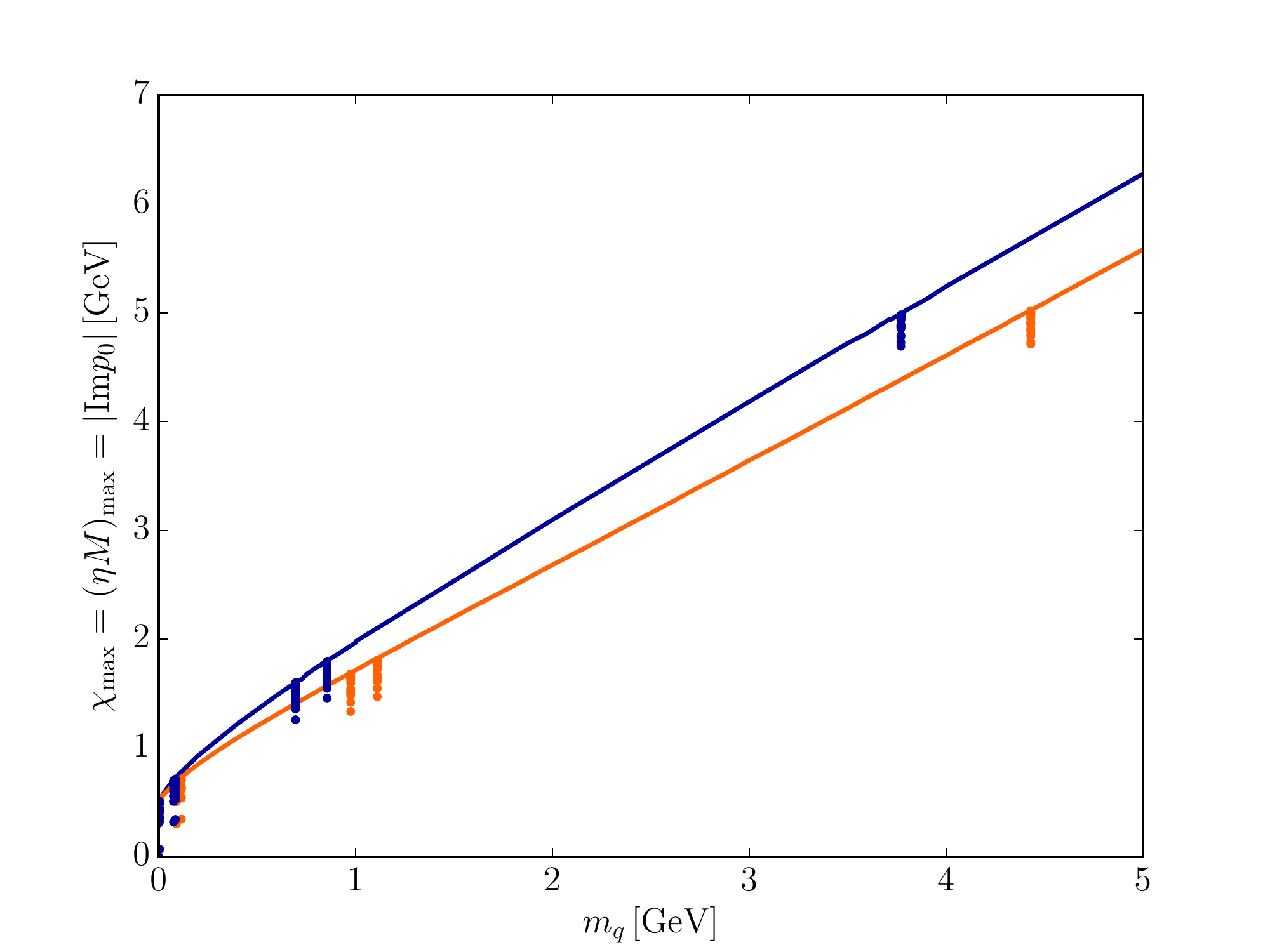}
\caption{The maximal bound-state parameter $\chi_{\mathrm{max}}(m_{\bar{q}^\prime},m_q) = \left( \eta M \right)_\mathrm{max}$ that is employable while solving the homogeneous \gls{BSE} for quarkonia (equal-flavour mesons) as a function of the participating quark masses.
Orange curve: \gls{AWW} model~\eqref{eq:AWW}; blue curve: \gls{MT} model~\eqref{eq:MT}.
The dots mark all quarkonia bound-state masses $M/2$ which have been found below the pole threshold at that particular quark-mass value.}
\label{fig:chimax}
\end{figure}

\begin{figure*}
\includegraphics[width=.48\textwidth]{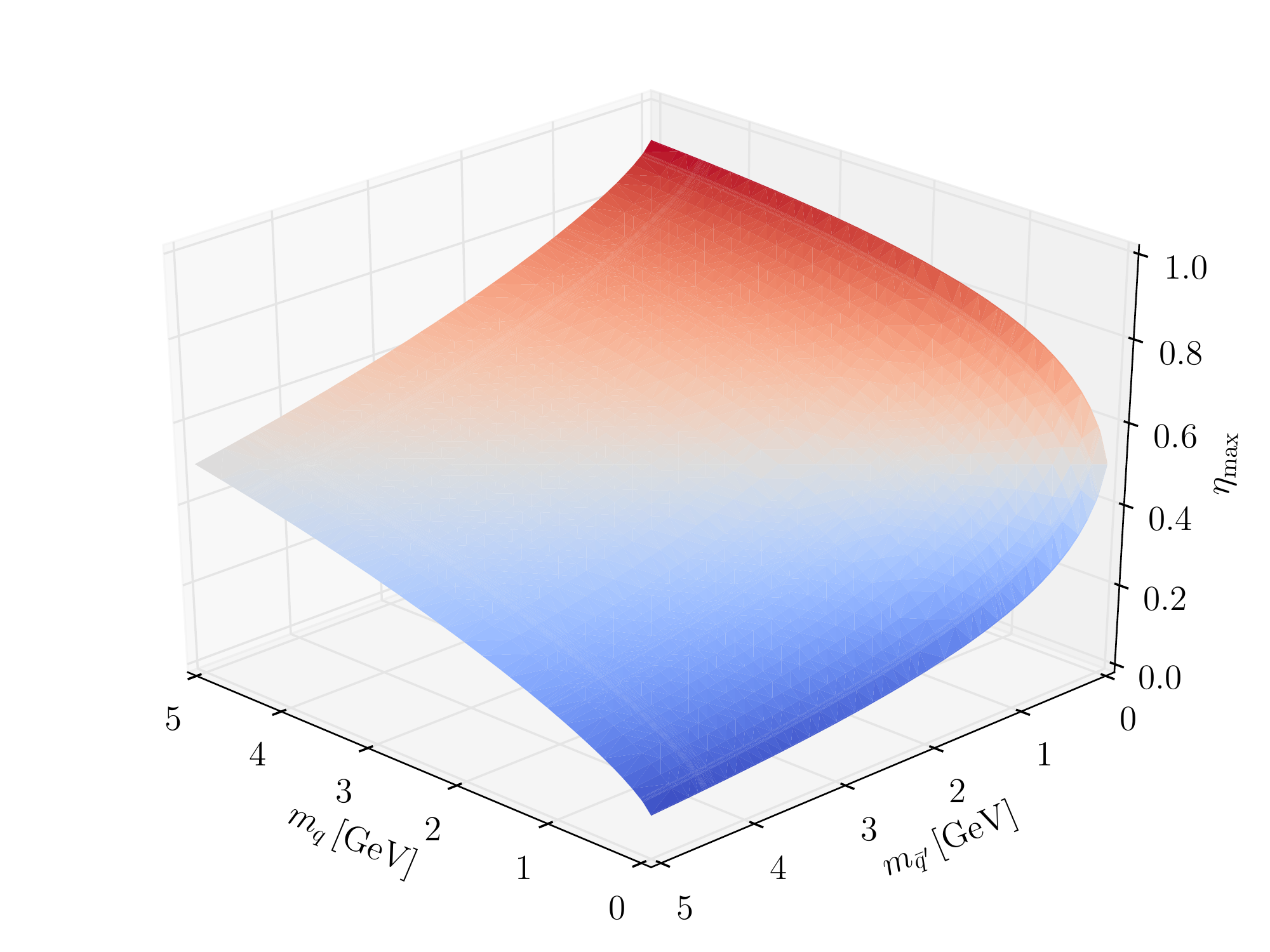}
\includegraphics[width=.48\textwidth]{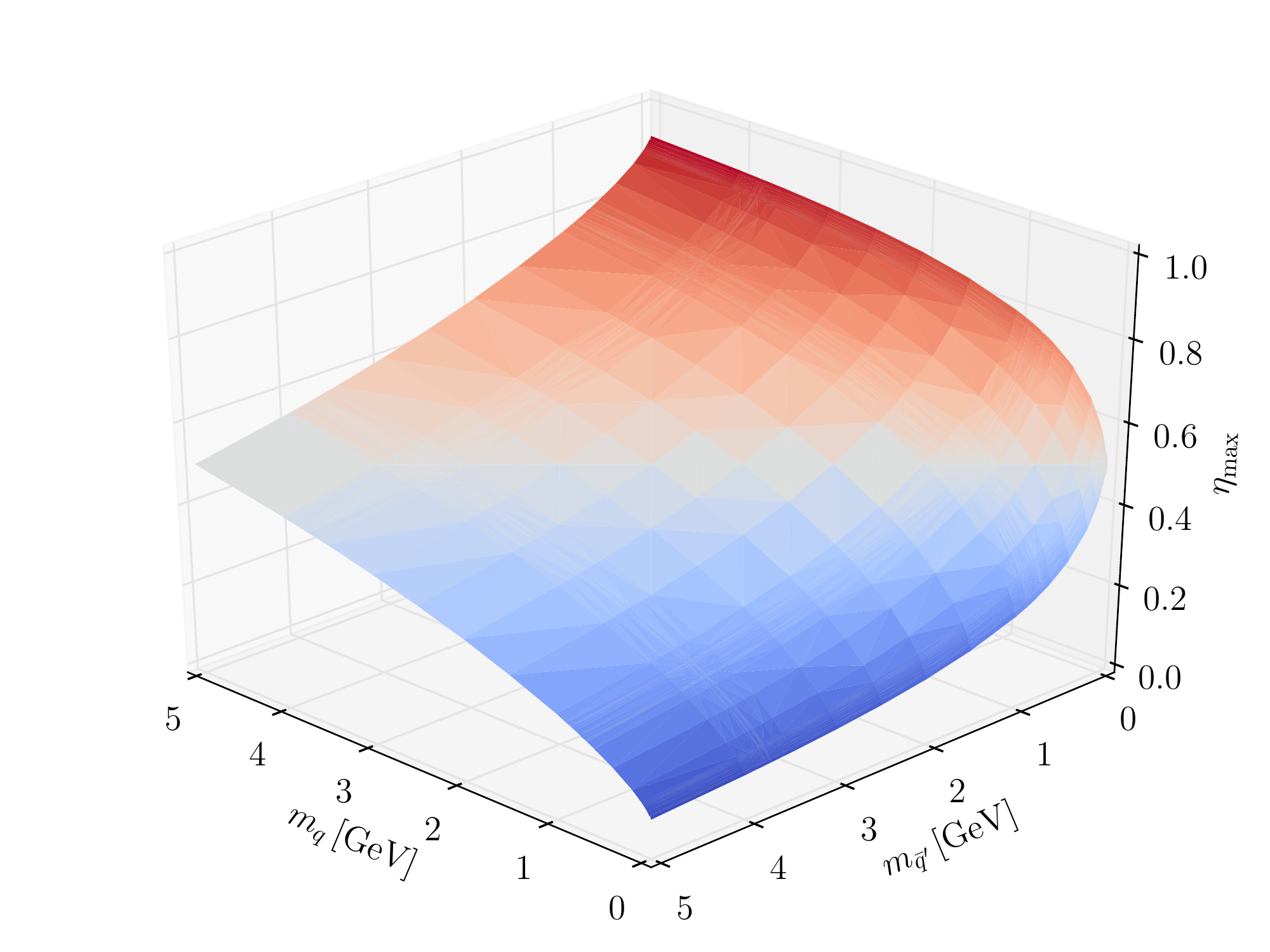}
\caption{The momentum-partitioning parameter $\eta_{\mathrm{max}}(m_{\bar{q}^\prime},m_q)$ as a function of the participating quark masses to be employed in order to maximize the pole threshold.
Left panel: \gls{AWW} model \eqref{eq:AWW}; right panel: \gls{MT} model \eqref{eq:MT}.}
\label{fig:etamax}
\end{figure*}

\begin{figure*}
\includegraphics[width=.48\textwidth]{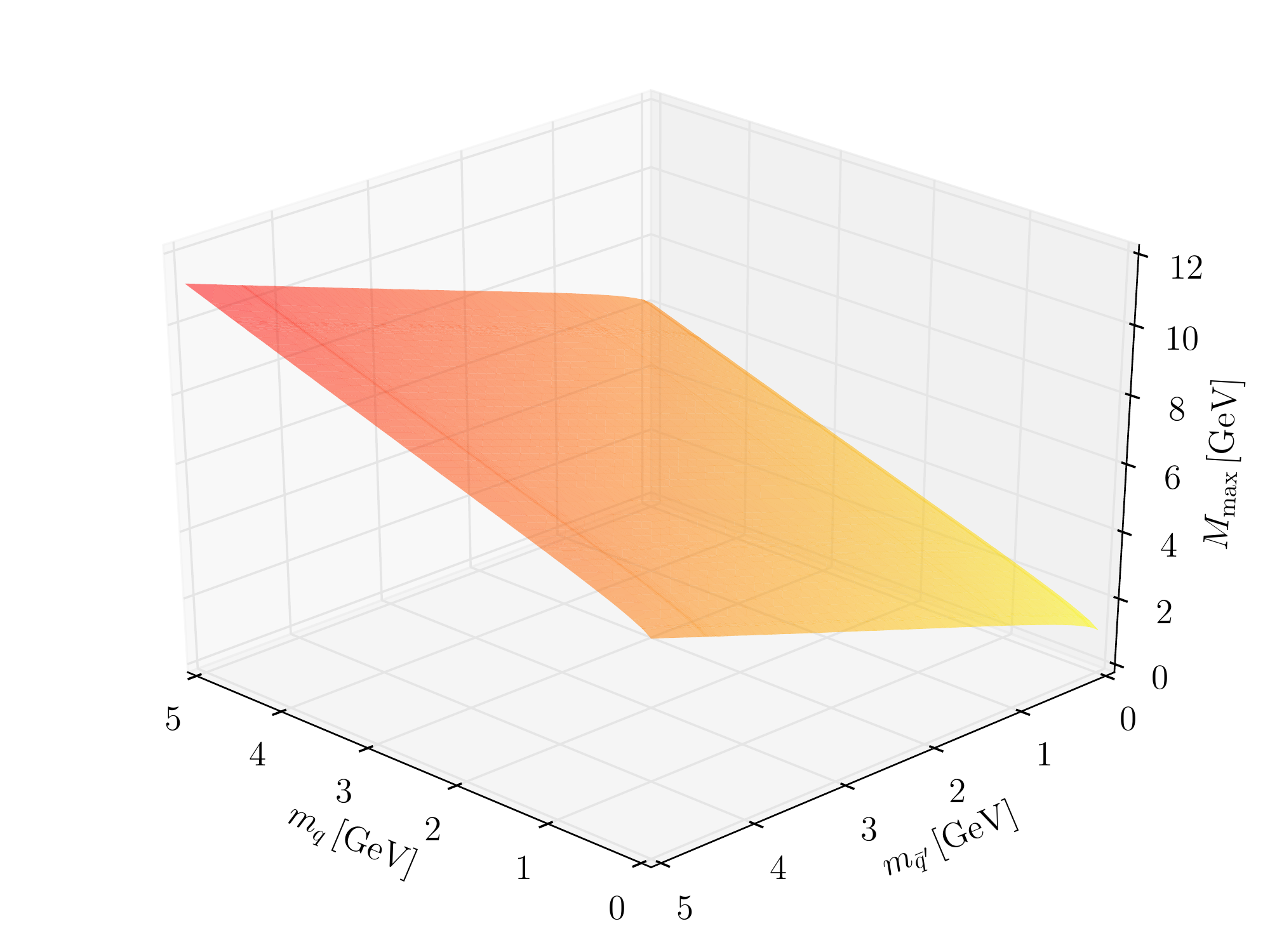}
\includegraphics[width=.48\textwidth]{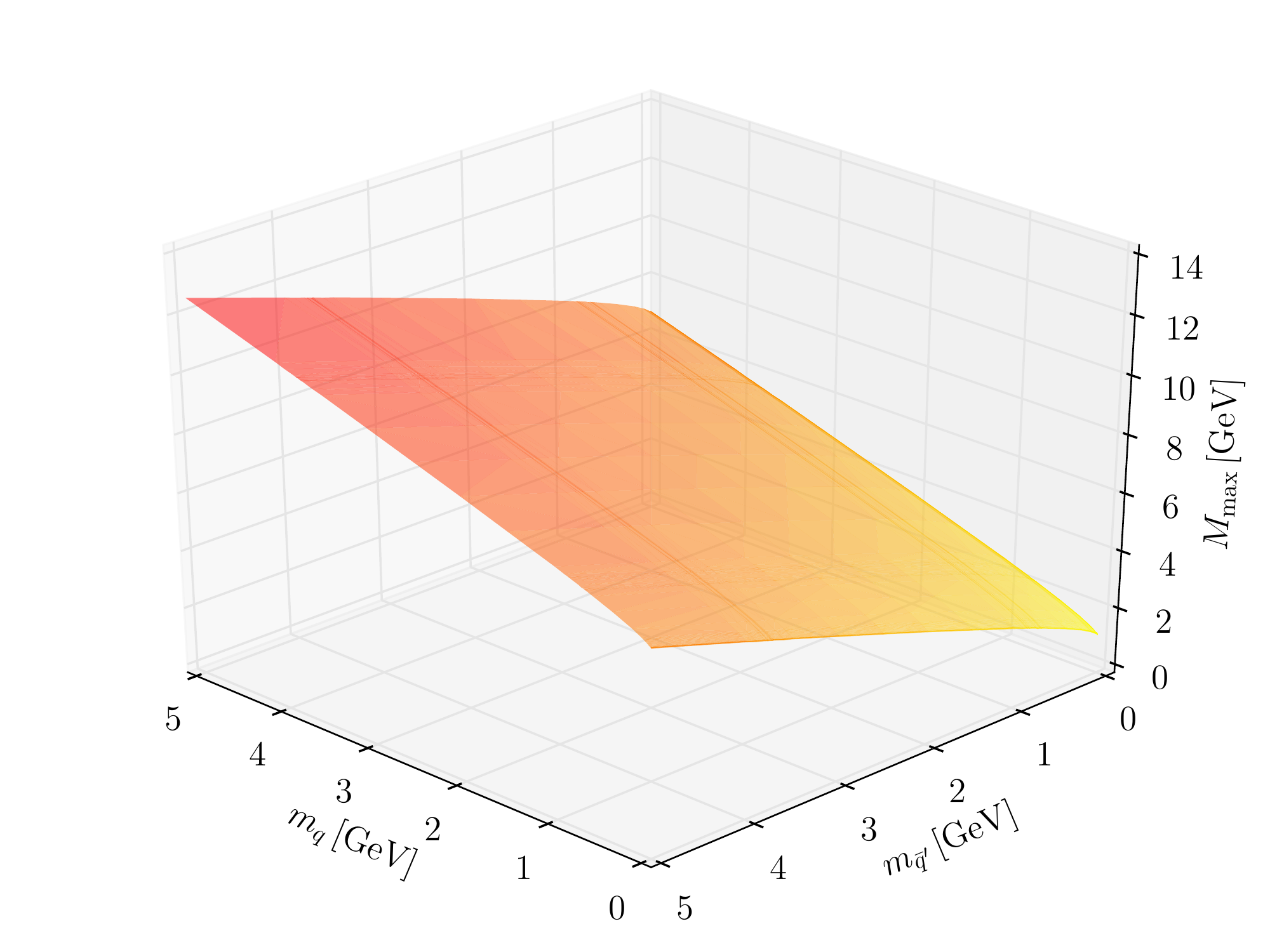}
\caption{The actual maximally accessible bound-state mass $M_{\mathrm{max}}(m_{\bar{q}^\prime},m_q)$ as a function of the participating quark masses.
Left panel: \gls{AWW} model \eqref{eq:AWW}; right panel: \gls{MT} model \eqref{eq:MT}.}
\label{fig:Mmax}
\end{figure*}

Before turning to an open-flavour case, we exhibit the maximal parabola parameter $\chi$ for the two employed models in Fig.~\ref{fig:chimax} as achievable for quarkonia.  
While solving the homogeneous \gls{BSE} for timelike bound-state momenta, $\eta \sqrt{-P^2} < \chi_\mathrm{max}$ and $(1-\eta) \sqrt{-P^2} < \chi_\mathrm{max}$ must hold in order to keep the quark-momentum domains free from non-analyticities.
Therefore, $\chi_\mathrm{max}$ uniquely determines the pole threshold for any quark-mass combination.
As visible in Fig.~\ref{fig:chimax}, the pole threshold for the \gls{MT} model \eqref{eq:MT} is larger over the whole range than for the \gls{AWW} model \eqref{eq:AWW}.
At the same time, also the bound-state masses are larger, which points to the intricate relation between genuine \gls{DSE} properties and solutions of the homogeneous \gls{BSE}.
Note that for each model, the fitted quarkonia are the $1^{--}$ mesons, which have the same bound-state mass up to the fitting accuracy related to a numerical tolerance for the fitted quark masses of $\Delta m_q = 5\,\mathrm{MeV}$.
Note as well that, since the $1^{--}$ states are neither the lightest nor the heaviest ones, they can not be clearly identified in this plot but are part of the tower of dots representing the spectra at each quark mass.
It is clearly demonstrated that with the presented setup one can indeed compute reliably very closely to the pole threshold.

In the open-flavour case, for each quark-mass combination, there is exactly one momentum-partitioning parameter, $\eta_{\mathrm{max}}(m_{\bar{q}^\prime},m_q)$, see Fig.~\ref{fig:etamax} and \ref{sec:DSEcontour}, which maximizes the pole threshold by extending the quark-momentum domains to the limit where poles of both quark propagators just do not enter the respective quark-momentum domain.
Both $\eta$-surfaces reveal the same qualitative behaviour, although curvature and steepness are enhanced for the \gls{MT} compared to the \gls{AWW} model, as one can expect from Fig.~\ref{fig:chimax}.
Similarly, the resulting maximal actually reachable bound-state mass, $M_{\mathrm{max}}(m_{\bar{q}^\prime},m_q)$, is depicted in Fig.~\ref{fig:Mmax}.
One can see that a bound-state mass far above the conventional pole threshold $M_{\mathrm{max}} = 2\,\mathrm{min}(\chi_1, \chi_2)$, where $\chi_i$ is the parabola parameter of quark $i$, is easily reachable for open-flavour mesons in this setup.

In the following, we present a comprehensive body of results for both interaction models as obtained below the respective pole thresholds.
Note, that the presented data set is the first to contain all states which are accessible below the pole threshold for this setup.
As such, this also represents a benchmark for what is currently available in this model.
More states and with better agreement to experimental data may be found for other parameters or more enhanced effective interactions.
Further details on the solution process of the quark \gls{DSE} in the complex plane of quark-momentum squared and aspects of handling the parabolic sampling domain can be found in \ref{sec:DSEcontour}.

\section{Results}\label{sec:results}

\subsection{Details of Presentation in Plots}\label{sec:plotdescription}
In the following subsections, we present plots comparing several sets of calculated data among each other as they are available for each flavour combination. 
The comparison of combined calculated values with appropriate error bars is presented below in Sec.~\ref{sec:exp}.
For each figure, we plot both the mass and the leptonic decay constant in $\mathrm{GeV}$ in terms of $J^{\mathcal{P(C)}}$ and excitation for all mesons that appear as solutions of the homogeneous \gls{BSE} below the respective pole threshold.
These thresholds are marked in each $J^{\mathcal{P(C)}}$ channel by small horizontal blue lines towards the top of the figure right above the bound-state mass of the corresponding quark-mass configuration.

There are two basic approaches to each flavour combination.
As a start, the light-quark mass is fixed by the pion mass.
Then, for the first approach, we fix the other quark masses by fitting the pseudoscalar ground-state mass of an open-flavour meson.
For the second approach, we fix the other quark masses to fit the vector quarkonium ground state in each case.
The results from these two quark-mass configurations are different and give us a handle on systematic errors due to the truncation.
In particular, while corrections to \gls{RL} truncation are expected to be more moderate in both the vector and the pseudoscalar quarkonium channels \cite{Bender:1996bb,Bhagwat:2004hn}, the imbalance in an open-flavour meson can play a large role regarding cancellation effects and thus lead to a pronounced truncation effect \cite{Gomez-Rocha:2015qga,Gomez-Rocha:2016cji}.
Note that results for combinations involving chiral quarks are presented as well to complete the picture and investigate the situation in the chiral limit.

For the results, we use the following systematics of presentation, also detailed in the legend on the right side in Fig.~\ref{fig:AWWqq}:
The bound-state masses are plotted in the left half of each quantum-number block against the left-axis scale with blue marker edges, leptonic decay constants are plotted in the right half of each quantum-number block against the logarithmic right-axis scale with red marker edges; note that this scale is linear below $0.1\,\mathrm{GeV}$.
Different marker shapes refer to the different quark-mass configurations as indicated in each case in the title of the figure.
The marker fill colours indicate the eigenvalue (ordered by their real parts, beginning with the largest) which generates the bound state in each $J^{\mathcal{P(C)}}$ channel: blue -- \nth{1} eigenvalue (ground state), orange -- \nth{2} eigenvalue (first excitation), white -- \nth{3} eigenvalue (second excitation), and yellow -- \nth{4} eigenvalue (third excitation).

The presentation of our results is divided into two larger parts, one for each effective interaction investigated here. 
The results obtained with the \gls{AWW} effective interaction, Eq.~(\ref{eq:AWW}), are presented and carefully discussed first in Sec.~\ref{sec:awwres}.
Thereafter, we present the results obtained with the \gls{MT} effective interaction, Eq.~(\ref{eq:MT}), in Sec.~\ref{sec:mtres}.
Within those parts, we report each flavour combination, one after the other.
Each figure title gives effective interaction (AWW or MT), flavor combinations (e.g.\,$\bar \upchi \mathrm{q}$) and quark-mass configurations (as subscripts for each flavor combination, indicating to which bound-state mass the quark mass was fixed, e.\,g., $\bar{\mathrm{s}}_{K^\pm} \mathrm{c}_{D^\pm}$ indicates a strange-quark mass fixed to the kaon and a charm-quark mass fixed to the $D$ meson masses) for increased clarity and quick reference.

The precise numbers plotted in the figures are also collected in tables in \ref{sec:numbers} for easy reference.

\subsection{AWW}\label{sec:awwres}

The employed model parameters for the \gls{AWW} model investigation are $\omega=0.5\,\mathrm{GeV}$ and $D=1.0\,\mathrm{GeV}^2$.
Note that the parametrisation of the intermediate-momentum part of the interaction is different in the original literature, but unified herein to enable a comparison of the parameter values used.

\subsubsection{Chiral and Light Quarks}

\begin{figure*}
\centerline{\includegraphics[width=.7\textwidth]{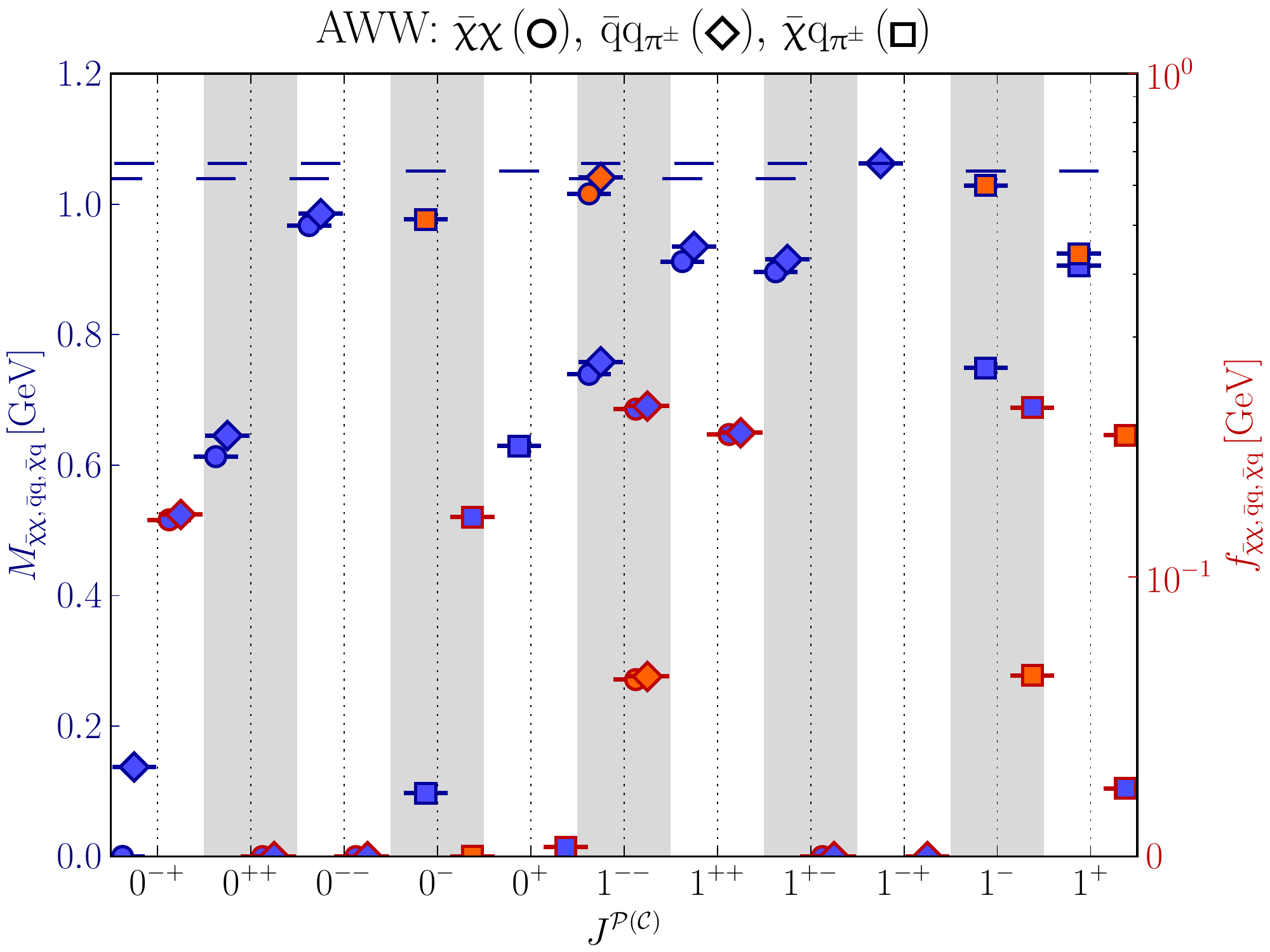}\hspace{2em}\includegraphics[width=.22\textwidth]{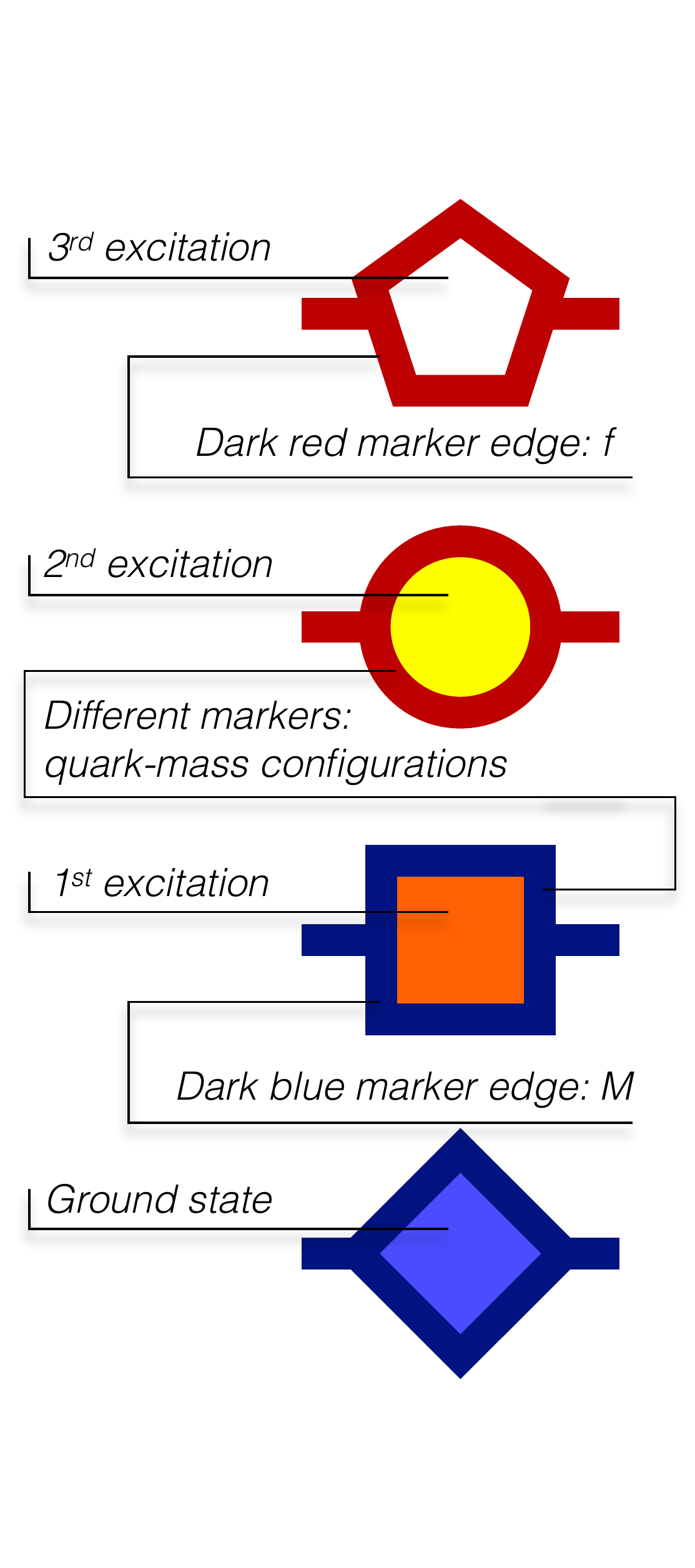}}
\caption{
Meson masses and decay constants for the flavour combinations $\bar \upchi \upchi$ (circles), $\bar{\mathrm{q}}\mathrm{q}$ (diamonds), and $\bar \upchi \mathrm{q}$ (squares), where the greek letter $\upchi$ denotes a chiral quark.
The horizontal lines at the top of this and subsequent figures are limits to the attainable masses from quark pole singularities, e.\,g., the two different methods to fit the heavier quark mass result in two such lines. In general, there is one such line for and above each marker shape. See also the detailed description of the figures as given in Sec.~\ref{sec:plotdescription} and by the legend on the right-hand side.
The light-quark mass $m_\mathrm{q}=5\,\mathrm{MeV}$ was fitted to a pion mass of $m_\uppi = 137\,\mathrm{MeV}$. 
}
\label{fig:AWWqq}
\end{figure*}

The results for quarkonia and combinations made out of chiral and light quarks are shown in Fig.~\ref{fig:AWWqq}.
In particular, we plot the combinations $\bar{\mathrm{q}}\mathrm{q}$, $\bar \upchi \upchi$, and $\bar \upchi \mathrm{q}$ together, which allows for a direct comparison and investigation of the quark-mass dependence of results close to the chiral limit.



For this plot, we make a number of interesting observations: First of all, the results are robust across the various light and chiral flavour combinations, which is an indication of the computational and conceptional stability of the results.

Next, the pseudoscalar channel is especially close to the chiral limit, since the pion is the would-be Goldstone boson related to \gls{DCSB}. 
This is exactly realised in our setup and clearly visible in the plot: The chiral pion is massless while its leptonic decay constant is nonzero and has the usual value in the chiral limit of $\sim 130$ MeV. 
The well-known behaviour of the ground-state mass as a function of the current-quark mass is a square-root behaviour, i.\,e., the pion mass rises rapidly away from zero. 
While for the excited $0^{-+}$ states the behaviour of $f$ close to the chiral limit is known and was documented elsewhere \cite{Holl:2004fr,Holl:2005vu}, namely, linearly rising with the current-quark mass and exactly zero in the chiral limit, we do not find such an excitation here below the pole threshold.

Also in other quantum-number channels and cases such as the scalar or exotic pseudoscalar ones, the chiral limit requires particular behaviour or values for $f$, as described above in Sec.~\ref{sec:spice}.
Wherever states appear that match these specific cases, all required qualitative features are realized exactly. 

The situation for open-flavour states was recently elucidated in detail with regard to the connection to $J^{\mathcal{PC}}$ exotics for the corresponding charmonia \cite{Hilger:2016efh,Hilger:2016drj}.
In particular, one observes continuous connections between meson masses as the quark masses change within the meson. 
Since no states should appear or disappear along such trajectories, a correspondence between, e.\,g., unflavoured $0^{-+}$ or $0^{--}$ quarkonium states to $0^{-}$ states for open flavours can be established.

The reasoning behind this is that, in the scope of a Poincar\'e-covariant calculation, $J^{\mathcal{P\pm}}$-states actually belong to the same tower of $J^{\mathcal{P}}$ excitations.
For example, in Fig.~\ref{fig:AWWqq} the first excitation in the chiral-light $0^{-}$ channel corresponds to the $0^{--}$ ground state in both the chiral-chiral as well as the light-light quarkonia.
As a result, this first excitation's leptonic decay constant is negligible.

The $1^{--}$-states have non-zero decay constants in general, with the ground state's decay constant being larger than the one of the first excitation.
This behaviour is nontrivial and can be related to the orbital-angular-momentum content of the states.
In particular, a smaller value for $f$ is correlated to a larger $D$-wave content of the state under consideration \cite{Hilger:2015ora}.

\begin{figure*}
\centerline{\includegraphics[width=.7\textwidth]{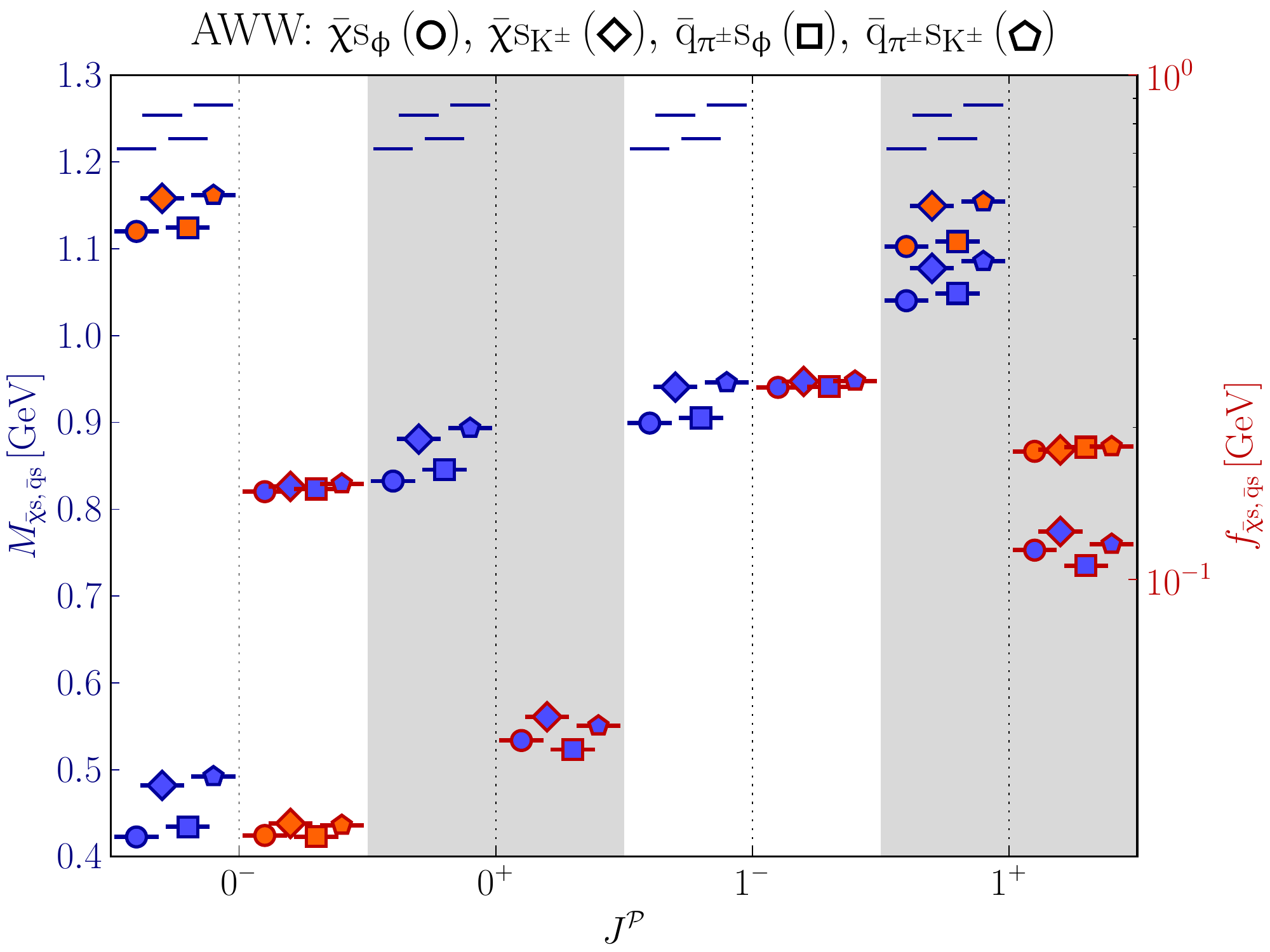}\hspace{2em}\includegraphics[width=.22\textwidth]{legend}}
\caption{
Same as Fig.~\ref{fig:AWWqq} for AWW strange mesons ($\bar{\mathrm{q}}\mathrm{s}$, $\bar \upchi \mathrm{s}$) with detailed description in Sec.~\ref{sec:plotdescription} and the legend on the right-hand side.
The quark-mass configurations are: strange-quark mass $m_\mathrm{s}=90\,\mathrm{MeV}$ fitted to the $\upphi(1020)$ mass (circles);
strange-quark mass $m_\mathrm{s}=115\,\mathrm{MeV}$ fitted to the kaon mass (diamonds) with light-quark mass $m_\mathrm{q}=5\,\mathrm{MeV}$;
$\bar{\upchi} \mathrm{s}$-states with strange-quark mass $m_\mathrm{s}=90\,\mathrm{MeV}$ (squares);
$\bar{\mathrm{q}}\mathrm{s}$-states with strange-quark mass $m_\mathrm{s}=115\,\mathrm{MeV}$ (pentagons).
}
\label{fig:AWWqs}
\end{figure*}

In the exotic $1^{-+}$-channel, one has to evaluate the \gls{BSE} very closely to the pole threshold, which requires an extremely efficient numerical treatment of the quark-pole issue in both the \gls{DSE} as well as the \gls{BSE} (cf.\ \ref{sec:DSEcontour}), and only one quark-mass configuration generates a sub-pole-threshold bound state.
Unfortunately, the corresponding quasi-exotic open-flavour $1^-$-state could not be found below the pole threshold.
It would be the second excitation in this channel due to the ordering of states in the combined $1^{-\pm}$-channels, where the first excitation is conventional and corresponds to a $1^{--}$ quarkonium. 
One would again expect a negligible decay constant for the quasi-exotic state due to the small difference between the light and chiral quark masses.

\subsubsection{Light and Strange Quarks}

The masses and leptonic decay constants for mesons with strangeness are presented in Fig.~\ref{fig:AWWqs}, where we collect both the light-strange as well as chiral-strange results.
In general, one finds considerably fewer states below the pole threshold than for the light-quark mesons.
However, the systematics are good, i.\,e., the masses and leptonic decay constants are robust across the various configurations and the variations follow the expected patterns.

The mass difference of the participating quarks results in non-zero leptonic decay constants for, among others, the (conventional) $0^{+}$-ground state which corresponds to the conventional $0^{++}$-ground states in the light-quark and $\bar{\mathrm{s}}\mathrm{s}$ quarkonia.
Also, both $1^{+}$-states have non-zero decay constants, where the ground states correspond to the $1^{+-}$ axialvector ground states in the light-quark-meson and $\bar{\mathrm{s}}\mathrm{s}$-sectors.

\subsubsection{Light and Charm Quarks}

\begin{figure}
\includegraphics[width=.48\textwidth]{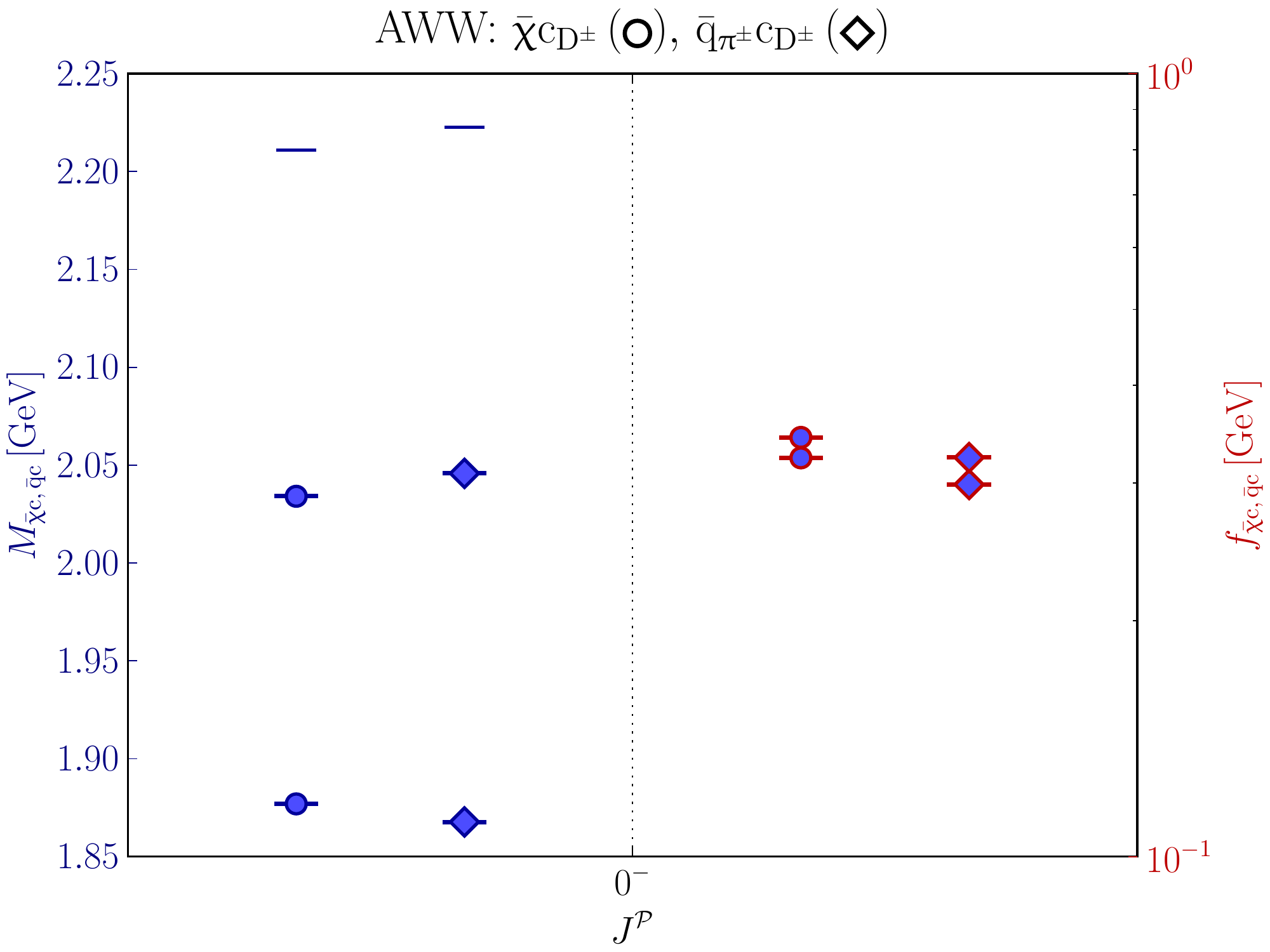}
\caption{
Same as Fig.~\ref{fig:AWWqq} for AWW charmed mesons $\bar{\mathrm{q}}\mathrm{c}$ (diamonds) and $\bar \upchi \mathrm{c}$ (circles) with detailed description in Sec.~\ref{sec:plotdescription} and the legend in Fig.~\ref{fig:AWWqs}.
Charm-quark mass $m_\mathrm{c}=975\,\mathrm{MeV}$ fitted to the $D$-meson mass with the light-quark mass $m_\mathrm{q}=5\,\mathrm{MeV}$.
A charm-quark mass being fitted to the $\mathrm{J}/\psi$ mass, $m_\mathrm{c}=1.11\,\mathrm{GeV}$, does not yield solutions below the relevant pole threshold of $M_\mathrm{max} = 2.211\,\mathrm{GeV}$, or $M_\mathrm{max} = 2.223\,\mathrm{GeV}$, respectively, for the chiral and non-chiral quark-mass configuration.
}
\label{fig:AWWqc}
\end{figure}

\begin{figure*}
\centerline{\includegraphics[width=.7\textwidth]{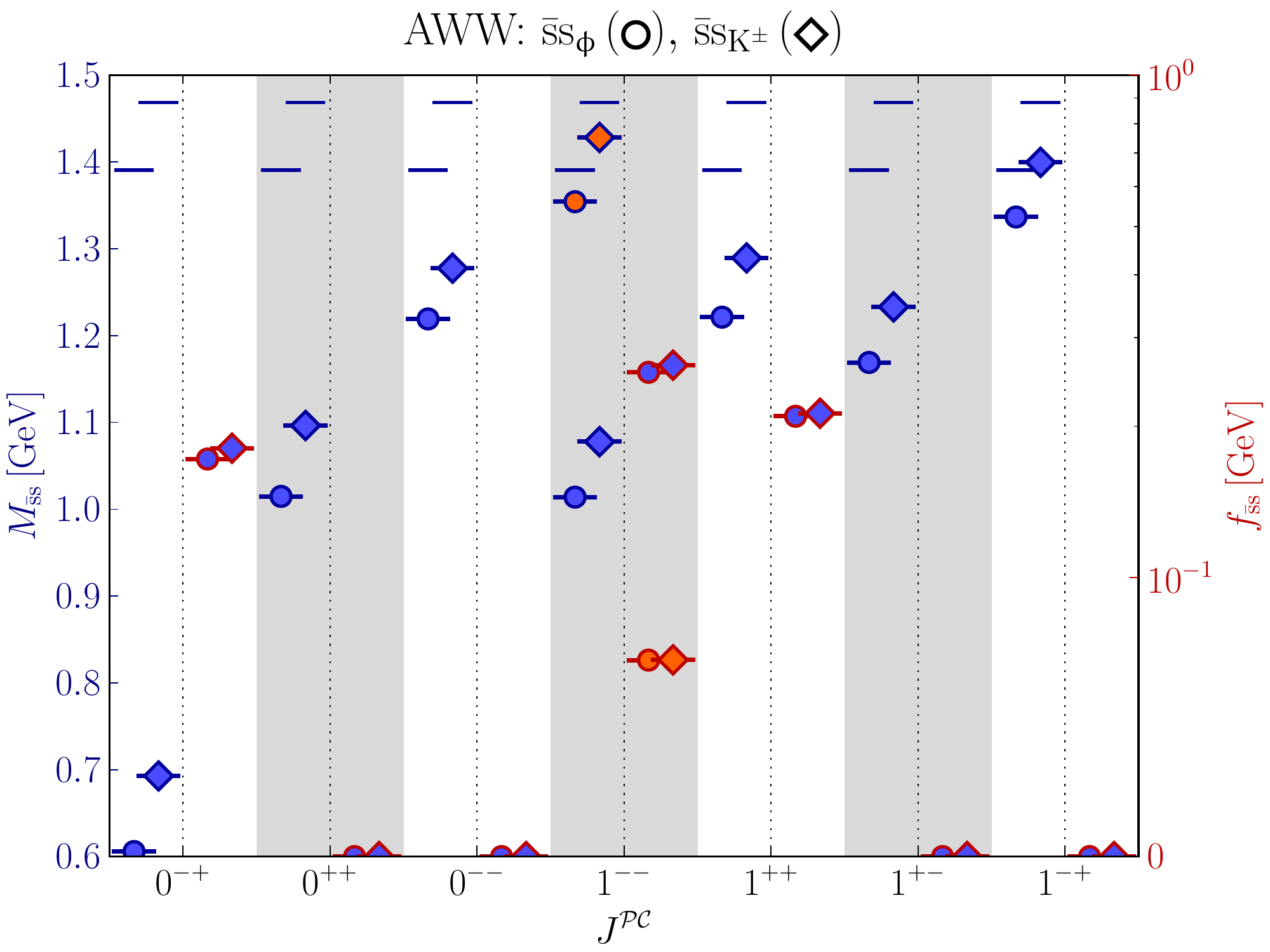}\hspace{2em}\includegraphics[width=.22\textwidth]{legend}}
\caption{Same as Fig.~\ref{fig:AWWqq} for AWW strangeonium ($\bar{\mathrm{s}}\mathrm{s}$) with detailed description in Sec.~\ref{sec:plotdescription}.
The quark-mass configurations are: strange-quark mass $m_\mathrm{s} = 90\,\mathrm{MeV}$ fitted to the $\upphi(1020)$ mass (circles);
strange-quark mass $m_\mathrm{s} = 115\,\mathrm{MeV}$ fitted to the kaon mass (diamonds) with the light-quark mass $m_\mathrm{q}=5\,\mathrm{MeV}$.
}
\label{fig:AWWss}
\end{figure*}

The $D$-meson mass spectrum and leptonic decay constants are presented in Fig.~\ref{fig:AWWqc}.
Our data here is rather sparse, since no \gls{BSE} solution below the pole threshold has been found in our setup using a charm-quark mass which has been fitted to the $J/\psi$ meson, $m_\mathrm{c} = 1.11\,\mathrm{GeV}$.
In addition, the $0^-$ channel is the only one where a bound-state solution below the pole threshold has been found. 
Overall, apart from the $B$ and $B_s$ cases, where no solution is found at all, this is the flavour combination with the fewest states below the pole threshold.
The depicted quark-mass configurations are $\bar{\mathrm{q}}\mathrm{c}$ and $\bar \upchi \mathrm{c}$ states with the charm-quark mass being $m_\mathrm{c} = 975\,\mathrm{MeV}$, adjusted to fit the $D(1870)$-meson mass (cf.\ Fig.~\ref{fig:AWWqc}).
For each the chiral- and light-quark configurations there are two solutions of the homogeneous \gls{BSE} and thus data points.
This happens because of the non-monotonicity of the eigenvalue curves in this case, where two bound states are generated by the first eigenvalue curve for both accessible quark-mass configurations.
It is possible that such additional states are abnormal in the sense of the canonical norm of the \gls{BSA} \cite{Smith:1969zk}.
This non-monotonicity also spoils the possible extraction of on-shell properties of the \gls{BSE} solutions beyond the pole threshold from off-shell eigenvalue curves below the pole threshold which are algorithmically used for finding numerical solutions of the \gls{BSE} \cite{Blank:2010bp}.
Therefore, any extension of the \gls{DSBSE} approach beyond the presented quantum numbers in the open-charm sector requires a sophisticated numerical treatment of the impeding mechanisms \cite{Eichmann:2016nsu}; however, such an advanced treatment is beyond the scope of the present study.
Finally, the resulting leptonic decay constants of all four solutions are very similar.

Another remark concerns the large imbalance of quark masses for heavy-light mesons such as $D$ or $B$. 
It has been known for some time in the literature that the ladder truncation of the \gls{BSE} is not well suited for the description of such unequal-mass systems, where a large part of the discussion usually takes place in the context of QED \cite{Gross:1982nz,Gross:1993,Eides:2000xc}. 
In particular, it is usually remarked that in the limit of the heavy constituent mass going to infinity, the Dirac equation is not reproduced in ladder truncation.
Similarly, one can ask the question in the context of our study, how strongly this behavior in the mentioned limit affects our results.
However, there is no a priori reason to consider any state more unreliable on the basis of such arguments than on the basis of truncation effects in general. 
Since truncation effects are certainly sizeable, we interpret our results for the heavy-light case with caution to begin with. 
The non-existence of the correct infinite-heavy-mass limit in ladder truncation must be viewed in the context of the model setup, much like each set of results. 
The model parameters capture some truncation effects and their variation makes others apparent, but there is no reason to judge a charm quark or even a bottom quark heavy by the standards of the limit in question. 
Thus it is unclear how influential the corresponding effects from the infinite-mass limit are in our results.

\subsubsection{Strangeonium}

The strangeonium mass spectrum and leptonic decay constants are shown in Fig.~\ref{fig:AWWss}.
While the mass spectrum exhibits a natural dependence on the sizeable quark-mass change, the leptonic decay constants only show remarkably small variations.
The latter is particularly true for the only excitation found below the pole threshold, in the $1^{--}$ channel.
Where available, leptonic decay constants follow the same pattern as for light quarkonia (cf.\ Fig.~\ref{fig:AWWqq}).
In particular, their values are zero in certain channels as required by symmetries.

\subsubsection{Strange and Charm Quarks}

\begin{figure}
\includegraphics[width=.48\textwidth]{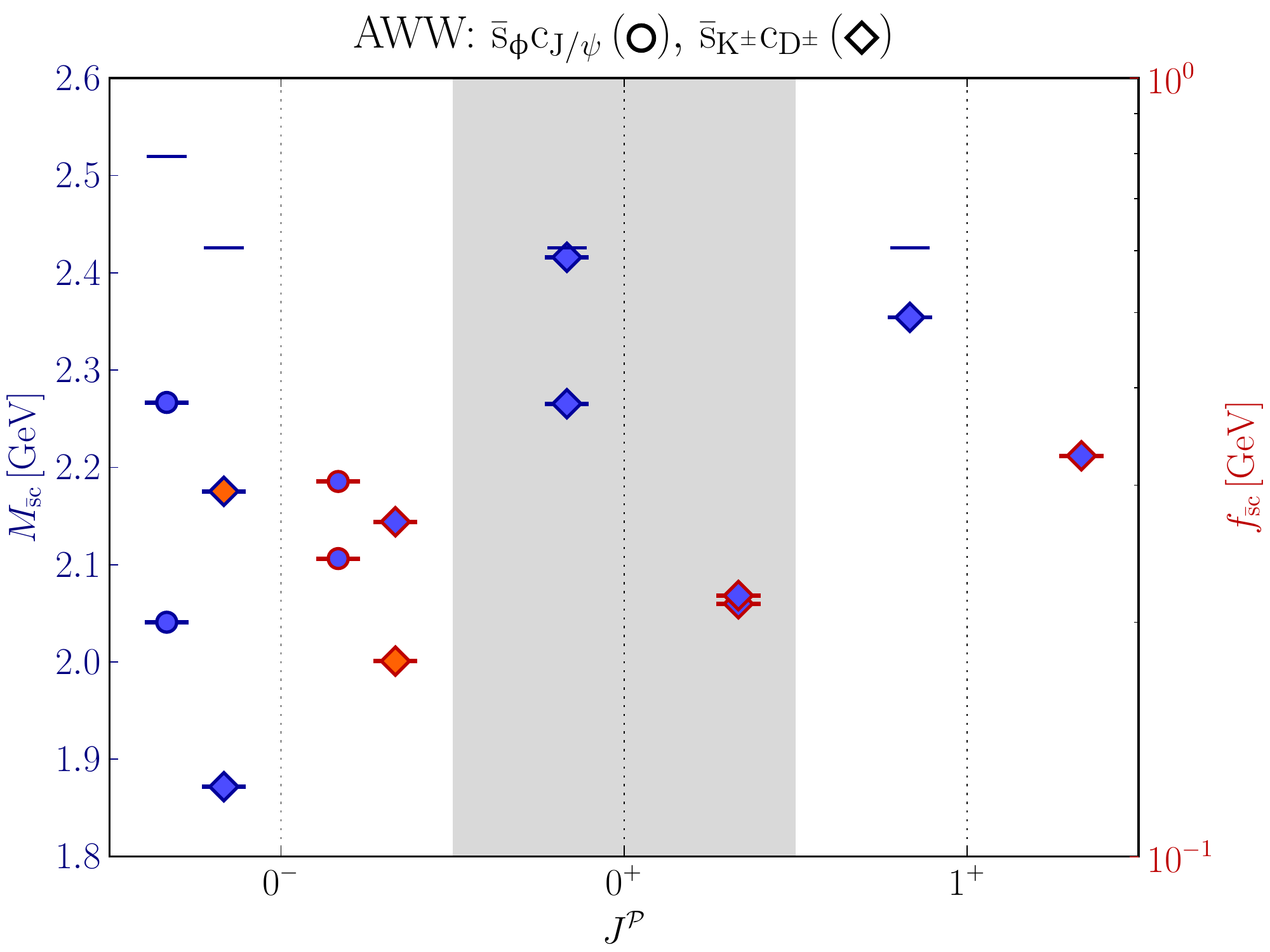}
\caption{
Same as Fig.~\ref{fig:AWWqq} for AWW charmed strange mesons ($\bar{\mathrm{s}}\mathrm{c}$) with detailed description in Sec.~\ref{sec:plotdescription}.
The quark-mass configurations are: $m_\mathrm{s} = 90\,\mathrm{MeV}$ and $m_\mathrm{c}=1.11\,\mathrm{GeV}$ both fitted to their $1^{--}$ ground states (circles);
$m_\mathrm{s} = 115\,\mathrm{MeV}$ and $m_\mathrm{c}=975\,\mathrm{MeV}$ both fitted to the open-flavour pseudoscalar ground states (diamonds).
}
\label{fig:AWWsc}
\end{figure}

For the $\bar{\mathrm{s}} \mathrm{c}$-spectrum, presented in Fig.~\ref{fig:AWWsc}, only $0^-$-states have been found below the pole threshold for the $1^{--}$-based configuration of quark masses.
While here also the leptonic decay constants appear strongly dependent on the employed quark masses, one must not forget the large change in the quark masses themselves for both the strange and the charm quarks.
Remarkably, the states of the $1^{--}$-quarkonium-based quark-mass configuration in the $0^-$-channel are both generated by the first eigenvalue curve, in contrast to the states found via the open-flavour-based quark-mass configuration in this channel.
This qualitative difference alone must be expected capable of producing large systematic errors.
However, the variation is well within a reasonable domain considering that also the underlying changes in the quark masses are sizeable.

The large mass difference of the participating quarks results in non-zero leptonic decay constants in the $1^+$- and $0^+$-channels.
The corresponding equal-flavour channels in the $\bar{\mathrm{s}}\mathrm{s}$- and $\bar{\mathrm{c}}\mathrm{c}$-sectors (cf.\ Figs.~\ref{fig:AWWss} and \ref{fig:AWWcc}) are the $1^{+-}$-channel for the $1^+$-state and the $0^{++}$-channel for the $0^+$-ground state, both with zero leptonic decay constants.
The assignment for the first excitation in the $0^+$-channel is unclear, as the next excitation with appropriate quantum numbers in the $\bar{\mathrm{s}}\mathrm{s}$-sector is the first (conventional) $0^{++}$-excitation (there are no $0^{+-}$-states below the pole threshold in the $\bar{\mathrm{s}}\mathrm{s}$-sector), while the associated state in the $\bar{\mathrm{c}}\mathrm{c}$-sector is the exotic $0^{+-}$-ground state, both of which have negligible leptonic decay constants.
Analogously, non-vanishing leptonic decay constants are found in the $0^+$-channel for both states, where the ground state corresponds to the $0^{++}$-ground state and the quasi-exotic first excitation to the exotic $0^{+-}$-ground state in the $\bar{\mathrm{c}}\mathrm{c}$-sector.

\subsubsection{Charmonium}

\begin{figure*}
\centerline{\includegraphics[width=.7\textwidth]{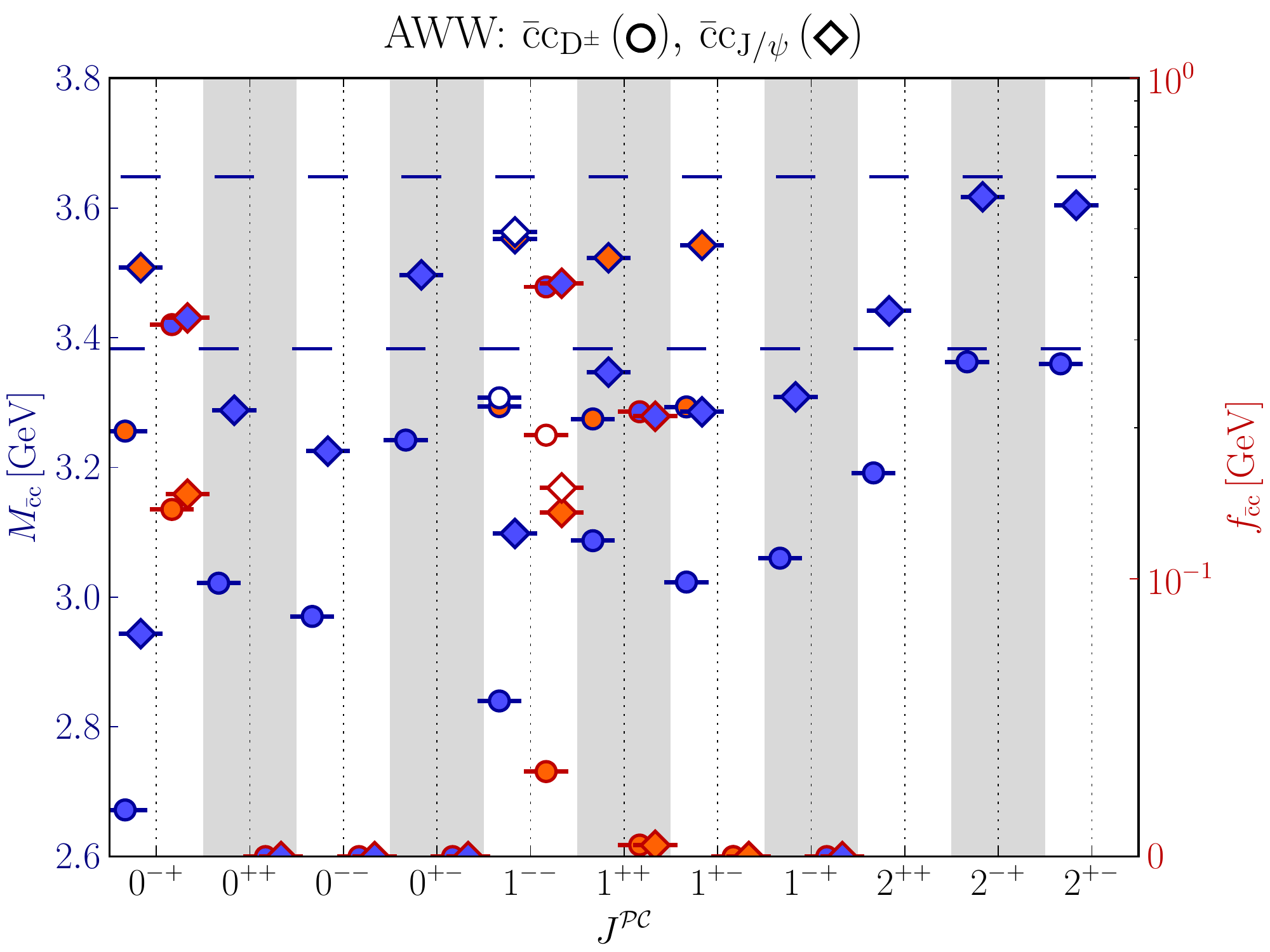}\hspace{2em}\includegraphics[width=.22\textwidth]{legend}}
\caption{
Same as Fig.~\ref{fig:AWWqq} for AWW charmonium ($\bar{\mathrm{c}}\mathrm{c}$)  with detailed description in Sec.~\ref{sec:plotdescription}.
The quark-mass configurations are: charm-quark mass $m_\mathrm{c} = 975\,\mathrm{MeV}$ fitted to the $D$-meson mass (circles);
charm-quark mass $m_\mathrm{c} = 1.11\,\mathrm{GeV}$ fitted to the $\mathrm{J}/\psi$ mass (diamonds).
}
\label{fig:AWWcc}
\end{figure*}

Figure \ref{fig:AWWcc} shows the $\bar{\mathrm{c}}\mathrm{c}$-meson mass spectrum and leptonic decay constants. 
Charmonium allows for more states to be found below the pole threshold than for the light-quark quarkonia, as discussed above.
In particular, genuine (exotic) $0^{+-}$ and tensor-meson states have been evaluated and are presented here.
It is also the first quarkonium spectrum presented here with first excitations in the $0^{-+}$-, $1^{++}$-, and $1^{+-}$-channels and a second excitation in the $1^{--}$-channel below the pole threshold.

The dependence of the pole threshold on the quark-mass configuration may seem large, but its relative size is comparable to the strangeonium sector.
The same is true for the dependence of the meson masses on the quark-mass configuration.
Note that the second $1^{--}$-excitation is almost degenerate in mass with the first excitation, which is not uncommon and an expected possibility for states with different predominant orbital angular momentum.

In particular, the excited-state values for the leptonic decay constant in the $1^{--}$-channel hint towards the interpretation of larger $f$ as being generated by an $S$-wave state and smaller $f$ by a $D$-wave state \cite{Hilger:2015ora}.
Note that leptonic decay constants of excitations in the axialvector channels are negligible, as are the ones of the $0^{++}$-channel, as well as those of the exotic $0^{--}$-, $0^{+-}$-, and $1^{-+}$-channels.
Also note that we do not define a leptonic decay constant for mesons with $J \geq 2$.

\subsubsection{Charm and Bottom Quarks}

\begin{figure}
\includegraphics[width=.48\textwidth]{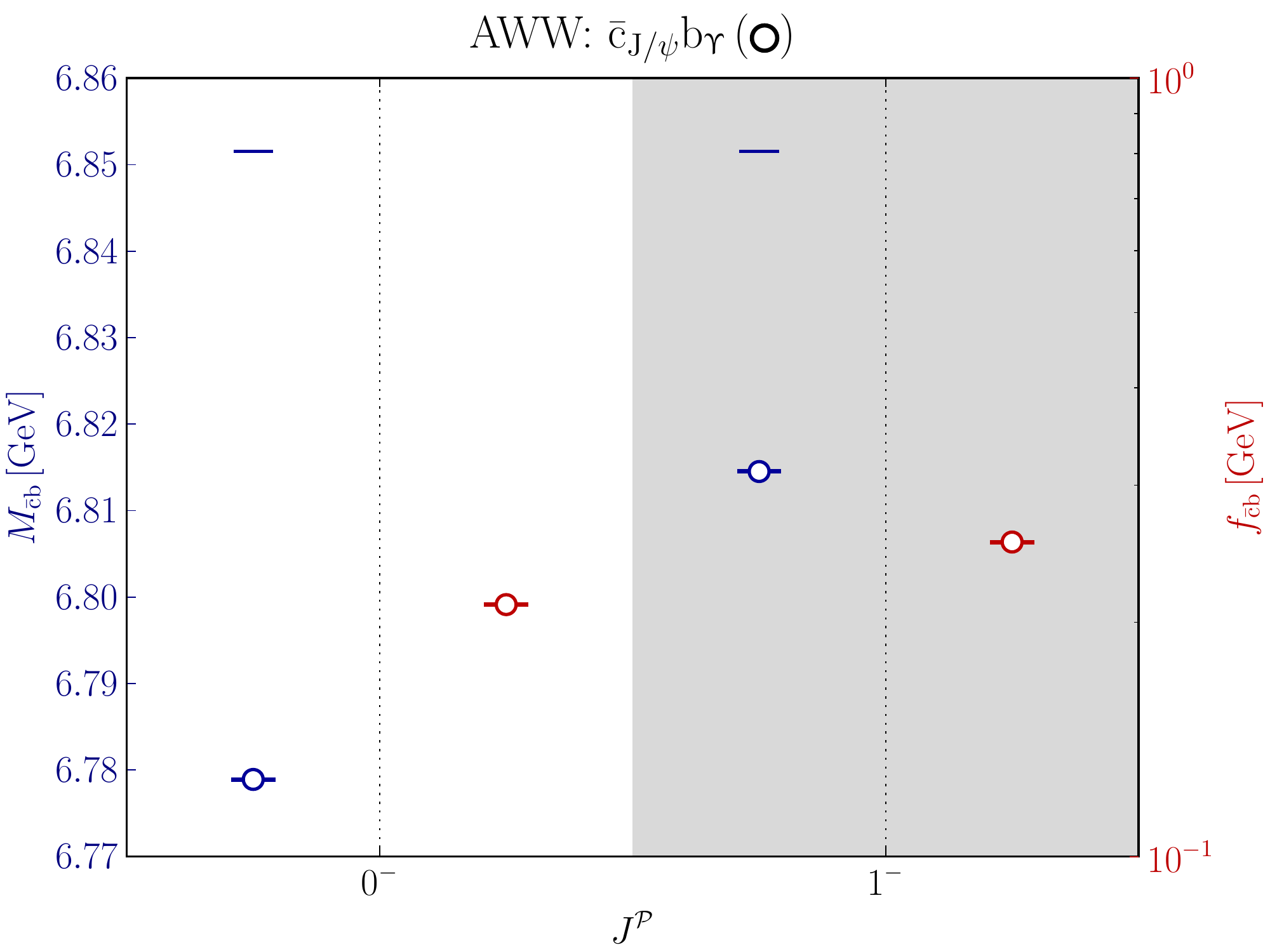}
\caption{
Same as Fig.~\ref{fig:AWWqq} for AWW bottom charmed mesons ($\bar{\mathrm{c}}\mathrm{b}$) with detailed description in Sec.~\ref{sec:plotdescription}.
Bottom-quark mass $m_\mathrm{b} = 4.43\,\mathrm{GeV}$ and charm-quark mass $m_\mathrm{c} = 1.11\,\mathrm{GeV}$ fitted to respective $1^{--}$ ground-state masses.
}
\label{fig:AWWcb}
\end{figure}

The $\bar{\mathrm{c}}\mathrm{b}$-meson pattern, presented in Fig.~\ref{fig:AWWcb}, exemplifies the interesting case where the lowest bound state is generated by the third eigenvalue due to the presence of complex conjugated first and second eigenvalues at the point where their real parts intersect the value one.
As argued in \cite{Hilger:2016drj}, one might suspect that genuine solutions of the \gls{BSE} with $\mathrm{Im} P^2 \neq 0$ also exist for these eigenvalues.
However, this contradicts the fact that the \gls{RL}-truncated \gls{BSE} has no decay channels and, therefore, does not provide resonances or a meson's width.
Implementation of weak decay channels is work in progress \cite{Mian:2016eel}. The analysis of this situation will be continued elsewhere.

\subsubsection{Bottomonium}

\begin{figure*}
\centerline{\includegraphics[width=.7\textwidth]{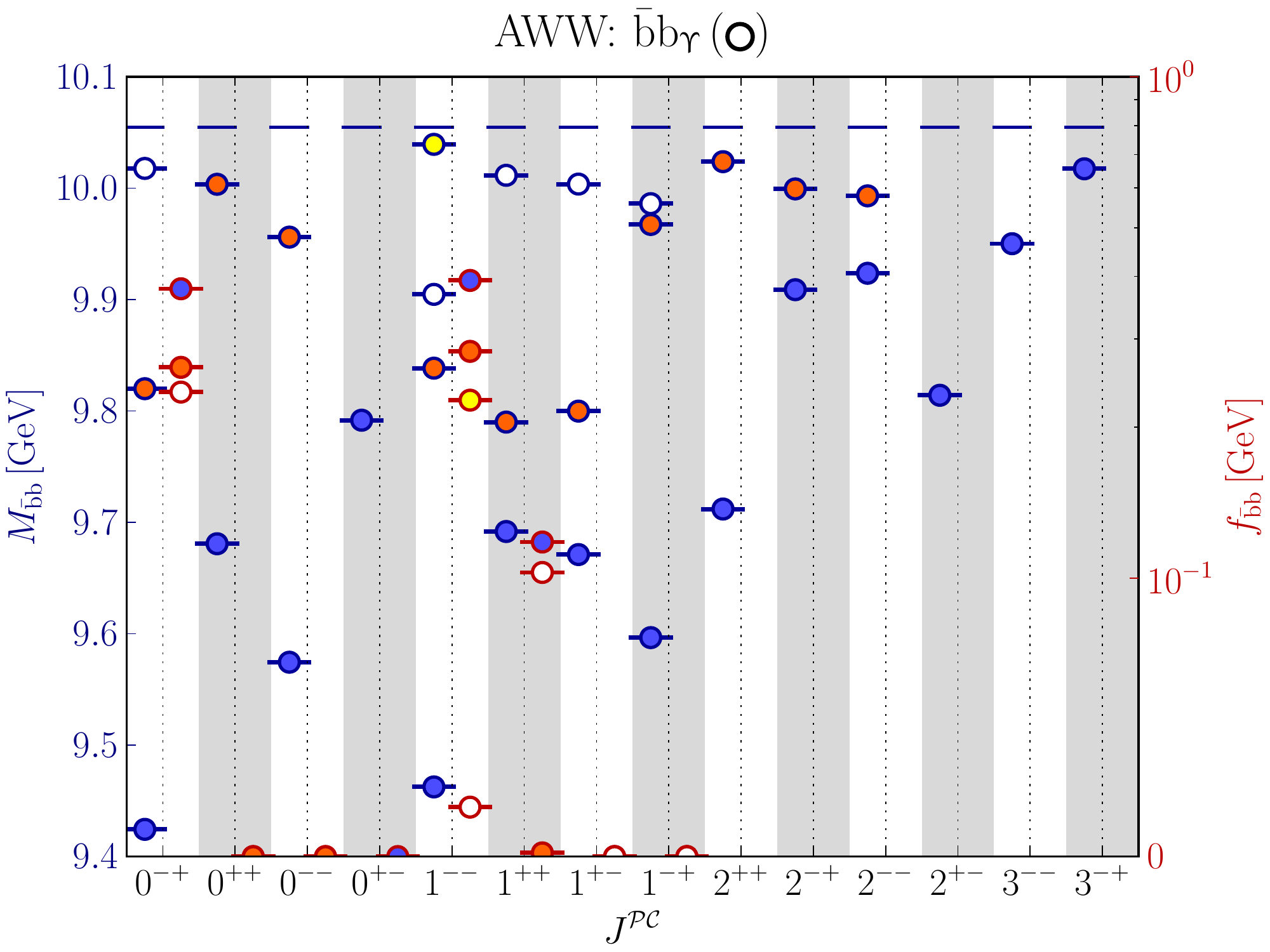}\hspace{2em}\includegraphics[width=.22\textwidth]{legend}}
\caption{
Same as Fig.~\ref{fig:AWWqq} for AWW bottomonium ($\bar{\mathrm{b}}\mathrm{b}$) with detailed description in Sec.~\ref{sec:plotdescription}.
Bottom-quark mass $m_\mathrm{b} = 4.43\,\mathrm{GeV}$ fitted to the $\Upupsilon$ mass.
}
\label{fig:AWWbb}
\end{figure*}

The $\bar{\mathrm{b}}\mathrm{b}$ results, depicted in Fig.~\ref{fig:AWWbb}, are the richest below the pole threshold in this investigation.
However, since no $B$-meson mass could be fitted, there is only one quark mass configuration with the bottom-quark mass fitted to the vector ground-state quarkonium.

Additional excitations compared to the $\bar{\mathrm{c}}\mathrm{c}$-pattern can be found in all channels apart from $0^{+-}$ and $2^{+-}$.
This is also the only case for the \gls{AWW} model \eqref{eq:AWW} where bound states below the pole threshold in the $3^{--}$ and $3^{-+}$ channels are found.

Different from the $\bar{\mathrm{c}}\mathrm{c}$-pattern, the leptonic decay constants in the $1^{--}$-channel are reversed in size for the first and second excitations, while the third excitation again has a larger decay constant.
This suggests an orbital angular momentum excitation pattern of $S$-wave for the ground state, the first and third excitations, while the second excitation appears to be $D$-wave.

\subsubsection{In-Hadron Condensates}

\begin{figure}
\includegraphics[width=.48\textwidth]{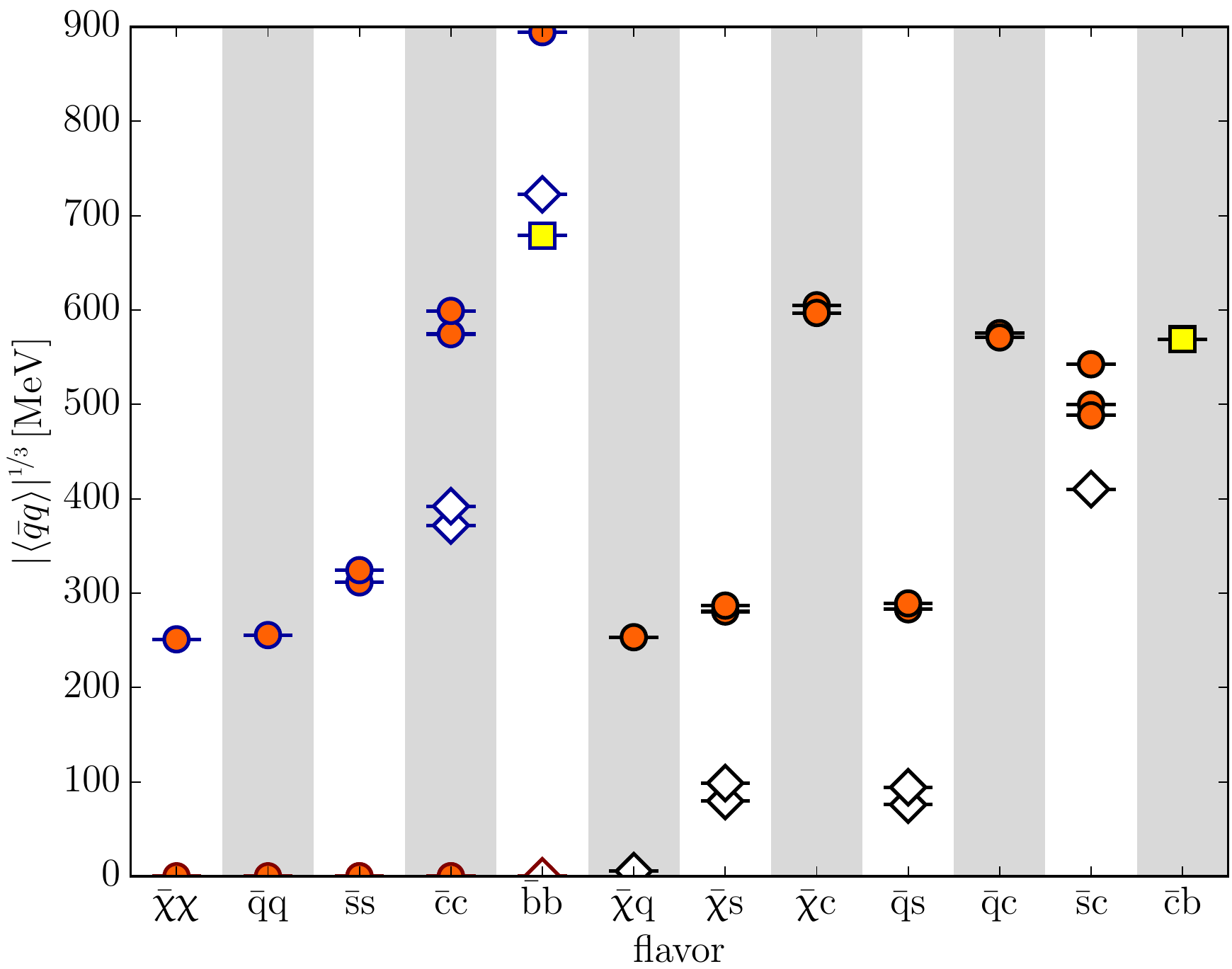}
\caption{
$J^\mathcal{P(C)} = 0^{-(\mathcal{C})}$ in-hadron condensates according to definition \eqref{eq:qq} in the \gls{AWW} model \eqref{eq:AWW} for all accessible flavour and charge-parity combinations.
Blue marker edges refer to $\mathcal{C}=+1$, red to $\mathcal{C}=-1$, and black to undefined $\mathcal{C}$ for open-flavour mesons.
Marker fill colours and marker shapes refer to eigenvalue numbering in that particular $\mathcal{C}$-channel (ground state -- red circles, first excitation -- white diamonds, second excitation -- yellow squares); quark mass configurations are not distinguished here.
}
\label{fig:AWWqbarq}
\end{figure}

To conclude the presentation of results from the \gls{AWW} model calculations, we present in-hadron condensates evaluated according to Eq.~\eqref{eq:qq} in Fig.~\ref{fig:AWWqbarq}.
For quarkonium in-hadron condensates, one generally observes that the condensates defined in this way increase with quark-mass content in modulus and decrease with excitation.
Furthermore, exotic in-hadron condensates are generally zero.
For open-flavour mesons, one observes that the in-hadron condensates are dominated by the heavy-quark content, e.\,g., the $\bar{\upchi}\mathrm{c}$-condensate can be approximated by the $\bar{\mathrm{c}}\mathrm{c}$-condensate.

As the $\bar{\mathrm{c}}\mathrm{b}$-in-hadron condensate is similar to the $\bar{\mathrm{c}}\mathrm{c}$-condensate, this indicates that one indeed misses $\bar{\mathrm{c}}\mathrm{b}$-states, in particular, the ground state.
The quasi-exotic open-flavour analogues to exotic quarkonia generally have rather small or even negligible in-hadron condensates.
Note that the second eigenvalue in the $\bar{\mathrm{s}}\mathrm{c}$-sector does not generate a quasi-exotic state, as can be seen from Figs.~\ref{fig:AWWss} and \ref{fig:AWWcc}.

\subsection{MT}\label{sec:mtres}

The second part of the presentation of our results deals with the \gls{MT} effective interaction \eqref{eq:MT}.
The difference between these two \emph{Ans\"atze} is the \gls{UV} part of the effective interaction, which is responsible for the behaviour in high-energy processes and also has an immediate impact on, e.\,g., the large-momentum behaviour of the components of the \gls{BSA}.
For the present investigation the main point is that the two different versions of the effective interaction allow us to study the model dependence of our setup, at least to some extent.

\begin{figure*}
\centerline{\includegraphics[width=.7\textwidth]{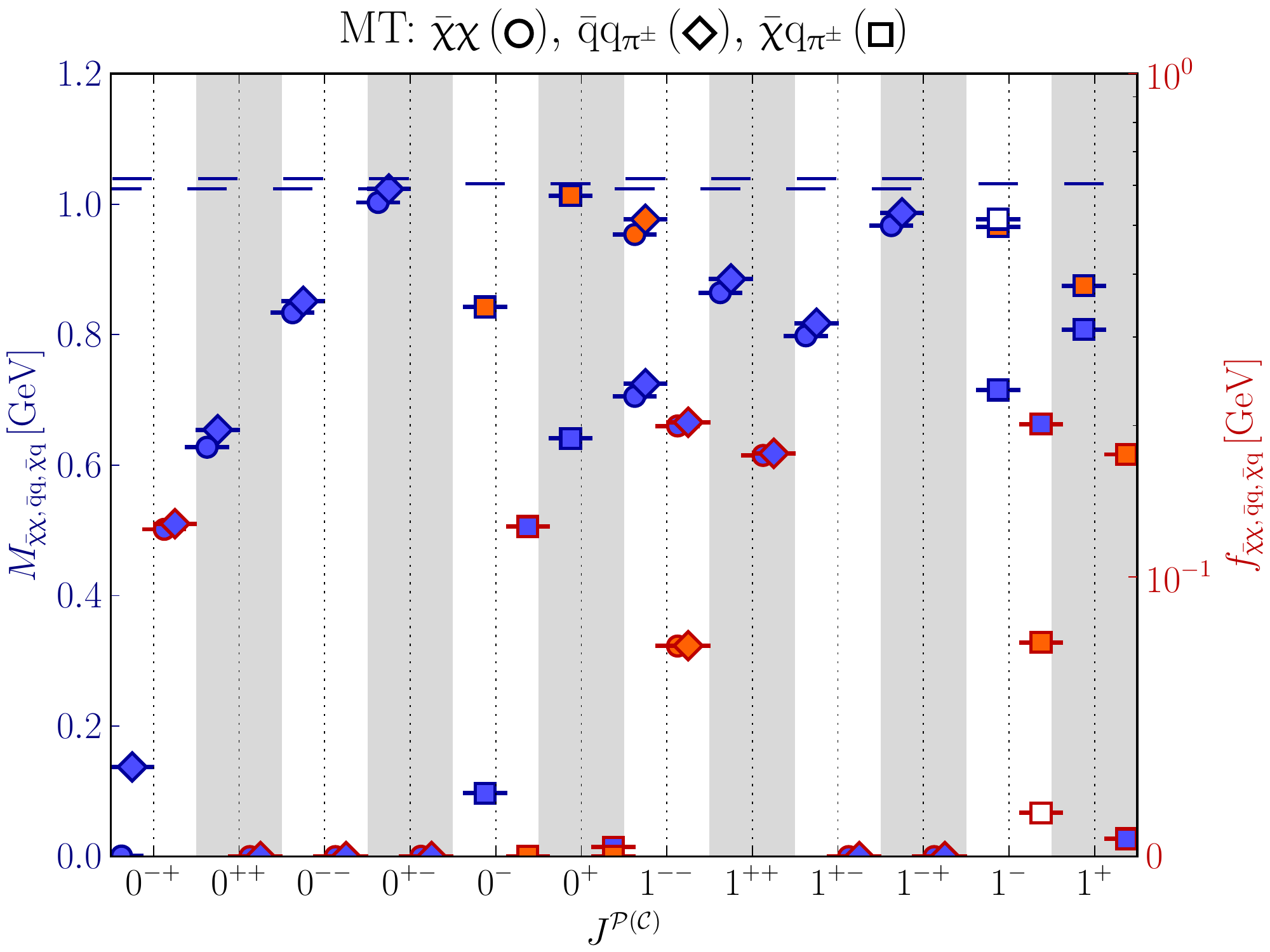}\hspace{2em}\includegraphics[width=.22\textwidth]{legend}}
\caption{Same as Fig.~\ref{fig:AWWqq} but for the \gls{MT} model \eqref{eq:MT}. 
Meson masses and decay constants for the flavour combinations $\bar \upchi \upchi$ (circles), $\bar{\mathrm{q}}\mathrm{q}$ (diamonds), and $\bar \upchi \mathrm{q}$ (squares), where the greek letter $\upchi$ denotes a chiral quark.
The light-quark mass there is $m_\mathrm{q}=3.8\,\mathrm{MeV}$.
Detailed figure description in Sec.~\ref{sec:plotdescription} and in the legend on the right-hand side.
}
\label{fig:MTqq}
\end{figure*}

In this context it would be desirable to have even more different forms of the effective interaction and compare the results. However, comparing two of them in detail is certainly enough in terms of a first step and does not overload the amount of data and figures presented here. We generally find that there are differences in the details of the various flavour-combination setups, but that the results for $M$ and $f$ are rather robust with respect to the difference in the \gls{UV} behaviour of the two effective interactions. The changes are expected to be largest in the excited states in each channel due to the strength variation in the intermediate-momentum range from the AWW to the MT case. Such a variation is brought about by the change in model parameters necessary to maintain appropriate fits of the data.

The employed model parameters for the \gls{MT} model investigation are $\omega=0.4\,\mathrm{GeV}$ and $D=0.372\,\mathrm{GeV}^3 / \omega$.
The presentation is analogous to the one above in Sec.~\ref{sec:awwres} and given in Figs.~\ref{fig:MTqq} through \ref{fig:MTbb}.

\subsubsection{Chiral and Light Quarks}

In analogy to Fig.~\ref{fig:AWWqq} for the \gls{AWW} model interaction, Fig.~\ref{fig:MTqq} shows the light-quark meson spectrum and leptonic decay constants for the \gls{MT} model \eqref{eq:MT}.
The overall pattern of results is very similar to the ones already presented above, a statement that can be made about the entire set of results.

Compared to the \gls{AWW} case, one finds more states below the pole threshold, e.\,g., in the $1^{-+}$-channel, and in particular an exotic $0^{+-}$-state.
Furthermore, the first excitation in the $0^{+}$-channel is found, which is the quasi-exotic open-flavour analogue to the exotic $0^{+-}$-ground state.
Thus, the open-flavour $0^{+}$-channel, with its conventional ground state and its quasi-exotic first excitation, clearly resembles the mass spectra and decay constants of the combined $0^{+\pm}$-channel, as it should in our line of argument.

Similarly, three sub-pole-threshold bound-state solutions to the \gls{BSE} can be found in the $1^{-}$-channel, thus providing us with a genuine quasi-exotic open-flavour meson.
The first (a conventional) and second (a quasi-exotic) excitations are almost mass-degenerate, as one would expect from the respective $1^{-\pm}$-channel spectra.
Contrary to the (quasi-) exotic states of the $1^{-(+)}$-channel with negligible decay constants, the conventional states have sizable decay constants, which provides a means to distinguish these two types of states~\cite{Hilger:2016efh}.

\subsubsection{Light and Strange Quarks}

\begin{figure*}
\centerline{\includegraphics[width=.7\textwidth]{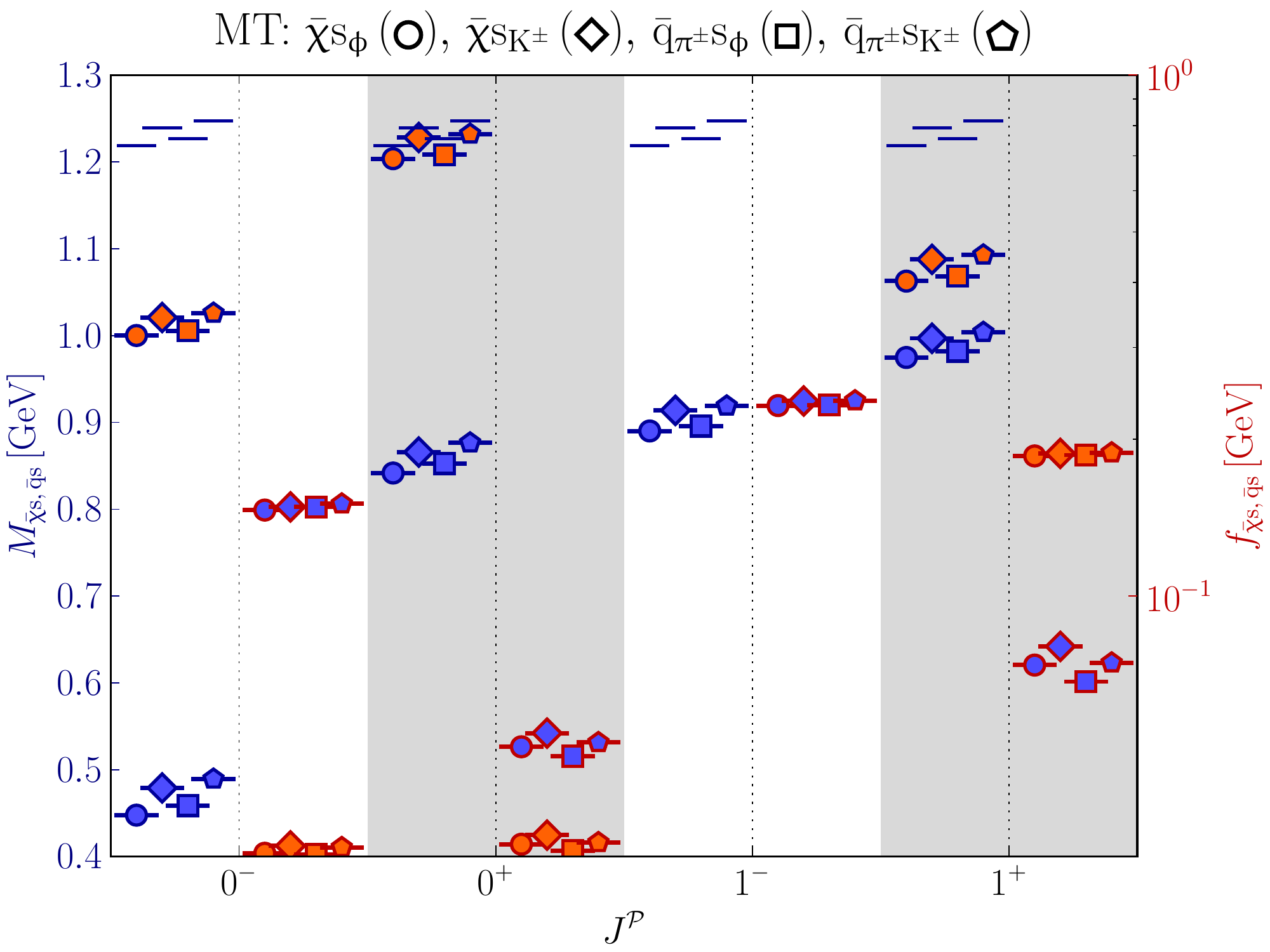}\hspace{2em}\includegraphics[width=.22\textwidth]{legend}}
\caption{
Same as Fig.~\ref{fig:MTqq} for MT strange mesons ($\bar{\mathrm{q}}\mathrm{s}$, $\bar \upchi \mathrm{s}$).
The quark-mass configurations are: 
$\bar{\upchi} \mathrm{s}$ with strange-quark mass $m_\mathrm{s}=75\,\mathrm{MeV}$ fitted to the $\upphi(1020)$ (circles);
$\bar{\upchi} \mathrm{s}$ with strange-quark mass $m_\mathrm{s}=85\,\mathrm{MeV}$ fitted to the kaon with the light-quark mass  $m_\mathrm{q}=3.8\,\mathrm{MeV}$ (diamonds);
$\bar{\mathrm{q}}\mathrm{s}$ with strange-quark mass $m_\mathrm{s}=85\,\mathrm{MeV}$ (squares);
$\bar{\mathrm{q}}\mathrm{s}$ with strange-quark mass $m_\mathrm{s}=75\,\mathrm{MeV}$ (pentagons).
Detailed figure description in Sec.~\ref{sec:plotdescription}.
}
\label{fig:MTqs}
\end{figure*}

The strange-meson spectra and decay constants, exhibited in Fig.~\ref{fig:MTqs}, feature an additional excitation in the $0^+$-channel compared to the  \gls{AWW} model spectra in Fig.~\ref{fig:AWWqs}.
It is found slightly below the pole threshold and provides a very nice example of bound states of one quark-mass configuration being found beyond the pole threshold of another configuration.
The additional excitation is a quasi-exotic open-flavour meson corresponding to two exotic $0^{+-}$ quarkonia.
Overall, the spectrum appears even more robust across the four configurations for both the masses and decay constants compared to the \gls{AWW} model, cf., e.\,g., the $0^+$-channel.

\subsubsection{Light and Charm Quarks}

\begin{figure}
\includegraphics[width=.48\textwidth]{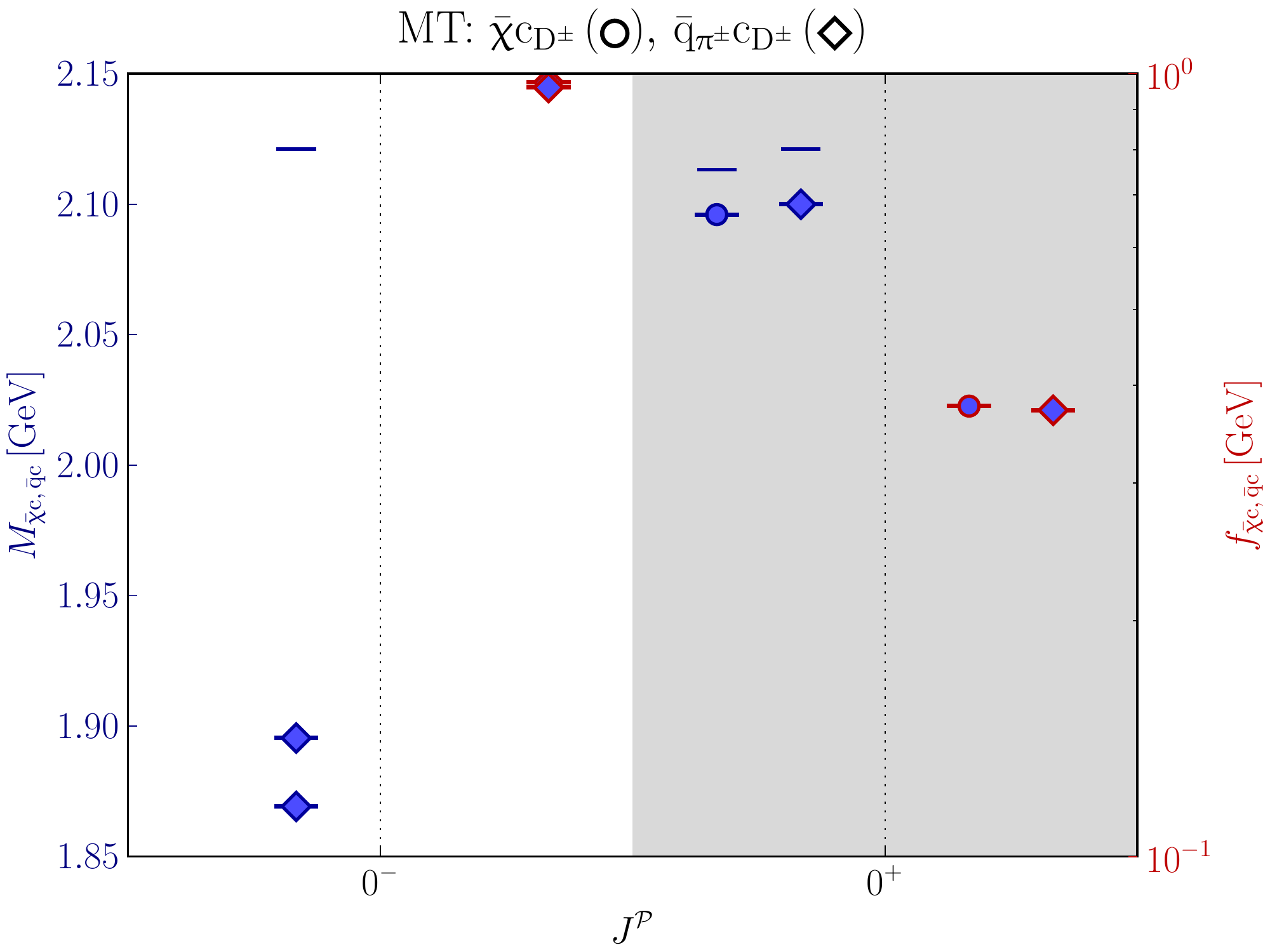}
\caption{Same as Fig.~\ref{fig:MTqq} for MT charmed mesons $\bar{\mathrm{q}}\mathrm{c}$ (diamonds), $\bar \upchi \mathrm{c}$ (circles).
Charm-quark mass of $m_\mathrm{c}=695\,\mathrm{MeV}$ fitted to the $D$-meson mass with light-quark mass $m_\mathrm{q}=3.8\,\mathrm{MeV}$.
No solutions are found below the pole threshold for the vector-ground-state adjusted charm-quark mass.
Detailed figure description in Sec.~\ref{sec:plotdescription}.
}
\label{fig:MTqc}
\end{figure}

The charmed-meson results are presented in Fig.~\ref{fig:MTqc}.
In contrast to the \gls{AWW} model (cf.\ Fig.~\ref{fig:AWWqc}), one finds sub-pole-threshold bound states in the $0^+$-channel.
On the other hand, in the $0^-$-channel sub-pole-threshold bound states can be found only for one quark-mass configuration.
Due to the non-monotonicity of the respective eigenvalue curve in this case, one formally obtains two bound-state solutions for the first \gls{BSE} eigenvalue, similar to the \gls{AWW} case in this channel.
The corresponding leptonic decay constants are again almost equal.
The $0^{+}$-state with a sizeable decay constant corresponds to the $0^{++}$-ground states in the charmonium and the light quarkonium, which both have zero decay constant.

\subsubsection{Strangeonium}

\begin{figure*}
\centerline{\includegraphics[width=.7\textwidth]{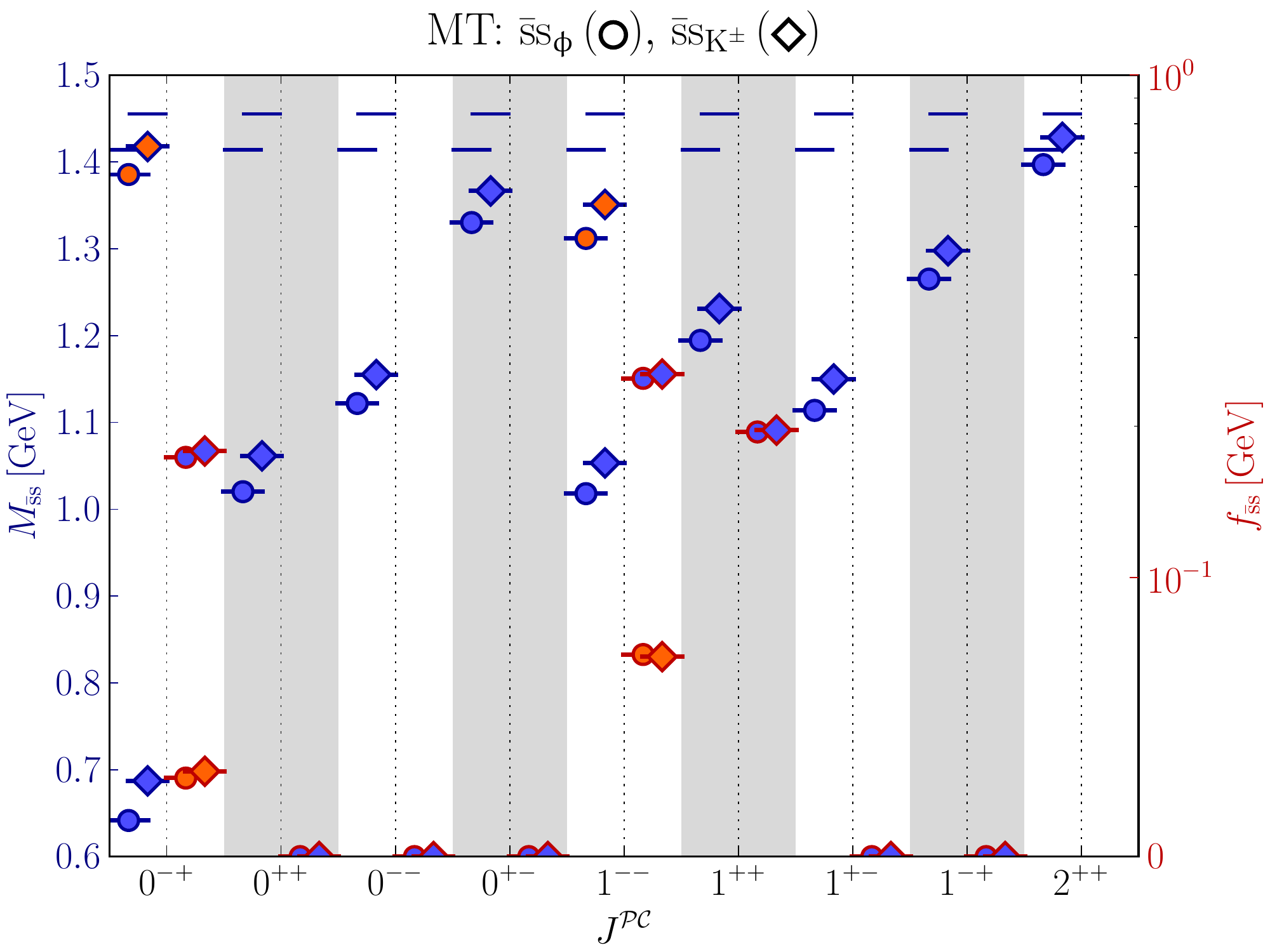}\hspace{2em}\includegraphics[width=.22\textwidth]{legend}}
\caption{Same as Fig.~\ref{fig:MTqq} for MT strangeonium ($\bar{\mathrm{s}}\mathrm{s}$).
The quark-mass configurations are: 
strange-quark mass $m_\mathrm{s} = 75\,\mathrm{MeV}$ fitted to the $\upphi(1020)$ mass (circles);
strange-quark mass $m_\mathrm{s} = 85\,\mathrm{MeV}$ fitted to the kaon with light-quark mass $m_\mathrm{q}=3.8\,\mathrm{MeV}$ (diamonds).
Detailed figure description in Sec.~\ref{sec:plotdescription}.
}
\label{fig:MTss}
\end{figure*}

Figure \ref{fig:MTss} shows the strangeonium mass spectrum and leptonic decay constants.
The similarities and differences to the case of the \gls{AWW} model (cf.\ Fig.~\ref{fig:AWWss}) remain consistent with observations made above in that there are more sub-pole-threshold states, more excitations, and the results appear to be even more robust with respect to the quark-mass configurations.

\subsubsection{Strange and Charm Quarks}

\begin{figure}
\includegraphics[width=.48\textwidth]{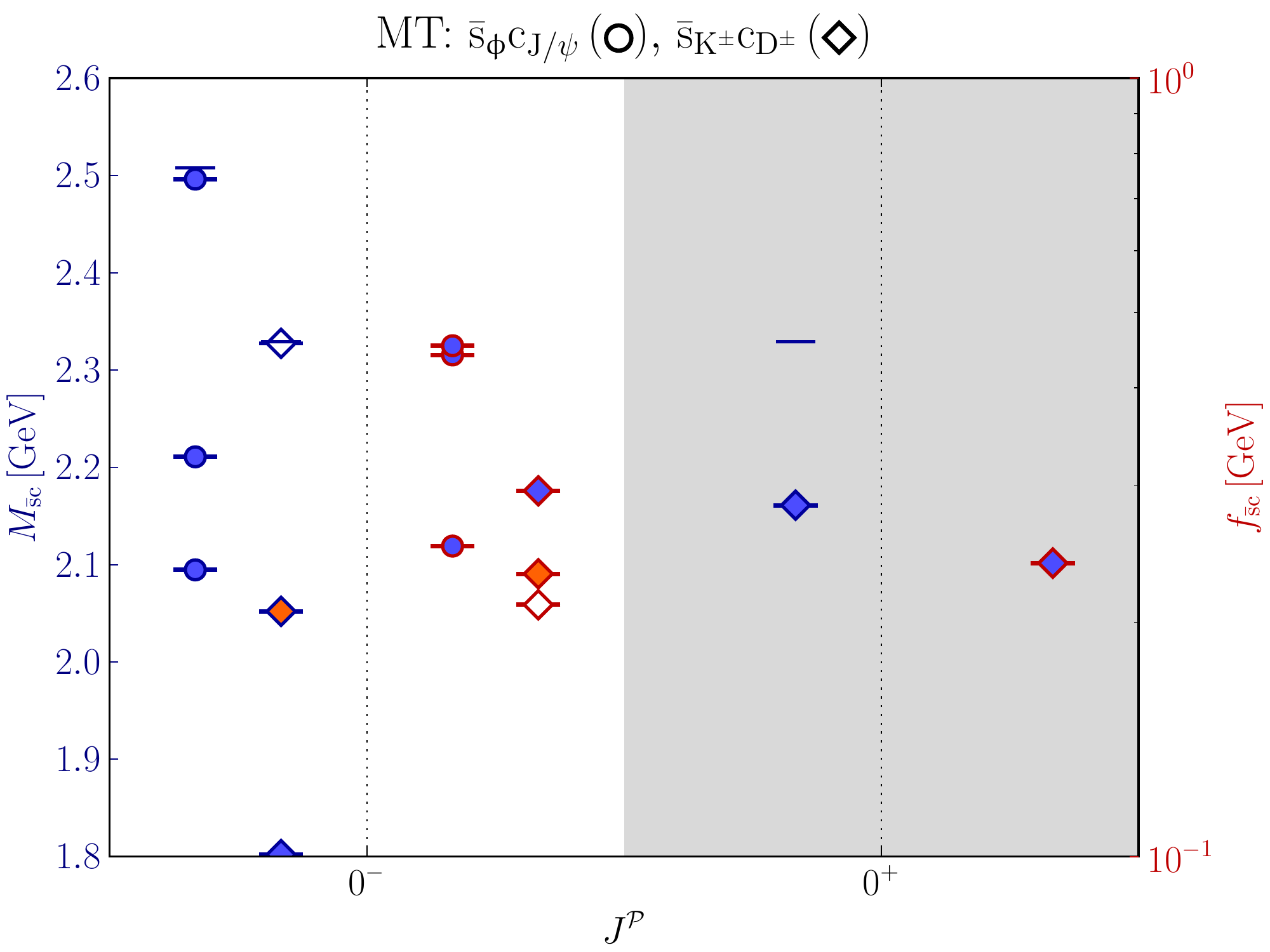}
\caption{Same as Fig.~\ref{fig:MTqq} for MT charmed strange mesons ($\bar{\mathrm{s}}\mathrm{c}$).
The quark-mass configurations are: 
strange-quark mass $m_\mathrm{s} = 75\,\mathrm{MeV}$ and charm-quark mass $m_\mathrm{c}=855\,\mathrm{MeV}$ each fitted to vector-quarkonium ground state (circles);
strange-quark mass $m_\mathrm{s} = 85\,\mathrm{MeV}$ and charm-quark mass $m_\mathrm{c}=695\,\mathrm{MeV}$ each fitted to open-flavour pseudoscalar ground state.
Detailed figure description in Sec.~\ref{sec:plotdescription}.
}
\label{fig:MTsc}
\end{figure}

Contrary to the patterns before, the charmed-strange-meson pattern from the \gls{MT} model, presented in Fig.~\ref{fig:MTsc}, is less populated by sub-pole-threshold states as compared to the \gls{AWW} model (cf.\ Fig.~\ref{fig:AWWsc}).
While both patterns have the same number of sub-pole-threshold states, no such states in the $1^+$-channel have been found.
Furthermore, the dependence w.\,r.\,t.\ the quark-mass configuration appears stronger than in the \gls{AWW} case.

As a curiosity, one observes that three states are generated by the first eigenvalue curve in the $0^-$-channel for the vector-quarkonium-based quark-mass configuration.
Inspecting the respective leptonic decay constants reveals that ground state and first excitation have very similar decay constants, both of which are well-separated from and much larger than that of the second excitation.
Apart from the similarity of leptonic decay constants of the ground state and the first excitation, there is no apparent reason to discredit one of these excitations.
Finally, the $0^{+}$-state corresponds to the conventional $0^{++}$-ground states of the charmonium (cf.\ Fig.~\ref{fig:MTcc}) and strangeonium (cf.\ Fig.~\ref{fig:MTss}) cases.

\subsubsection{Charmonium}

\begin{figure*}
\centerline{\includegraphics[width=.7\textwidth]{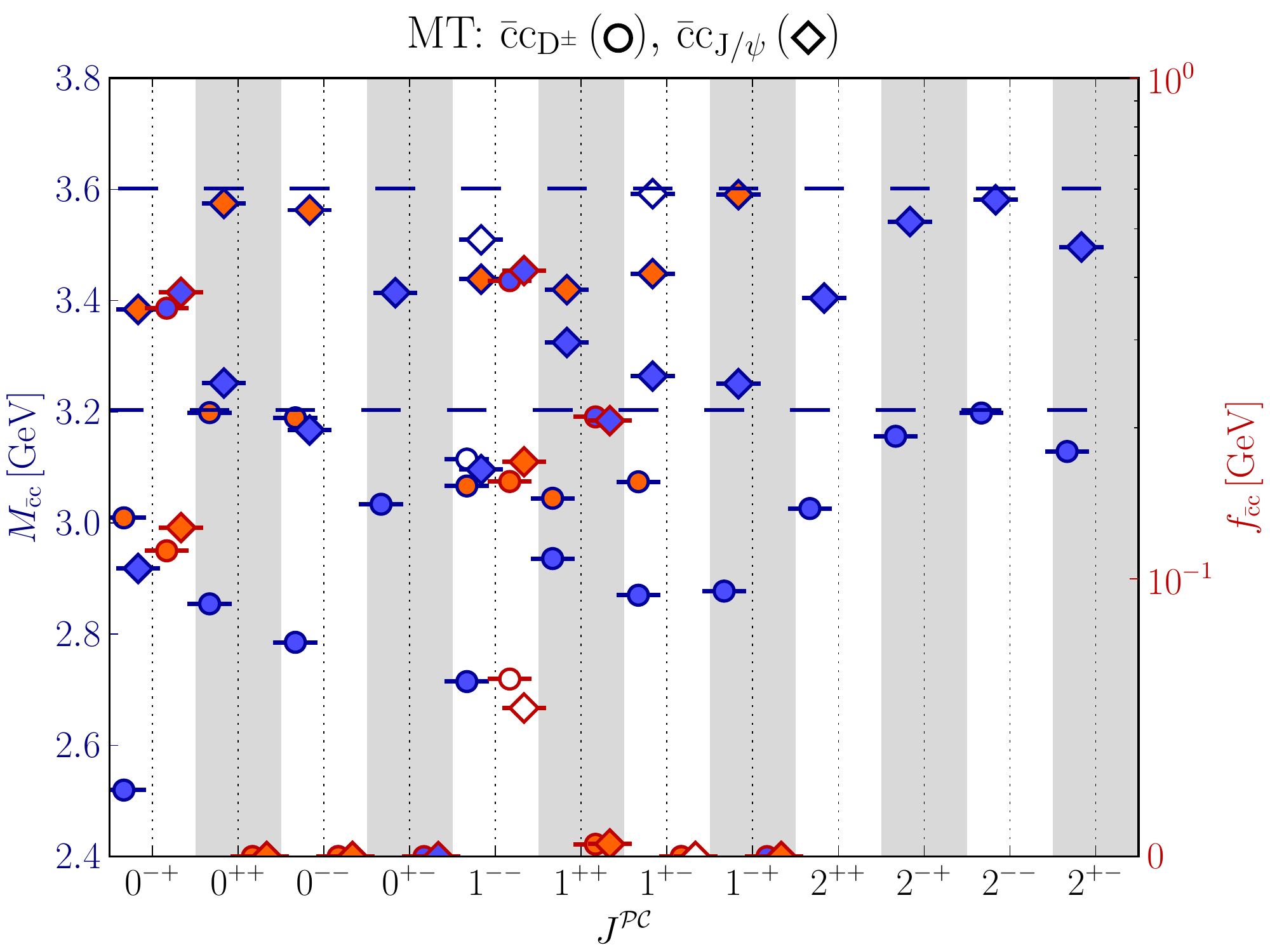}\hspace{2em}\includegraphics[width=.22\textwidth]{legend}}
\caption{Same as Fig.~\ref{fig:MTqq} for MT charmonium ($\bar{\mathrm{c}}\mathrm{c}$).
The quark-mass configurations are: 
charm-quark mass $m_\mathrm{c} = 695\,\mathrm{MeV}$ fitted to the $D$-meson mass (circles);
charm-quark mass $m_\mathrm{c} = 855\,\mathrm{MeV}$ fitted to the $\mathrm{J}/\psi$ mass (diamonds).
Detailed figure description in Sec.~\ref{sec:plotdescription}.
}
\label{fig:MTcc}
\end{figure*}

In the charmonium \gls{MT} results, presented in Fig.~\ref{fig:MTcc}, more sub-pole-threshold states have been found than in the \gls{AWW} case (cf.\ Fig.~\ref{fig:AWWcc}), in accordance with the usual pattern.
In particular, there are first excitations in the $0^{++}$, $0^{--}$, and $1^{-+}$ channels, a ground state in the $2^{--}$ channel, and a second $1^{+-}$ excitation for the vector-charmonium-based quark-mass configuration.
Similarly to other settings, the mass spectrum shows a larger dependence on the quark-mass configuration than the leptonic decay constants.

An interesting issue is the level ordering of the first and second $1^{--}$ excitations with respect to the size of the leptonic decay constant, which is reversed compared to the corresponding \gls{AWW} setting. 
Together with a larger mass splitting, this hints at a certain model dependence of the ordering of $S$- and $D$-wave states in the vector channel of charmonium in our setup.

\subsubsection{Charm and Bottom Quarks}

\begin{figure}
\includegraphics[width=.48\textwidth]{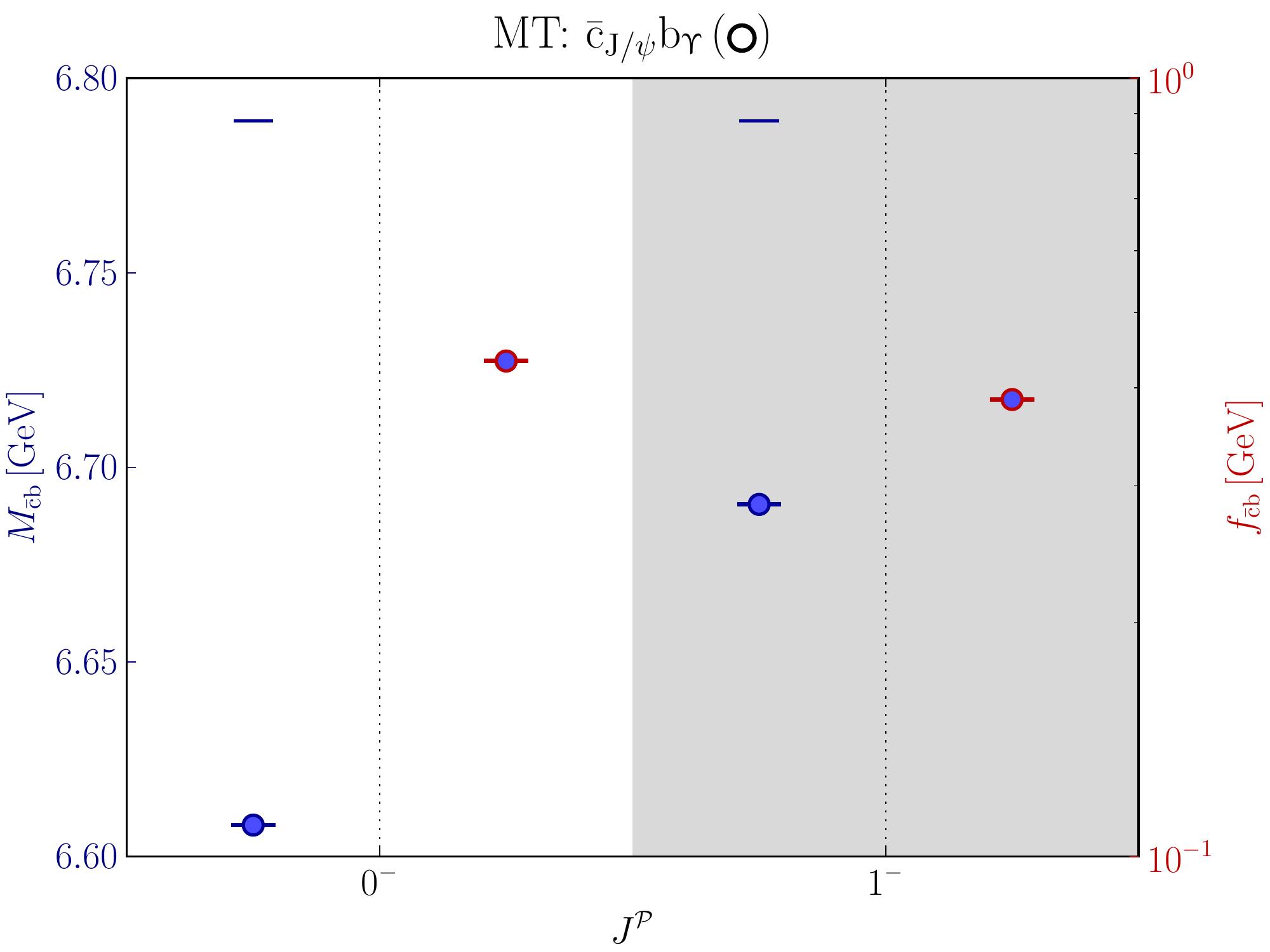}
\caption{Same as Fig.~\ref{fig:MTqq} for MT bottom charmed mesons ($\bar{\mathrm{c}}\mathrm{b}$).
Bottom-quark mass $m_\mathrm{b} = 3.77\,\mathrm{GeV}$ and charm-quark mass $m_\mathrm{c} = 855\,\mathrm{MeV}$ each fitted to the vector-quarkonium ground state.
Detailed figure description in Sec.~\ref{sec:plotdescription}.
}
\label{fig:MTcb}
\end{figure}

The bottom-charmed sector, shown in Fig.~\ref{fig:MTcb}, is as sparsely populated with sub-pole-threshold states as in the \gls{AWW} case (cf.\ Fig.~\ref{fig:AWWcb}).
The main difference is that the accessible sub-pole-threshold states are generated by the first eigenvalue and not the third as for the \gls{AWW} model.
However, it is hard to tell if the found states can actually be uniquely identified by mere inspection of the bound-state masses, which differ only at the percent level ($\leq 5\,\%$).
On the other hand, the discrepancy for the leptonic decay constants is sizable, by a factor $2$ to $3$.
If one assumes that the \gls{AWW} states are second excitations, it is plausible to suspect that the found \gls{AWW} $1^-$ state is a quasi-exotic state, because the second excitation in the $1^-$ channel can be associated with the exotic $1^{-+}$ state.
The \gls{AWW} $0^-$ state, instead, would plausibly be conventional.

\subsubsection{Bottomonium}

\begin{figure*}
\centerline{\includegraphics[width=.7\textwidth]{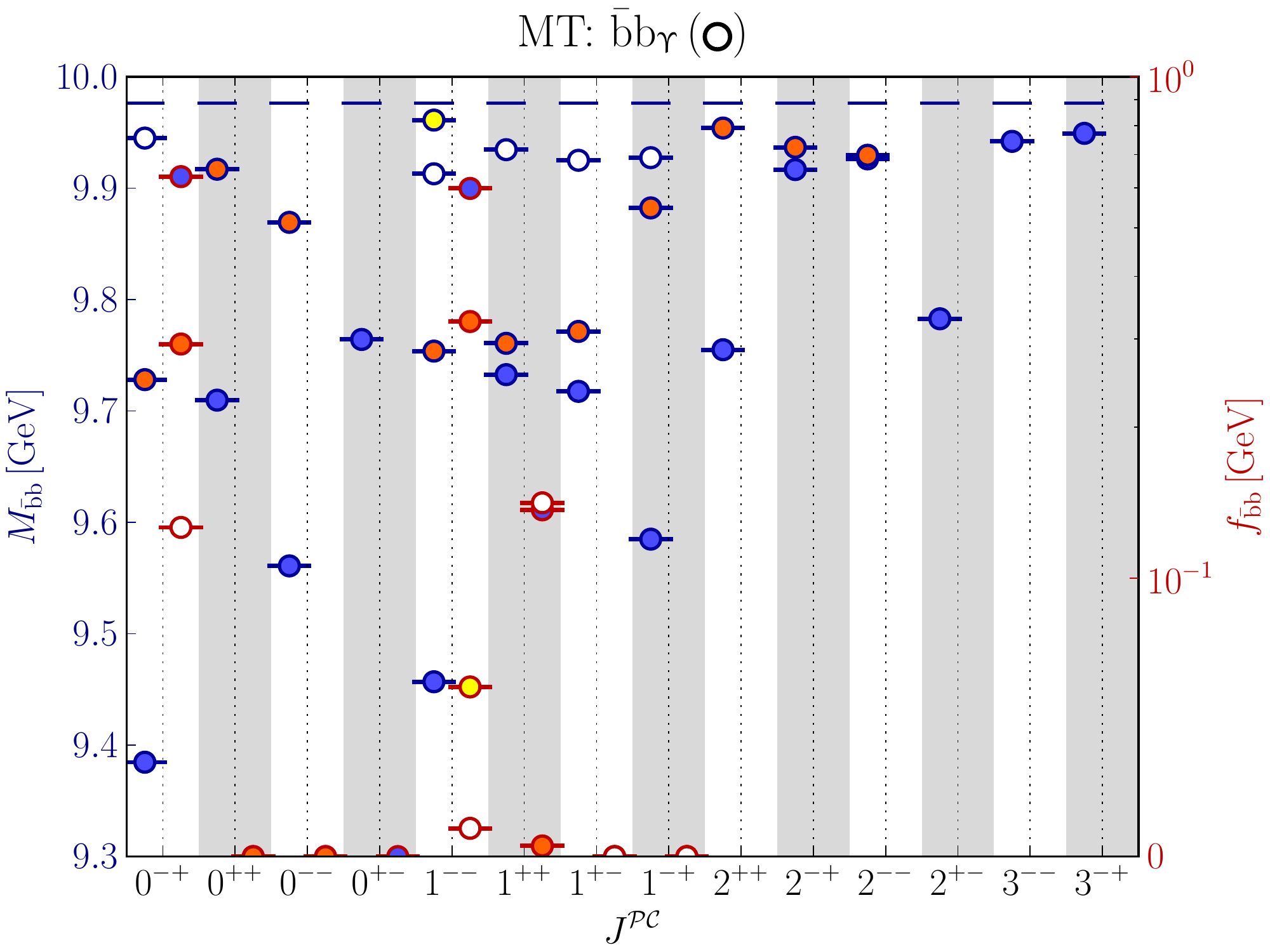}\hspace{2em}\includegraphics[width=.22\textwidth]{legend}}
\caption{Same as Fig.~\ref{fig:MTqq} for MT bottomonium ($\bar{\mathrm{b}}\mathrm{b}$).
Bottom-quark mass of $m_\mathrm{b} = 3.77\,\mathrm{GeV}$ fitted to the $\Upupsilon$ mass.
Detailed figure description in Sec.~\ref{sec:plotdescription}.
}
\label{fig:MTbb}
\end{figure*}

The bottomonium results, depicted in Fig.~\ref{fig:MTbb}, contain the same channels and sub-pole-threshold states as the corresponding \gls{AWW} set shown in Fig.~\ref{fig:AWWbb}.
Since no solution can be found for the $B$-meson case, the only quark-mass configuration is fitted to the vector-bottomonium mass.

An interesting difference can be found in the $2^{--}$ channel, where ground state and first excitation are almost degenerate.
Furthermore, the leptonic decay constants in the $1^{--}$ channel are monotonically decreasing in size with excitation, contrary to the \gls{AWW} case, where the third excitation has a larger decay constant than the second one. 
This again hints at a certain model-sensitivity of the level orderings of $S$- and $D$-wave excitations.

\subsubsection{In-Hadron Condensates}

\begin{figure}
\includegraphics[width=.48\textwidth]{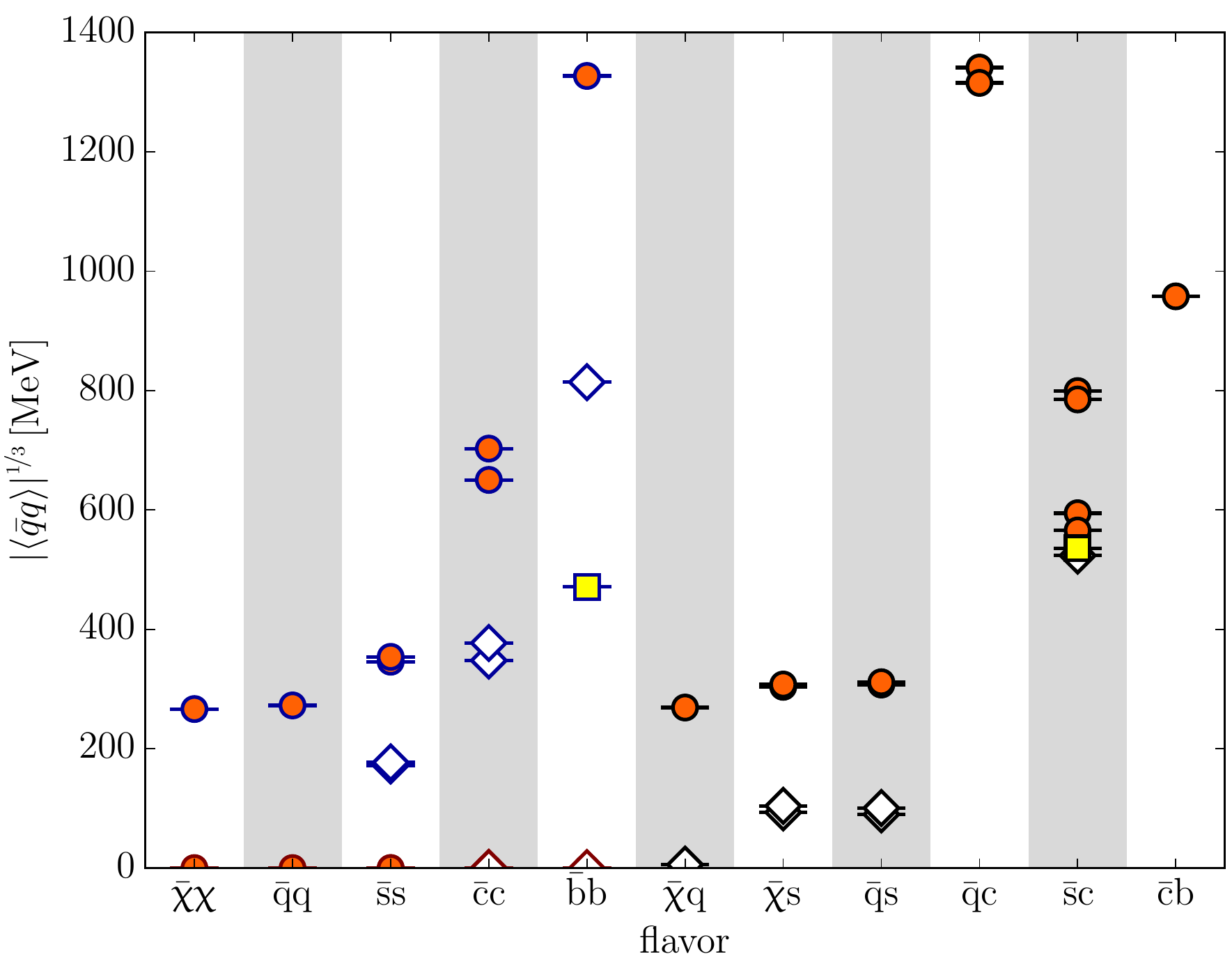}
\caption{$J^\mathcal{P(C)} = 0^{-(\mathcal{C})}$ in-hadron condensates according to definition \eqref{eq:qq} in the \gls{MT} model \eqref{eq:MT} for all accessible flavour and charge-parity combinations.
Blue marker edges refer to $\mathcal{C}=+1$, red to $\mathcal{C}=-1$, and black to undefined $\mathcal{C}$ for open-flavour mesons.
Marker face colours and marker shapes refer to eigenvalue numbering in that particular $\mathcal{C}$-channel (ground state -- red circles, first excitation -- white diamonds, second excitation -- yellow squares), the quark-mass configurations are not distinguished here.}
\label{fig:MTqbarq}
\end{figure}

\begin{figure*}
  \centering
 \begin{subfigure}[t]{0.8\textwidth}
  \centering
	\includegraphics[width=\textwidth]{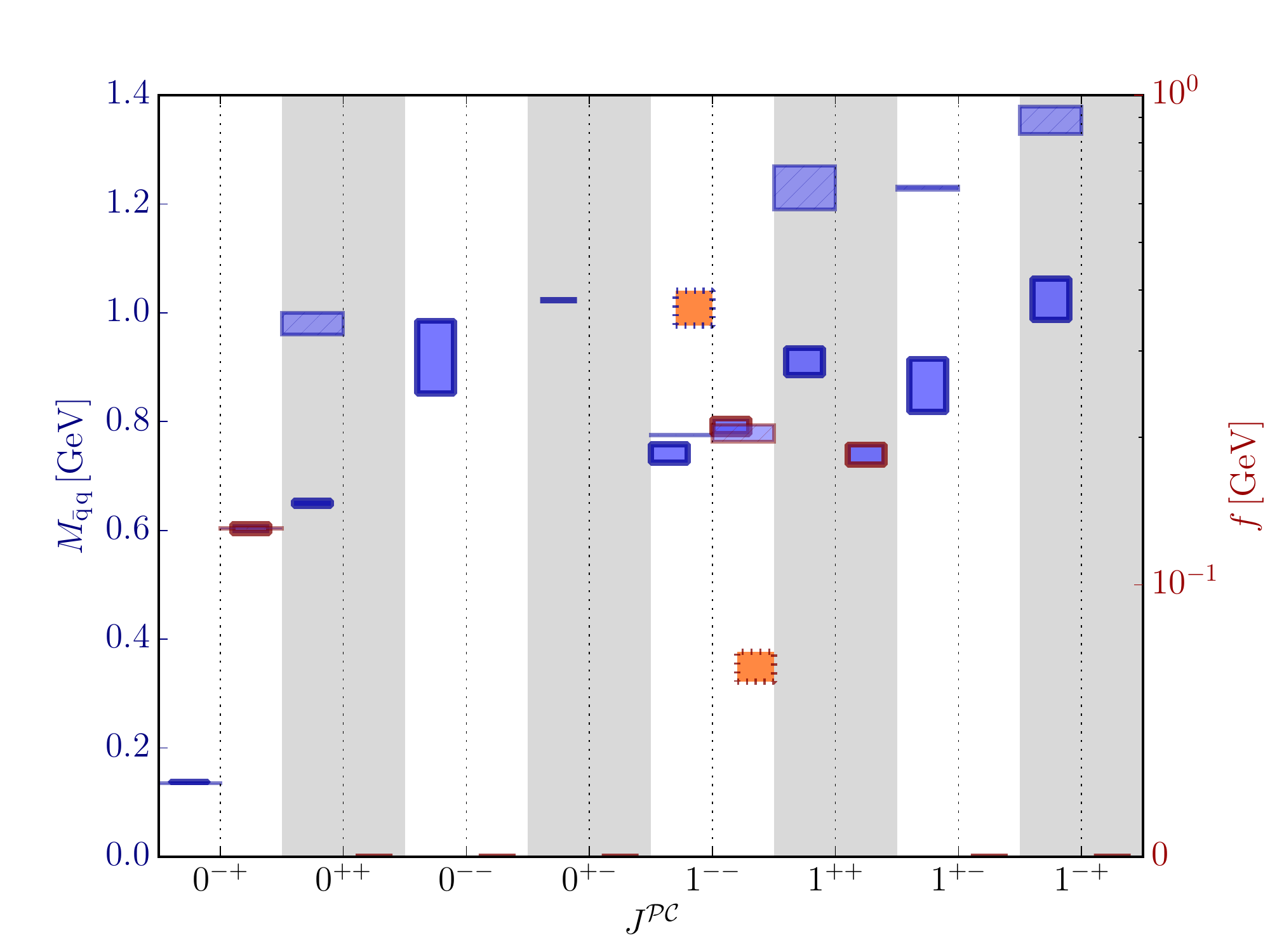}
	\caption{$\bar{\mathrm{q}} \mathrm{q}$}		
 \end{subfigure}
 \begin{subfigure}[t]{0.8\textwidth}
  \centering
	\includegraphics[width=\textwidth]{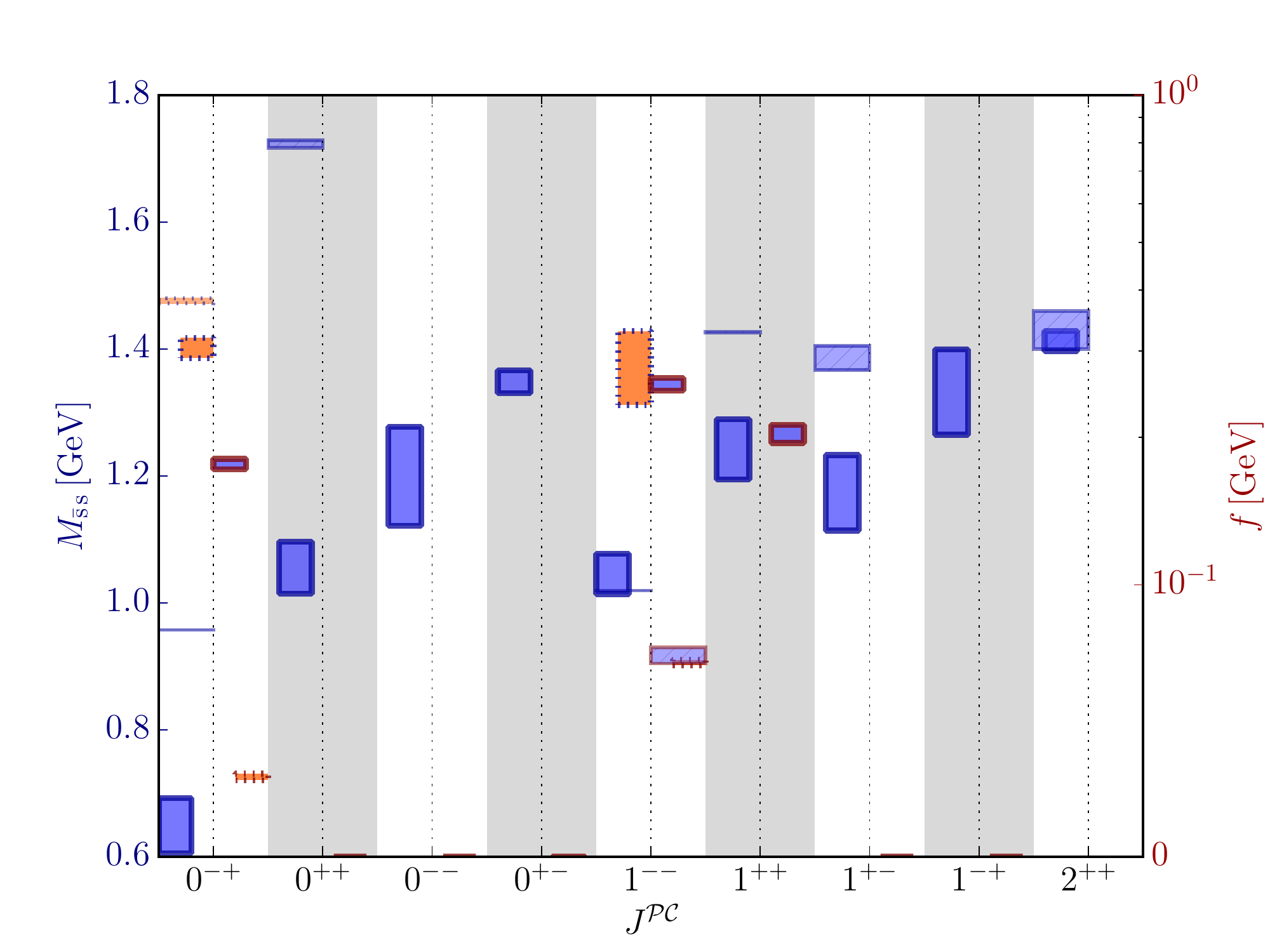}
	\caption{$\bar{\mathrm{s}} \mathrm{s}$}		
 \end{subfigure}
 	\caption{Light quarkonia masses and leptonic decay constants in comparison to experimental data.
	Both the AWW and MT interactions have been employed to set the vertical height of the narrow boxes. 
	The height of the wide boxes is set by experimental uncertainties. The fill colours of the boxes are blue (ground state), orange (\nth{1} excitation), green (\nth{2} excitation), and magenta (\nth{3} excitation).
	Blue borders mark meson masses, red borders mark meson leptonic decay constants.
}
\label{fig:expqq1}
\end{figure*}

\begin{figure*}
  \centering
 \begin{subfigure}[t]{0.8\textwidth}
  \centering
	\includegraphics[width=\textwidth]{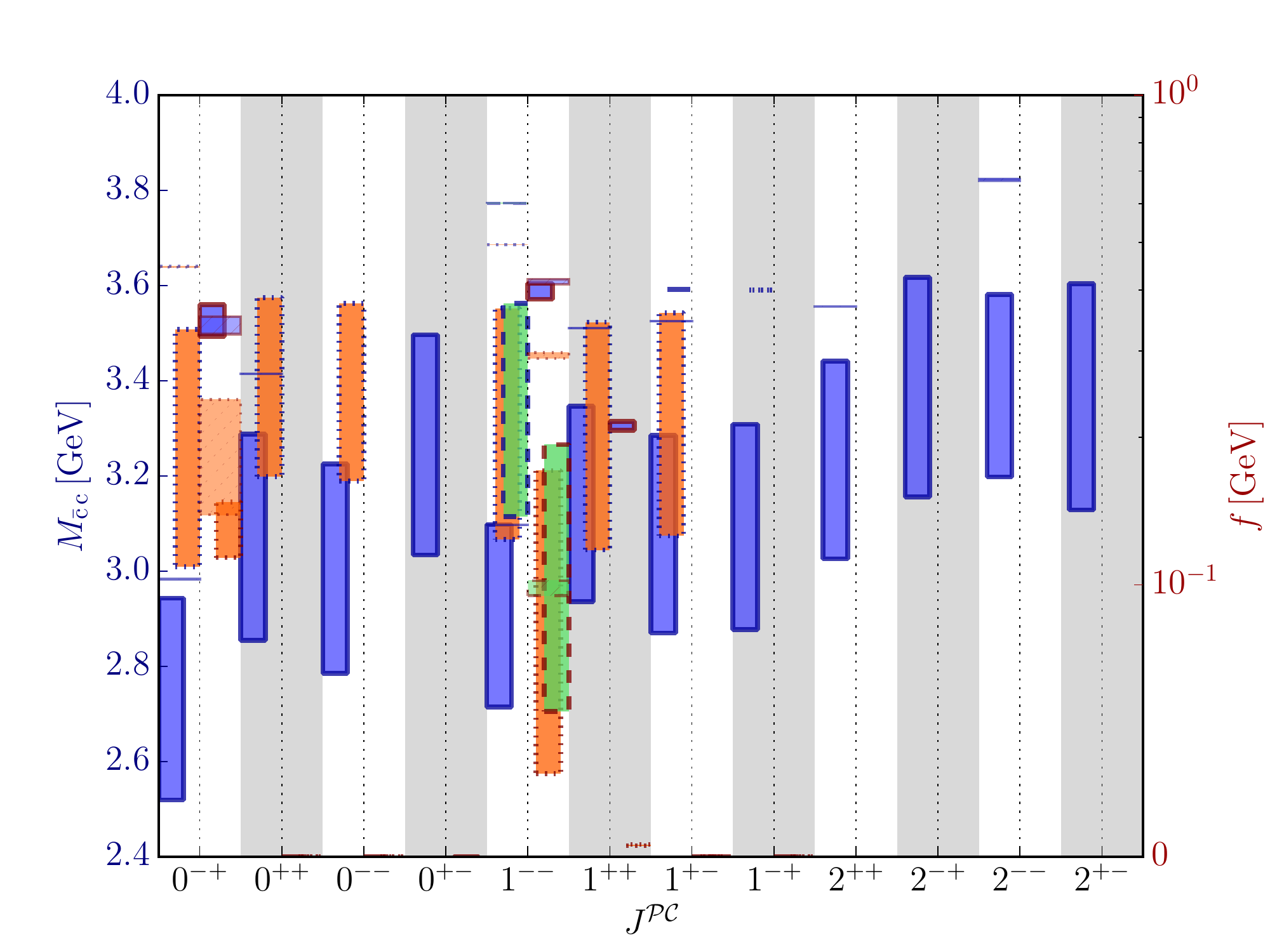}
	\caption{$\bar{\mathrm{c}} \mathrm{c}$}		
 \end{subfigure}
 \begin{subfigure}[t]{0.8\textwidth}
  \centering
	\includegraphics[width=\textwidth]{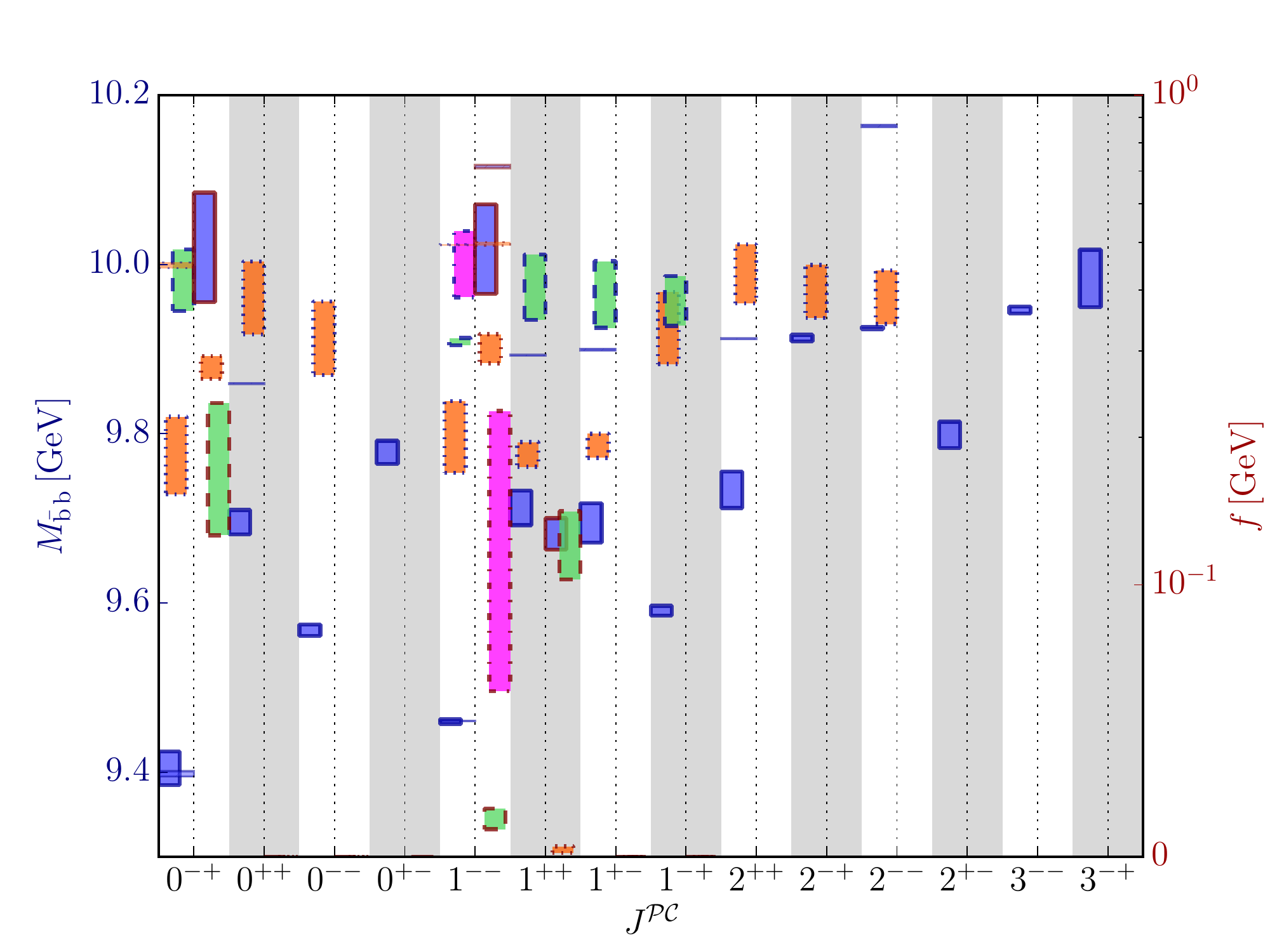}
	\caption{$\bar{\mathrm{b}} \mathrm{b}$}		
 \end{subfigure}
 	\caption{Heavy quarkonia masses and leptonic decay constants in comparison to experimental data.
	Both the AWW and MT interactions have been employed to set the vertical height of the narrow boxes. 
	The height of the wide boxes is set by experimental uncertainties. The fill colours of the boxes are blue (ground state), orange (\nth{1} excitation), green (\nth{2} excitation), and magenta (\nth{3} excitation).
	Blue borders mark meson masses, red borders mark meson leptonic decay constants.
}
\label{fig:expqq2}
\end{figure*}

\begin{figure*}
  \centering
 \begin{subfigure}[t]{0.8\textwidth}
  \centering
	\includegraphics[width=\textwidth]{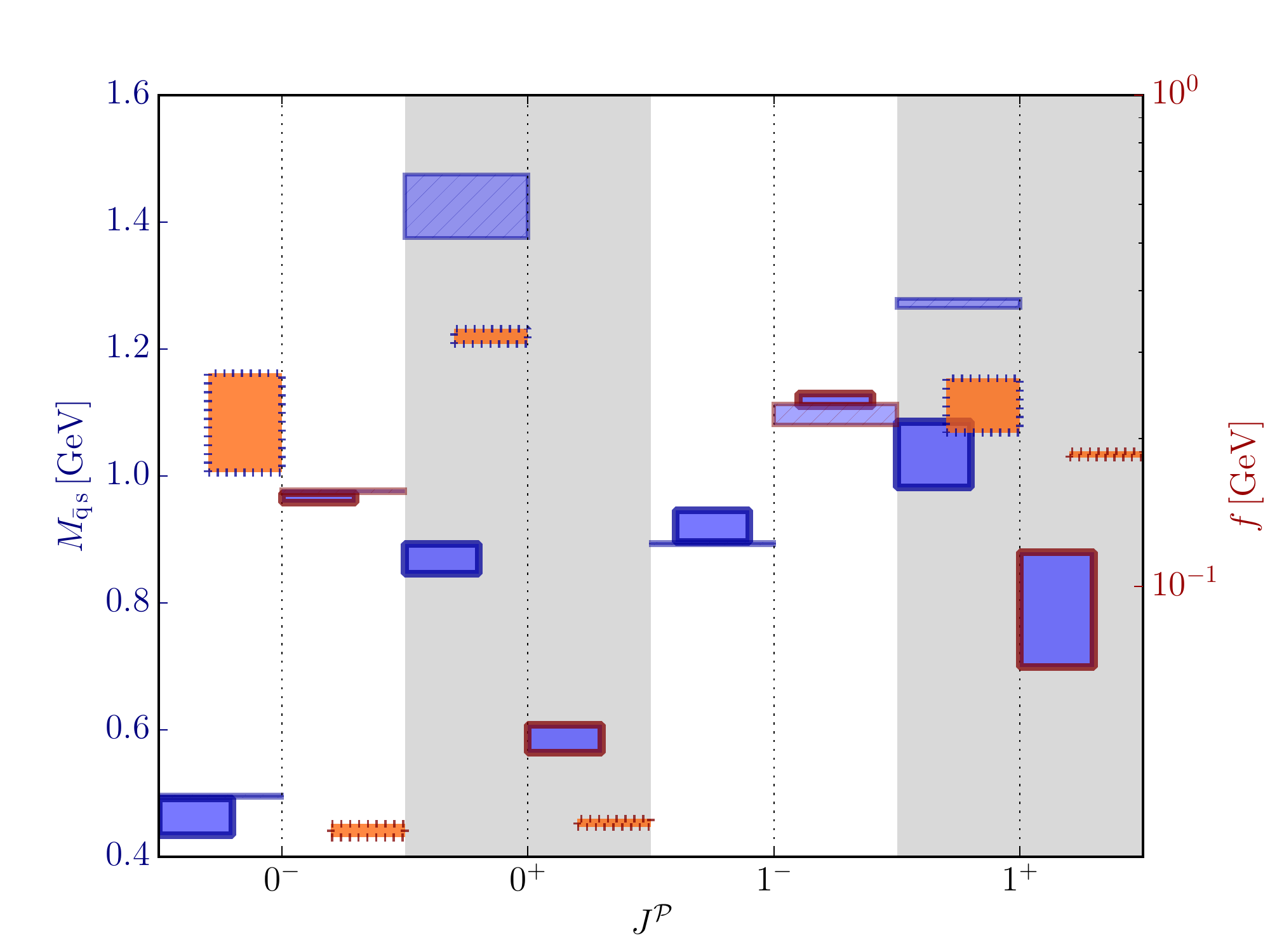}
	\caption{$\bar{\mathrm{q}} \mathrm{s}$}		
 \end{subfigure}
 \begin{subfigure}[t]{0.8\textwidth}
  \centering
	\includegraphics[width=\textwidth]{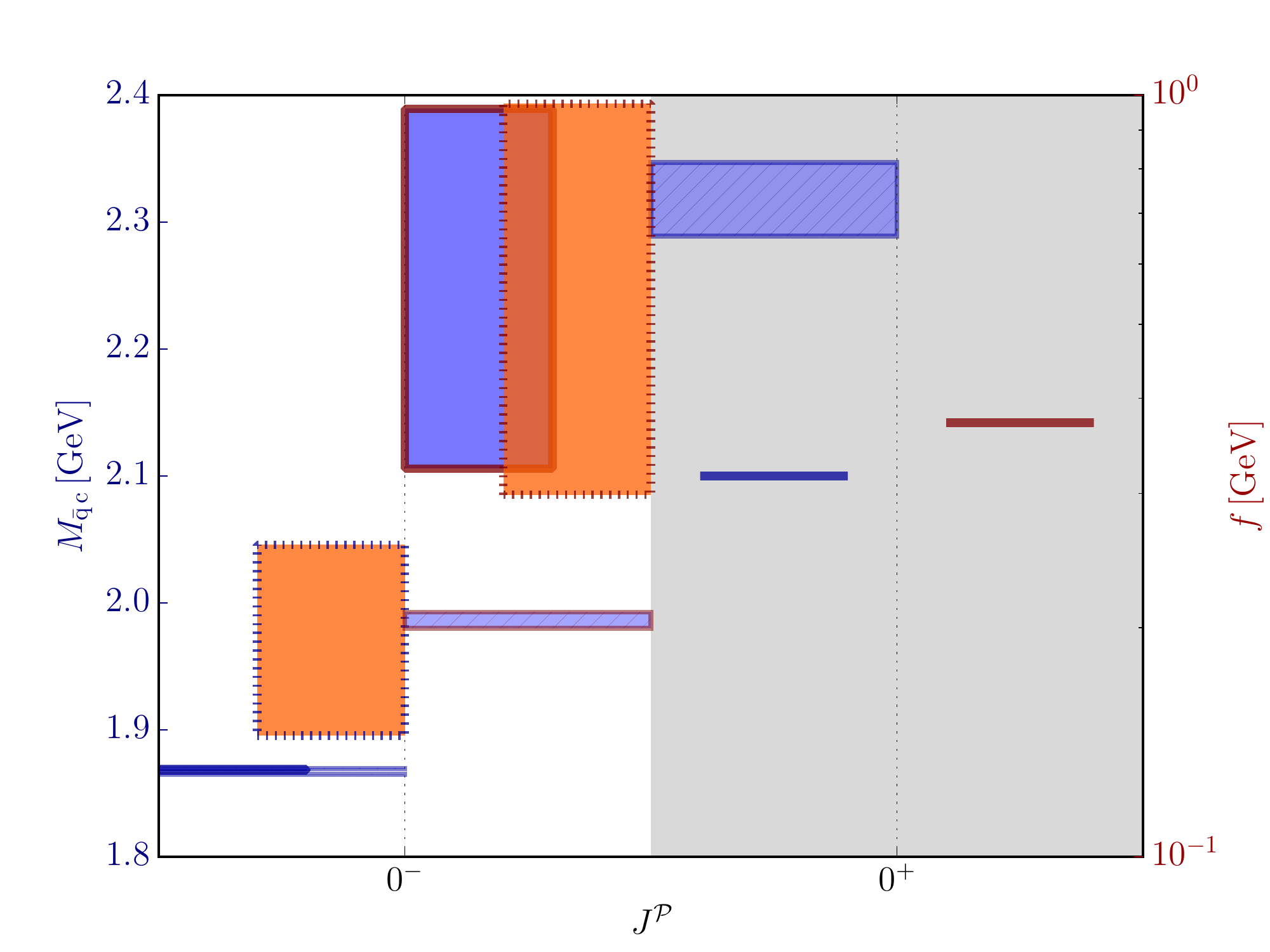}
	\caption{$\bar{\mathrm{q}} \mathrm{c}$}		
 \end{subfigure}
 	\caption{Open-flavour meson masses and leptonic decay constants as we find them in comparison to experimental data.
	Both the AWW and MT interactions have been employed to set the vertical height of the narrow boxes. 
	The height of the wide boxes is set by experimental uncertainties. The fill colours of the boxes are blue (ground state), orange (\nth{1} excitation), green (\nth{2} excitation), and magenta (\nth{3} excitation).
	Blue borders mark meson masses, red borders mark meson leptonic decay constants.
}
\label{fig:expof1}
\end{figure*}

\begin{figure*}
  \centering
 \begin{subfigure}[t]{0.8\textwidth}
  \centering
	\includegraphics[width=\textwidth]{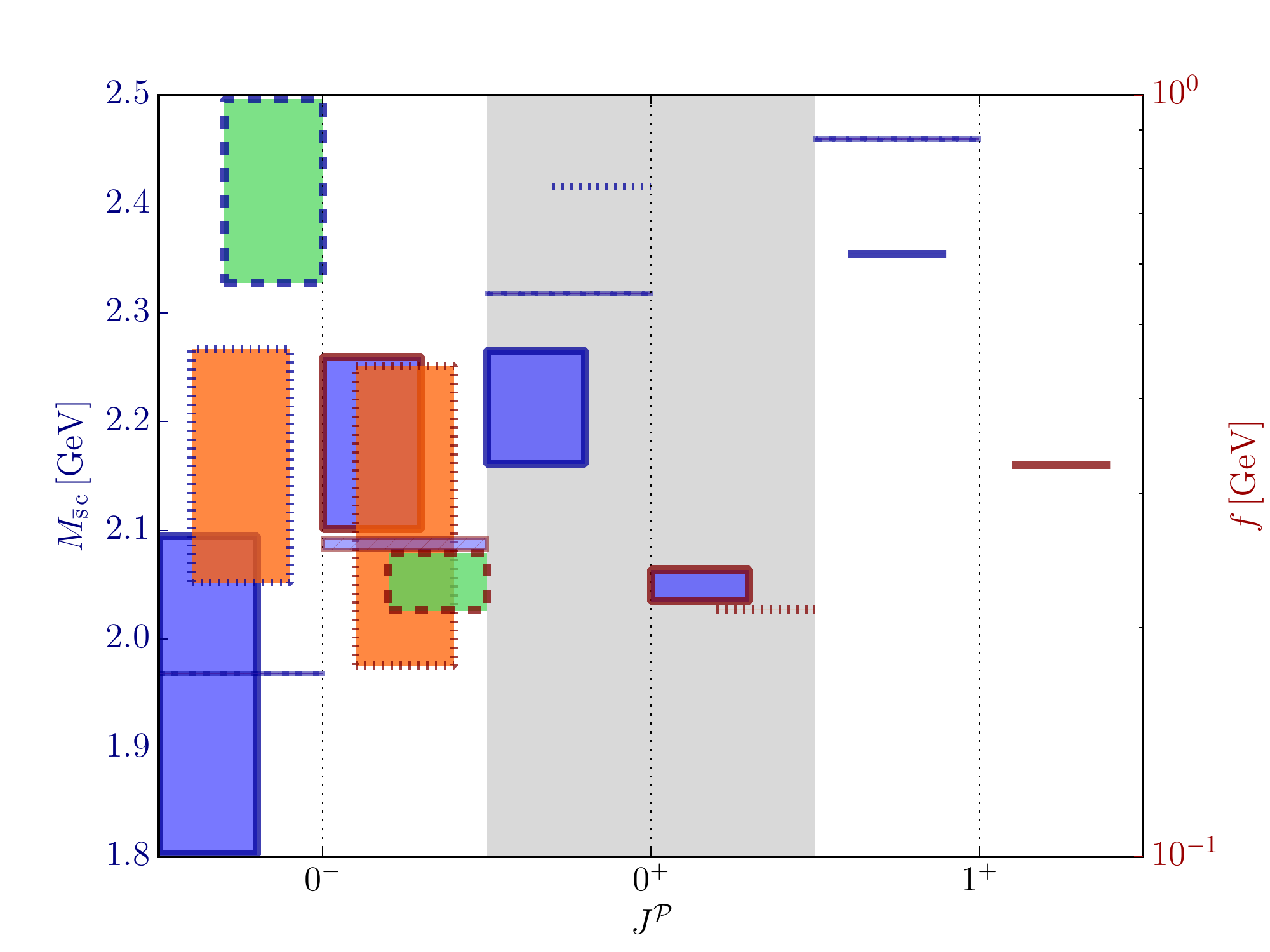}
	\caption{$\bar{\mathrm{s}} \mathrm{c}$}		
 \end{subfigure}
 \begin{subfigure}[t]{0.8\textwidth}
  \centering
	\includegraphics[width=\textwidth]{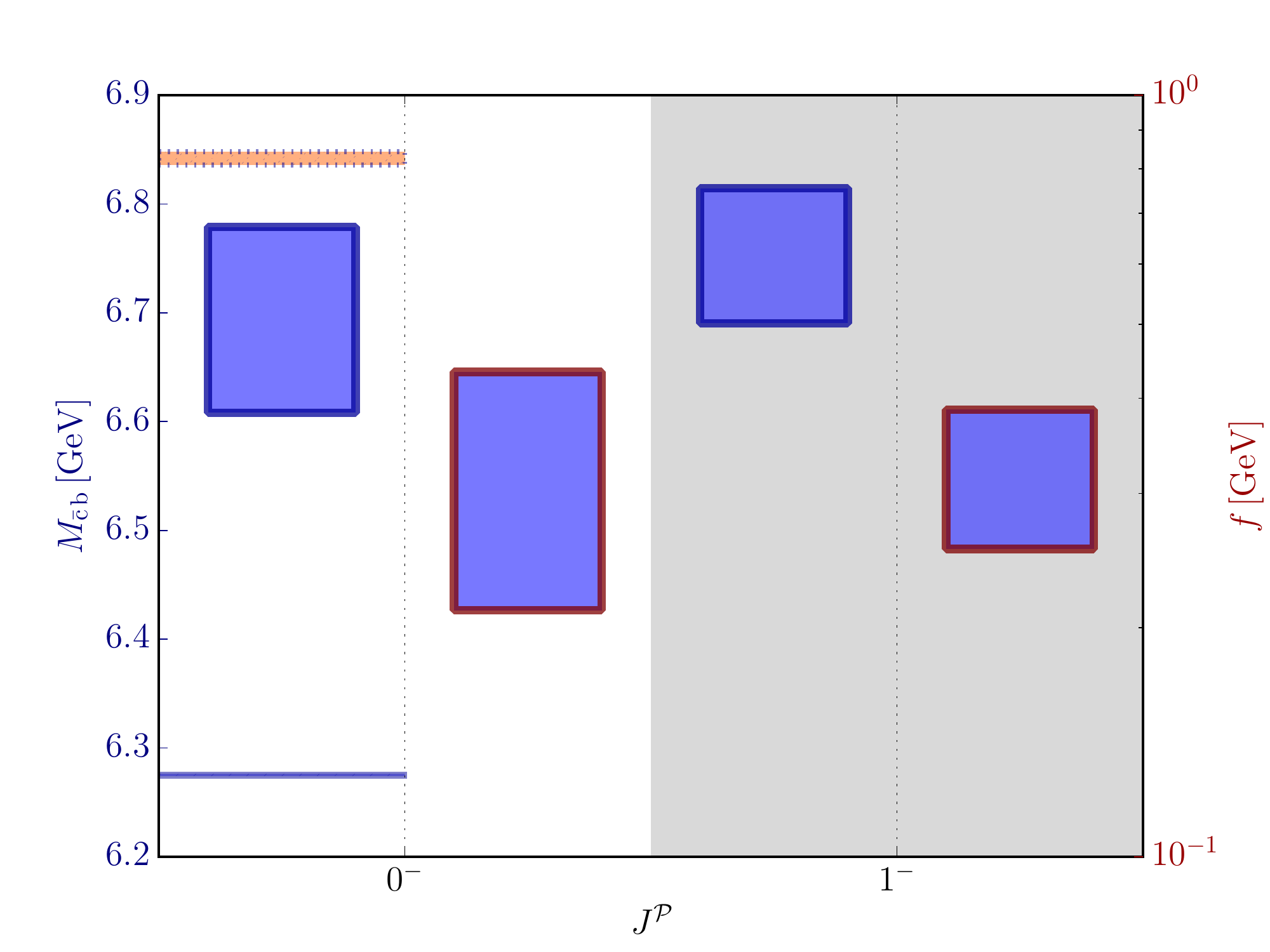}
	\caption{$\bar{\mathrm{c}} \mathrm{b}$}		
 \end{subfigure}
 	\caption{Open-flavour meson masses and leptonic decay constants as we find them in comparison to experimental data.
	Both the AWW and MT interactions have been employed to set the vertical height of the narrow boxes. 
	The height of the wide boxes is set by experimental uncertainties. The fill colours of the boxes are blue (ground state), orange (\nth{1} excitation), green (\nth{2} excitation), and magenta (\nth{3} excitation).
	Blue borders mark meson masses, red borders mark meson leptonic decay constants.
}
\label{fig:expof2}
\end{figure*}

Figure \ref{fig:MTqbarq} depicts the  \gls{MT} in-hadron condensates evaluated according to Eq.~\eqref{eq:qq}.
Qualitatively, the pattern looks similar to the \gls{AWW} in-hadron condensates in Fig.~\ref{fig:AWWqbarq}.
However, the bottomonium ground state generates a significantly larger condensate in the \gls{MT} model.
Similarly, the $D$-meson condensates are rather large as compared to the bottomonium and the charmonium ground-state condensates.
In the $D_s$-channel, the upper two condensate values correspond to the lowest two states of the vector-quarkonium-based quark-mass configuration.
They are well separated from their third state and all states of the $0^-$-open-flavour-based quark-mass-configuration condensates.

\section{Comparison to Experimental Data}\label{sec:exp}

After the technical and methodological analysis of our results in Sec.~\ref{sec:results}, we now make a comparison to experimental data.
A few comments are in order at this point. 
First of all, our study is an as-comprehensive-as-possible presentation of masses and leptonic decay constants in our setup.
As such, it has facets of a survey which aims at a bigger picture and not at a precise reproduction of every detail.
On the other hand, it is limited by technicalities which are objects of ongoing investigation and development and is thus incomplete. 
Still, it can be viewed as a baseline study which we aim to improve upon in the future.

The model parameters for the effective interactions are preset by earlier studies and taken to be representative of a typical effort of this kind.
They are fitted to meson properties as well as \gls{DCSB} itself, i.\,e., mainly anchored to the light-quark domain.
The quark masses have been fitted in two different ways, as explained above, to investigate in part the model dependence of our setup.
The other model dependence we allow for is the \gls{UV} behaviour of the effective interaction, which is detailed in Sec.~\ref{sec:effint}.
Both of these are a measure of the truncation effects in our approach, since the role played by various \emph{Ans\"atze} diminishes with the level of sophistication of the truncation within the systematic truncation scheme employed.
Note that no fine-tuning has been performed at all, neither at the level of the quarkonium cases, nor after fixing the quark masses in the open-flavour cases.
In summary, there are many ways to improve our results beyond the scope of the present analysis.

To illustrate this and to provide an instructive picture as well as an as-is comparison with experimental data, we summarize our results from the above discussion in such a way that all data with the same quark content are combined over different models as well as quark-mass configurations and plotted as a box which visualizes the minimum-to-maximum range of calculated values.

These boxes are plotted in different colours for different excitations in each $J^\mathcal{P(C)}$ channel, slightly offset from each other, in analogy to the figures in Sec.~\ref{sec:results}, with the following colour coding: 
blue borders -- meson masses; red borders -- meson leptonic decay constants.
Fill colours are blue (ground state), orange (\nth{1} excitation), green (\nth{2} excitation), and magenta (\nth{3} excitation).
Note that our results are now ordered and marked according to their level of excitation, rather than the index of the respective eigenvalue.

Experimental data are plotted as wider boxes spanning the entire width of each channel and in the colour code corresponding to the calculated results.
The data are taken from the PDG review \cite{Olive:2016xmw}; box heights represent experimental uncertainties. 
If the experimental value is yet unknown, we simply do not plot a wide box; a prominent example for this is the mass of the $B_c^*$ meson \cite{Gomez-Rocha:2016cji}.

In Figs.~\ref{fig:expqq1} and \ref{fig:expqq2} we present the comparison for all (light, strange, charm, and bottom) quarkonia.
In essence, this shows both strengths and weaknesses of the not-fine-tuned setup, in particular, since we use one unified set of model parameters for all quark masses, in contrast to previous studies \cite{Hilger:2014nma,Hilger:2015hka,Hilger:2015ora}. In fact an possible additional step in parameter freedom, which we did not emply here, and which is not easily applicable to the open-flavour case, is sketched below in \ref{sec:intdependent}. However, such an Ansatz remains inconclusive at this point.

Nonetheless, the landmarks are well reproduced, i.\,e., we have good descriptions of the pion, $\uprho$, and $\upphi$ masses and leptonic decay constants as well as, for the heavy pseudoscalar and vector quarkonia, of both their masses and leptonic decay constants, where available. 
The quality of the description for orbital or radial excitations varies as expected, due to the missing flexibility of the combination of truncation and model freedom. In summary, this is neither new nor surprising.

The set of actually new results are the collected masses and decay constants for the open-flavour cases in Figs.~\ref{fig:expof1} and \ref{fig:expof2}.
While the description is excellent in the strange sector, in particular in the pseudoscalar and vector channels, the heavy-light systems as well as the bottom-charmed case is, at least, promising for future studies.
As always in the \gls{RL} truncation, one can argue for cancellation effects in the pseudoscalar and vector channels, but, as already mentioned above, the imbalance of a heavy-light system is an important factor.
In addition, we find a rather pronounced model dependence in our set of results, as illustrated by the box sizes in Figs.~\ref{fig:expof1} and \ref{fig:expof2}.

While our results are often consistent with experiment, our model uncertainties are easily of the order of 10\%, which is rather sizable, particularly if considering the precision achieved for heavy quarkonia.
This observation is valid for both masses and leptonic decay constants, but still not enough to rule out sophisticatedly tuned \gls{RL} studies as a means to study open-flavour mesons.
In particular, it will be interesting to see how technical improvements can influence the amount of both model dependence and variation due to quark-mass changes.

\section{Conclusions and Outlook}
\label{sct:conclucions}

We present an as-comprehensive-as-possible account of quarkonia and open-flavour meson states in the \gls{DSBSE} approach together with a thorough discussion of technical issues surrounding this topic.
The study uses two forms and parameter sets for the effective model dressed-gluon interaction in the \gls{RL}-truncated setup, in order to estimate the model dependence of our results. 
However, for each set, both interaction and parameter values are kept the same for all quark-mass values.
We also investigate the effect of using different anchoring and fitting strategies for fixing the quark-mass values to either heavy vector quarkonia or open-flavour pseudoscalar-meson ground states.

In addition to meson masses and leptonic decay constants, we present an account of in-hadron quark condensates together with an extended discussion of issues surrounding this definition of condensates in \gls{QCD}.
Our results are first analysed and then compared to experimental data, where available. 
This comparison is encouraging in that we have not made any attempt at fine-tuning model parameters, in particular as functions of the quark mass, as was done successfully in previous studies for quarkonia.
We elucidate part of the reasons in an (unsuccessful) attempt to accommodate different model-parameter sets in an \gls{RL}-truncated setup.

We detail the three major technical issues for open-flavour studies in the \gls{DSBSE} approach, namely, the pole threshold induced by non-analyticities of the quark propagators in the complex squared-momentum plane, complex conjugated eigenvalues, and the non-monotonicity of eigenvalues as functions of $P^2$.
We exemplify in which sectors of the meson spectrum these issues become relevant and to what extent they can be circumvented.
While limitations induced by the pole threshold are well under control either by staying below it, as demonstrated in this work, or by employing well constrained extrapolation methods and extracting beyond-pole-threshold information from below-threshold eigenvalue curves, the occurrence of complex conjugated eigenvalues turns out to be a major issue as it blurs the fate of excitations.
Future investigations will reveal if the promising approach of \cite{Eichmann:2016nsu} will solve this issue.
It might well be that also the issue of non-monotonicity is resolved by this approach.

We identify $J=3$ as the largest spin that can be analysed below the pole threshold.
We also exemplify that in-hadron condensates may be utilized as a means of state identification.
We anticipate that our next steps will improve the results, namely, a set of fine-tuning procedures, which can also better explore the meson-mass regions beyond the reach of the present investigation. 
Furthermore, a natural ingredient of follow-up studies is an \gls{OAMD} of open-flavour states.

\begin{acknowledgements}
This work was supported by the Austrian Science Fund (FWF) under Grant No.\ P25121-N27.
\end{acknowledgements}

\appendix
\section{Renormalisation in the Chiral Limit}\label{sec:renorm}

To derive equations for the explicit renormalisation in the chiral limit, we write the quark \gls{DSE} in the form
\begin{subequations}
\label{eq:RenDSE}
\begin{align}
    A(p^2) &= Z_2 + \genfrac{\{}{\}}{0pt}{0}{1}{{Z_2}^2} \Sigma_\mathrm{V}(p^2) \, ,
    \\
    B(p^2) &= Z_4 m(\mu^2) + \genfrac{\{}{\}}{0pt}{0}{1}{{Z_2}^2} \Sigma_\mathrm{S}(p^2) \, ,
\end{align}
\end{subequations}
where $\Sigma_\mathrm{V,S}$ denote respective traces of the quark self-energy with extracted renormalisation constants.
For the upper (linear) approach, $Z_{1\mathrm{F}}$ has been absorbed in the effective interaction model $\mathcal{G}(q^2)$, while the lower (non-linear) approach explicitly accounts for the renormalisation of the \gls{DSE} kernel \cite{Eichmann:2009zx}.
The renormalisation condition at the renormalisation scale $\mu$ reads
\begin{align*}
    A(\mu^2) &= 1 \, ,
    \\
    B(\mu^2) &= m(\mu^2) \, .
\end{align*}
In the linear approach, the renormalisation condition translates to
\begin{subequations}
\begin{align}
    Z_2 &= 1 - \Sigma_\mathrm{V}(\mu^2) \, ,
    \\
    \label{eq:LinRenZ4}
    Z_4 &= 1 - \frac{\Sigma_\mathrm{S}(\mu^2)}{m(\mu^2)} \, .
\end{align}
\end{subequations}
Insertion into \eqref{eq:RenDSE} eliminates the renormalisation constants, yielding
\begin{subequations}
\label{eq:LinRenDSE}
\begin{align}
    A(p^2) &= 1 + \Sigma_\mathrm{V}(p^2) - \Sigma_\mathrm{V}(\mu^2) \, ,
    \\
    B(p^2) &= m(\mu^2) + \Sigma_\mathrm{S}(p^2) - \Sigma_\mathrm{S}(\mu^2) \, ,
\end{align}
\end{subequations}
which can be solved the same way as an un-renormalized \gls{DSE}.
This is a very intuitive form, as any solution to it clearly satisfies the renormalisation condition by subtracting an appropriate constant.
However, any subtracted function with the correct boundary condition would do due to linearity.

For the non-linear approach, this is not possible.
Equating \eqref{eq:RenDSE} and \eqref{eq:LinRenDSE} for the non-linear approach yields a relation that cannot be solved for constant $Z_{2,4}$ and all $p^2$.
The expressions for $Z_{2,4}$ become coupled:
\begin{subequations}
\begin{align}
    Z_2 &= -\frac{1}{2 \Sigma_\mathrm{V}(\mu^2)} \left( 1 - \sqrt{1 + 4 \Sigma_\mathrm{V}(\mu^2)} \right) \, ,
    \\
    \label{eq:QuadRenZ4}
    Z_4 &= 1 - {Z_2}^2 \frac{\Sigma_\mathrm{S}(\mu^2)}{m(\mu^2)} \, .
\end{align}
\end{subequations}
Thus, a form analogous to Eqs.~\eqref{eq:LinRenDSE} cannot be obtained.

In the chiral limit, $m \to 0$, Eqs.~\eqref{eq:LinRenZ4} and \eqref{eq:QuadRenZ4} are ill-defined.
Applying the theorem of de l'Hospital thus gives
\begin{equation} \label{eq:chiralZ4}
\begin{split}
    Z_4(m(\mu^2)=0)
        & = \lim\limits_{m(\mu^2) \rightarrow 0}{Z_4(m(\mu^2))}
        \\
        & = 1 - \genfrac{\{}{\}}{0pt}{0}{1}{{Z_2}^2} \lim\limits_{m(\mu^2) \rightarrow 0}{\frac{\partial\Sigma_\mathrm{S}(\mu^2)}{\partial m(\mu^2)}} \, ,
\end{split}
\end{equation}
which relates $Z_4$ to the quark-mass derivative of the quark self-energy's scalar projection.
With the r.\,h.\,s.\ of Eq.~\eqref{eq:RenDSE} being linear in the quark propagator,
\begin{align}
    A(p^2) &= Z_2 + \genfrac{\{}{\}}{0pt}{0}{1}{{Z_2}^2} C_\mathrm{F} \int_q \mathcal{G}((p-q)^2) K_\mathrm{V}(p,q) \sigma_\mathrm{V}(q^2) \, ,
    \\ \nonumber
    B(p^2) &= Z_4 m(\mu^2) \\&+ \genfrac{\{}{\}}{0pt}{0}{1}{{Z_2}^2} C_\mathrm{F}\int_q \mathcal{G}((p-q)^2) K_\mathrm{S}(p,q) \sigma_\mathrm{S}(q^2) \, ,
\end{align}
and the quark-mass derivatives of the dressing functions 
\begin{subequations}
\begin{align}
    \nonumber
    \frac{\partial \sigma_\mathrm{V}}{\partial m}
        & =  \left( \frac{\sigma_\mathrm{V}}{A} - 2 p^2 \sigma_\mathrm{V}^2\right) \frac{\partial A}{\partial m(\mu^2)}
            - 2 \sigma_\mathrm{V} \sigma_\mathrm{S} \frac{\partial B}{\partial m(\mu^2)}
    \\
        & \equiv M_\mathrm{AA} \frac{\partial A}{\partial m(\mu^2)} + M_\mathrm{AB} \frac{\partial B}{\partial m(\mu^2)} \, ,
    \\
    \nonumber
    \frac{\partial \sigma_\mathrm{S}}{\partial m}
        & = \left( \frac{\sigma_\mathrm{S}}{B} - 2 \sigma_\mathrm{S}^2 \right) \frac{\partial B}{\partial m(\mu^2)}
            - 2 p^2 \sigma_\mathrm{S} \sigma_\mathrm{V} \frac{\partial A}{\partial m(\mu^2)}
    \\
        & \equiv M_\mathrm{BB} \frac{\partial B}{\partial m(\mu^2)} + M_\mathrm{BA} \frac{\partial A}{\partial m(\mu^2)} \, ,
\end{align}
\end{subequations}
the quark-mass derivatives are given by
\begin{align}
    \frac{\partial A(p^2)}{\partial m(\mu^2)}
        &=\genfrac{\{}{\}}{0pt}{0}{1}{{Z_2}^2} C_\mathrm{F} \int_q \mathcal{G}((p-q)^2) K_\mathrm{V}(p,q) \frac{\partial \sigma_\mathrm{V}(q^2)}{\partial m(\mu^2)} \, ,
    \\
    \frac{\partial B(p^2)}{\partial m(\mu^2)}
        &= Z_4 + \genfrac{\{}{\}}{0pt}{0}{1}{{Z_2}^2} C_\mathrm{F} \nonumber
        \\&\times \int_q \mathcal{G}((p-q)^2) K_\mathrm{S}(p,q) \frac{\partial \sigma_\mathrm{S}(q^2)}{\partial m(\mu^2)} \, .
\end{align}
$K_\mathrm{V}$ and $K_\mathrm{S}$ are the integration kernels in $\Sigma_\mathrm{V}$ and $\Sigma_\mathrm{S}$, see \ref{sec:DSEcontour}.
For known propagator functions $A$, $B$, and renormalisation constants $Z_2$, $Z_4$, this is a linear, inhomogeneous, coupled integral equation for $\frac{\partial A(p^2)}{\partial m(\mu^2)}$ and $\frac{\partial B(p^2)}{\partial m(\mu^2)}$.
It can be solved by virtue of standard integral equation methods, such as fixed-point iteration, Newton-Krylov optimisation, matrix inversion, or, which often is more suitable, a direct linear equation solver.
We employed all of these, tested and confirmed their applicability.
For the scalar projection of the quark self-energy, one obtains
\begin{equation}
    \frac{\partial \Sigma_\mathrm{S}(\mu^2)}{\partial m(\mu^2)} =
    C_\mathrm{F} \int_q \mathcal{G}((p-q)^2) K_\mathrm{S}(p,q) \frac{\partial \sigma_\mathrm{S}(q^2)}{\partial m(\mu^2)} \, .
\end{equation}
In the chiral limit, $Z_4$ is not known and, therefore, a solution is not possible.
However, inserting \eqref{eq:chiralZ4} gives
\begin{equation}
    \frac{\partial B(p^2)}{\partial m(\mu^2)} = 1 \, .
\end{equation}
Hence, the equations decouple in the chiral limit and one is left with
\begin{equation} \label{eq:chiralmDrvtA}
\begin{split}
    \frac{\partial A(p^2)}{\partial m(\mu^2)}
        =\genfrac{\{}{\}}{0pt}{0}{1}{{Z_2}^2} C_\mathrm{F} \int_q \mathcal{G}((p-q)^2) K_\mathrm{V}(p,q)
        \\\times\left(M_\mathrm{AA} \frac{\partial A}{\partial m(\mu^2)} + M_\mathrm{AB} \right) \, ,
\end{split}
\end{equation}
a linear, inhomogeneous integral equation for $\frac{\partial A(p^2)}{\partial m(\mu^2)}$.
As now all functions and constants are known and the solution is unique, we choose matrix inversion to solve this equation.
The final expression for $Z_4$ in the chiral limit is thus given by
\begin{multline}
    Z_4(m(\mu^2)=0) = 1 - \genfrac{\{}{\}}{0pt}{0}{1}{{Z_2}^2} C_\mathrm{F}
        \left[ \int_q \mathcal{G}((p-q)^2) K_\mathrm{S}(p,q) \right.
        \\ \times \left.
        \left( M_\mathrm{BB} + M_\mathrm{BA} \frac{\partial A}{\partial m(\mu^2)} \right) \right]_{p^2 = \mu^2} \, ,
\end{multline}
with $\frac{\partial A}{\partial m(\mu^2)}$ the solution of \eqref{eq:chiralmDrvtA}.

\section{Review, Systematisation, and Generalisation of, and Pedagogical Introduction to Solving the DSE in the Complex Plane with Variable Momentum Routing}
\label{sec:DSEcontour}

In order to apply the combined \gls{DSE}-\gls{BSE} approach to open-flavour mesons involving charm and bottom quarks, precise knowledge of the analytic properties of the propagators is mandatory.
Furthermore, to extend its applicability to its maximal domain, the propagators have to be known very precisely and sophisticated methods of analytic continuation are necessary.
While for $A(p^2)$ and $B(p^2)$ no non-analyticities have been found so far, the functions $\sigma_\mathrm{S,V}(p^2)$ exhibit complex-conjugated pole pairs in the left-half plane, which spoil the numerical integration in the \gls{DSE} as well as the \gls{BSE} as soon as such a pole enters the sampled domain \cite{Windisch:2016iud}.
Similarly, the effective gluon propagator may also exhibit non-analyticities.
For example, the \gls{UV} dominating part of the \gls{MT} model, cf.\ Eq.~\eqref{eq:MT}, introduces branch cuts in the left-half complex-$p^2$ plane.
In order to ensure numerical stability, these structures have to be avoided, or more sophisticated integration algorithms have to be used \cite{Dorkin:2013rsa,Dorkin:2014lxa}.

One way to determine the pole positions of the dressing functions is to fit a complex-conjugate pole pair representation to the solution along the positive real axis \cite{Souchlas:2010boa}.
Another way is to determine $A(p^2)$ and $B(p^2)$ in the complex plane by inserting complex momenta, $\mathrm{Im}p^2 \neq 0$, in Eq.~\eqref{eq:dse} and search for roots of the propagator's denominator or root factors thereof, $p^2 A^2 + B^2 = (p A + {\rm i} B)(pA-{\rm i}B)$, by virtue of Cauchy's argument principle or sophisticated optimisation routines such as the Newton-Krylov algorithm \cite{Knoll2004357}.
Similarly, the basin-hopping algorithm \cite{Wales:1997} with some local minimisation algorithm, such as the bounded, limited-memory Broyden-Fletcher-Goldfarb-Shanno algorithm or the Nelder-Mead algorithm have been employed.
Despite of being very fast, numerical reliability of the methods mentioned so far worsens the larger the distance to the positive real $p^2$-axis becomes.
An alternative is described in \cite{Fischer:2005en,Krassnigg:2008gd}, which aims at solving the quark \gls{DSE} directly in the complex plane.

\begin{figure}[t]
\centering
\includegraphics[width=.48\textwidth]{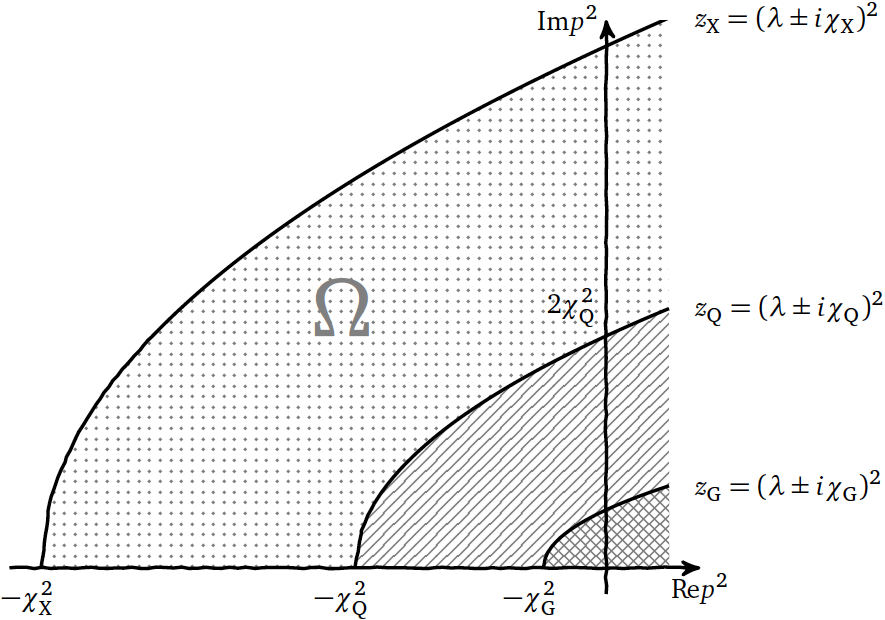}
\caption{The upper left part of the complex-$p^2$ plane with the parabola setting of the \gls{DSE} solution in the complex plane.
         The lower part is equivalent to the upper one, the right half-plane has no non-analyticities in the quark propagator and is therefore irrelevant.}
\label{fig:DSEparabola}
\end{figure}

To this end, symmetric and non-symmetric momentum routings where introduced in Eq.~\eqref{eq:dse}.
As a result, the quark-momentum domain sampled by the \gls{DSE} integration domain is bounded by a parabola, see Fig.~\ref{fig:DSEparabola}.
Introducing a contour which closes that parabola, the propagator functions can then be evaluated via Cauchy's formula from the propagator along that contour.
Usually, the combination of particular momentum routings and Cauchy's formula is complemented by employing some asymptotic fit for that part of the propagator which lies outside of the contour in the \gls{UV} region.
It is sometimes argued to be a weakness of this approach to employ some model for the asymptotic dependence of the propagator functions.
Thereby, the self-consistent solution approach inherent to the elegant \gls{DSBSE} framework is somehow spoiled.
In principle, this deficit may completely be circumvented by evaluating the propagator functions in this region from a \gls{DSE} integration over the quark momentum.
The latter is numerically stable because phenomenologically interesting models do not introduce numerical instabilities in that region.
We have also implemented such a callback algorithm, compared to the canonically employed \gls{UV}-fit \cite{Alkofer:2002bp} and confirmed the numerical reliability of the latter.
However, one must be sure to exclude all propagator poles from the integration domains of the quark and the gluon.

For large $\mathrm{Im}p^2$ and $\mathrm{Re}p^2 \leq 0$, it is difficult to determine the position of the poles in advance.
Also, it is computationally very expensive to iteratively expand the contour and self-consistently solve for each contour anew without a-priori knowledge of poles which might show up within the contour.
We explored all these methods and ended up relying on an algorithm with dynamical momentum routing.

To discuss the solution of the \gls{DSE} in the complex plane with arbitrary momentum routing, we rewrite the quark self-energy, Eq.~\eqref{eq:dse_sigma}, as 
\begin{equation}
    \Sigma(p) = C_\mathrm{F} \, \int^\Lambda_k\!\! \mathcal{G}\!\left(k_\mathrm{G}^2\right) \, D_{\mu\nu}^\mathrm{f}(k_\mathrm{G}) \,\gamma_\mu \,S(k_\mathrm{Q})\, \gamma_\nu \, .
\end{equation}
With the momentum routing parameter $\vartheta$, the internal quark momentum $k_\mathrm{Q}$ and gluon momentum $k_\mathrm{G}$ are parametrized as
\begin{equation}
    k_\mathrm{G} = (1-\vartheta) p \pm k \, ,
    \quad
    k_\mathrm{Q} = \vartheta p \mp k \, .
\end{equation}
Both substitutions, the upper and the lower sign, are equivalent and result in a symmetric momentum routing for $\vartheta = 0.5$ where quark and gluon propagator are sampled on the same domain bounded by their respective parabolas, see below and Fig.~\ref{fig:DSEparabola}.
For $\vartheta \in \{0,1\}$, the routing is asymmetric, with $\vartheta = 1$ resulting in an integration over the gluon momentum and $\vartheta = 0$ an integration over the quark momentum, i.\,e., no substitution w.\,r.\,t.\ Eq.~\eqref{eq:dse}.
For the upper sign and $\vartheta = 0$, the integration variable has to be mirrored, $k \to -k$; for the lower sign, the same is true for $\vartheta = 1$.

The projected coupled integral equations for the propagator functions in this parametrization read
\begin{subequations}\label{eq:projectedDSE}
\begin{align}
    A(p^2) &= Z_2 + \genfrac{\{}{\}}{0pt}{0}{1}{{Z_2}^2} C_\mathrm{F} \int_k \mathcal{G}(k_\mathrm{G}^2) K_\mathrm{V}(k_\mathrm{G},k_\mathrm{Q}) \sigma_\mathrm{V}(k_\mathrm{Q}^2) \, ,
    \\\nonumber
    B(p^2) &= Z_4 m(\mu^2) 
            \\
                &+ \genfrac{\{}{\}}{0pt}{0}{1}{{Z_2}^2} C_\mathrm{F} \int_k \mathcal{G}(k_\mathrm{G}^2) K_\mathrm{S}(k_\mathrm{G},k_\mathrm{Q}) \sigma_\mathrm{S}(k_\mathrm{Q}^2) \, .
\end{align}
\end{subequations}
With the transversal projector, $P_{\mu\nu}^\mathrm{T}(k) = \left( \delta_{\mu\nu} - \frac{k_\mu k_\nu}{k^2} \right)$, the reduced propagator projections $K_\mathrm{S,V}$ are given in rainbow truncation by
\begin{subequations}
\begin{align}
    \nonumber
    &K_\mathrm{V}(k_\mathrm{G},k_\mathrm{Q}) \sigma_\mathrm{V}(k_\mathrm{Q}^2)
    \\ &\equiv\nonumber
        -\frac{{\rm i}}{4p^2} P_{\mu\nu}^\mathrm{T}(k_\mathrm{G}) \mathrm{Tr}\left[ \gamma \cdot p \gamma_\mu S(k_\mathrm{Q}) \gamma_\nu \right]
    \\
        & = -\frac{\sigma_\mathrm{V}(k_\mathrm{Q}^2)}{p^2} \left( p\cdot k_\mathrm{Q} + 2 \frac{p\cdot k_\mathrm{G} k_\mathrm{G}\cdot k_\mathrm{Q}}{k_\mathrm{G}^2} \right) \, ,
    \\
    \nonumber
    &K_\mathrm{S}(k_\mathrm{G},k_\mathrm{Q}) \sigma_\mathrm{S}(k_\mathrm{Q}^2)
    \\ &\equiv\nonumber
        \frac{1}{4} P_{\mu\nu}^\mathrm{T}(k_\mathrm{G}) \mathrm{Tr}\left[ \gamma_\mu S(k_\mathrm{Q}) \gamma_\nu \right]
    \\&
         = 3 \, ,
\end{align}
\end{subequations}
where the $p$-dependence is neglected due to $p = k_\mathrm{G} + k_\mathrm{Q}$.
Solving these equations for an external parabola $z_\mathrm{X}$, cf.\ Fig.~\ref{fig:DSEparabola}, means
\begin{align*}
    p^2 \to z_\mathrm{X} \, , \quad
    k_\mathrm{G}^2~\text{bounded by}~z_\mathrm{G} \, , \quad
    k_\mathrm{Q}^2~\text{bounded by}~z_\mathrm{Q} \, ,
\end{align*}
where the respective parabola parameters $\chi$ are related by
\begin{equation}
    \chi_\mathrm{X} = \frac{\chi_\mathrm{Q}}{\vartheta} = \frac{\chi_\mathrm{G}}{1-\vartheta} = \chi_\mathrm{G} + \chi_\mathrm{Q} \, .
\end{equation}
Note that, for any point $z$ in the complex plane, there exists only one parabola of the form $(\lambda \pm \chi)^2$ on which $z$ lies, parametrised by
\begin{subequations}
\begin{align}
    \chi(z) & = \sqrt{\frac{\left|z\right| - \mathrm{Re}z}{2}} && = \left| \mathrm{Im} \sqrt{z} \right| \, ,
    \\
    \lambda(z) & = \frac{\mathrm{Im}z}{\sqrt{2(\left|z\right| - \mathrm{Re}z)}} && = \frac12 \frac{\mathrm{Im}z}{\left|\mathrm{Im}\sqrt{z}\right|}\;.
\end{align}
\end{subequations}
Also note that all parabolas of this kind have unit slope at the intersection point with the imaginary $z$ axis, i.\,e., they are all parallel along this axis.
The momentum variables $k_\mathrm{Q}$, $k_\mathrm{G}$ only sample the respective domains bounded by their respective parabolas $z_\mathrm{Q}$, $z_\mathrm{G}$ and the \gls{UV} closing line of the Cauchy contour.

Provided the analytic domain of the gluon propagator is known and extends up to
\begin{equation}
    \chi_\mathrm{G} = \chi_\mathrm{G}^\mathrm{max} \, ,
\end{equation}
setting
\begin{equation}
    \chi_\mathrm{Q}^{(0)} = 0 \, ,
\end{equation}
we proceed with the following iteration:
\begin{equation}
    \chi_\mathrm{X}^{(n)} = \chi_\mathrm{Q}^{(n)} + \chi_\mathrm{G}^\mathrm{max} = (n+1) \chi_\mathrm{G}^\mathrm{max} \, ,
\end{equation}
i.\,e.\
\begin{equation}
    \chi_\mathrm{Q}^{(n)} = n \chi_\mathrm{G}^\mathrm{max} \, .
\end{equation}
This allows us to safely and systematically expand the quark-propagator para\-bola by successive \gls{DSE} integration without iteration and without poles entering the sampled quark- or gluon-momentum domain. The steps and features in this regard are:
\begin{itemize}
 \item choose $\chi_\mathrm{G}^\mathrm{max}$ such that numerical problems (oscillations) or non-analytic structures (branch cuts) are avoided;
 
 \item with the propagator being known at $z_\mathrm{Q}(\lambda)$ and regular inside, the \gls{DSE} can be integrated with $\vartheta^{(n)} = \chi_\mathrm{Q}^{(n)} / \chi_\mathrm{X}^{(n)} = n/(n+1)$ regardless of the analytic structure between $z_\mathrm{X}(\lambda)$ and $z_\mathrm{Q}(\lambda)$ (domain $\Omega$ in Fig.~\ref{fig:DSEparabola};
 this gives the propagator along $z_\mathrm{X}(\lambda)$;
 
 \item
 check the analytic structure inside the contour defined by $z_\mathrm{X}(\lambda)$ via Cauchy's argument principle;
 
 \item
 repeat with $z_\mathrm{Q}(\lambda) \to z_\mathrm{X}(\lambda)$ or use some $\chi$ in between if non-analyticities are found;
 
 \item
 optionally after each (or some) expansion(s): iterate along the parabola via $\vartheta = 1$ to improve the numerical accuracy.
\end{itemize}
This allows to expand the parabola without iterating at each expansion step, on the one hand, but consequently also to remain in the numerically safe (or cheap) integration domain of the gluon propagator.
To apply Cauchy's argument principle, one needs the derivative of the propagator function along the external contour.
This derivative can be obtained analogously to the propagator function itself by integrating the first derivative of Eq.~\eqref{eq:projectedDSE}:
\begin{equation}
\begin{split}
    \frac{\partial F(p^2)}{\partial p^2}
        & = \genfrac{\{}{\}}{0pt}{0}{1}{{Z_2}^2} C_\mathrm{F} 
            \\
            & \int_k \Bigg[\frac{\partial \mathcal{G}({k_\mathrm{G}}^2)}{\partial {k_\mathrm{G}}^2} \frac{{k_\mathrm{G}}^2-k^2}{p^2} K_F(k_\mathrm{G},k_\mathrm{Q}) \sigma({k_\mathrm{Q}}^2)
        \\
            & + \mathcal{G}({k_\mathrm{G}}^2) \frac{\partial K_F(k_\mathrm{G},k_\mathrm{Q})}{\partial p^2} \sigma({k_\mathrm{Q}}^2)
        \\
            & + \mathcal{G}({k_\mathrm{G}}^2) K_F(k_\mathrm{G},k_\mathrm{Q}) \frac{\partial \sigma({k_\mathrm{Q}}^2)}{\partial {k_\mathrm{Q}}^2} \frac{{k_\mathrm{Q}}^2-k^2}{p^2}
            \Bigg] \, ,
\end{split}
\end{equation}
where ($F$, $\sigma$) refer to ($A$, $\sigma_\mathrm{V}$) or ($B$, $\sigma_\mathrm{S}$).
The derivatives of the reduced propagator projections are
\begin{subequations}
\begin{align}
    \frac{\partial K_A(k_\mathrm{G},k_\mathrm{Q})}{\partial p^2} & =
        - \frac{K_A}{p^2} + \frac{1}{p^4} \Bigg[ p\cdot k_\mathrm{Q}+ 2 \frac{ p\cdot k_\mathrm{G}}{{k_\mathrm{G}}^2}
        \Bigg( k_\mathrm{G} \cdot k_\mathrm{Q}
        \nonumber\\&
         + k^2 + k^2 \frac{ k_\mathrm{G} \cdot k_\mathrm{Q}}{{k_\mathrm{G}}^2} \Bigg) \Bigg]\;,
\\
    \frac{\partial K_B(k_\mathrm{G},k_\mathrm{Q})}{\partial p^2} & = 0\;. 
\end{align}
\end{subequations}
As the derivative of the propagator function is needed for this integration only inside the contour and not along the contour, it can be obtained from Cauchy's formula.

\section{Numerical Setup}
\label{sec:numeric}

For the sake of completeness, clarity, and reproducibility, we detail the numerical set-up used for the present investigation.

The radial momentum integration measure of, e.\,g., Eqs.~\eqref{eq:dse} and \eqref{eq:BSE}, as well as for all other momentum integrations, $p^2 \in [0,\infty)$, is mapped onto the canonical Gau\ss{} integration measure $[-1,1]$ via \cite{Maris:2004pc}
$$p^2 = c \frac{\mathrm{e}^{a (x+1)/2} - 1}{\mathrm{e} - \mathrm{e}^{(x+1)/2}}\,,$$
with $c=0.1$ and $a$ chosen such that the maximum momentum is $p^2_\mathrm{max} = 10^5\,\mathrm{GeV}^2$,
$$a = \frac{\ln\left(1 + \frac{p^2_\mathrm{max}}{c} \left(\mathrm{e}-\mathrm{e}^{\mathrm{max}\left((x+1)/2\right)}\right)\right)  }{ \mathrm{max}((x+1)/2) }\,,$$
with $\mathrm{max}(x)$ being the largest root of the Legendre polynomial of order $n$, the number of integration points.
A different and often used alternative parametrization is given by
$$p^2 = c \frac{1 + x^a}{1-x^a}\,.$$
Again, the parameter $a$ may be chosen to reach a certain maximum momentum $p^2_\mathrm{max}$,
$$a = \frac{\ln\left(\frac{p^2_\mathrm{max}-c}{p^2_\mathrm{max}+c}\right) }{ \ln\left(\mathrm{max}(x)\right)}\,.$$

For the \gls{DSE} solution, cf.\ \ref{sec:DSEcontour}, in the \gls{AWW} (\gls{MT}) model we use $512$ ($1024$) points for the Gau\ss{}-Legendre integration of the radial momentum and $512$ ($256$) points for the Gau\ss{}-Chebyshev integration of the hyperpolar angle.
The parabolic Cauchy contour in the complex $p^2$ plane has $128$ points along the parabola and $64$ points along the line, with a linear mapping of the Gau\ss{}-integration measure and $\mathrm{Re}p^2 \leq 10\,\mathrm{GeV}^2$ along the contour.
We solve the \gls{BSE} using $48$ points for the radial $p^2$ integration, and $32$ points for the hyperpolar-angle integration $P\cdot q$ and the polar-angle integration $p\cdot q$.

\section{Quark-Mass-Dependent Interaction Model}\label{sec:intdependent}

The following argument is based on the idea of changing model parameters in the effective interaction for each quark mass, in order to account for truncation effects and enable a better description of, as well as a better-suited anchoring to, experimental data in the different domains of the current-quark mass.

In order to ensure a neat anchoring of the interaction model in \gls{QCD} and \gls{DCSB}, the central equation to be satisfied is the \gls{AVWTI}, which reads in \gls{RL} truncation
\begin{multline}
    \Sigma_1 (k_+) \gamma_5 + \gamma_5 \Sigma_2 (k_-)
    \\
     = - \int_q^\Lambda  \gamma_\nu \frac{\lambda^a}{2} 
    \left[   S_1 (q_+) \gamma_5 + \gamma_5 S_2 (q_-) \right]
    \\\times
    {\cal G} (q,k;P) D^{\rm f}_{\mu\nu} (k-q) 
    \gamma_\mu \frac{\lambda^a}{2} \, ,
\label{eq:AVWTI}
\end{multline}
where the effective interaction $\mathcal{G}$ on the r.\,h.\,s.\ enters the \gls{RL}-truncated meson \gls{BSE}, and with the quark self-energy
\begin{multline}
    \Sigma_i[{\cal G}_i, S_i] (k)
    =
    \int_q^\Lambda {\cal G}_i((k-q)^2) D_{\mu\nu}^{\rm f} (k-q)
    \\\times\frac{\lambda^a}{2} \gamma_\mu 
     S_i(q)\frac{\lambda^a}{2} \gamma_\nu  
\end{multline}
being  linear in the effective interaction ${\cal G}_i$ and the quark propagator $S_i$.
If the effective interactions contain an effective quark-mass dependence, i.\,e., $\mathcal{G}_1 \neq \mathcal{G}_2$ for $m_1 \neq m_2$, Eq.~\eqref{eq:AVWTI} is the boundary condition to be fulfilled by the effective interaction $\mathcal{G}$ entering the \gls{BSE} if \gls{DCSB} is properly reflected by the combined \gls{DSBSE} approach.
Due to the simple and symmetric form of the \glspl{QGV} in \gls{RL} truncation, Eq.~\eqref{eq:AVWTI} resembles a clear similarity to the quark \gls{DSE}
\begin{equation}
    S^{-1}_i(k) =  S^{-1}_{i,0}(k) + \Sigma_i(k) \, .
\end{equation}
In vacuum, the quark propagator has the following projections
\begin{subequations}
\begin{align}
    A_i(k) :=& - \frac{\rm i}{4k^2}\mathrm{Tr}[k\cdot \gamma S^{-1}_i(k)] \, ,
    \\
    B_i(k) :=& \;\frac{1}{4} \mathrm{Tr}[S^{-1}_i(k)] \, .
\end{align}
\end{subequations}
Projecting the quark \gls{DSE} in vacuum thus gives
\begin{subequations}\label{eq:DSEprojected}
\begin{align}
    A_i(k) = & - \frac{\rm i}{4k^2}\mathrm{Tr}[k\cdot \gamma S_{i,0}^{-1}(k)]
    - \frac{\rm i}{4k^2}\mathrm{Tr}[k\cdot \gamma \Sigma_i(k)] 
    \nonumber \\
    \equiv &\;
    A_{i,0} + \Sigma_{\mathrm{V},i} [\mathcal{G}_i,\sigma_{\mathrm{V},i}] \;,
    \\
    B_i(k) = & \, \frac{1}{4} \mathrm{Tr}[S_{i,0}^{-1}(k)] + \frac{1}{4} \mathrm{Tr}[\Sigma_i(k)] \nonumber \\
    \equiv &\;
    B_{i,0} + \Sigma_{\mathrm{S},i} [\mathcal{G}_i,\sigma_{\mathrm{S},i}] \;.
\end{align}
\end{subequations}
Analogously, the \gls{AVWTI} has two non-zero projections,
\begin{subequations}\label{eq:AVWTIprojected}
\begin{multline}
    \mathrm{Tr}\left[ \Sigma_1 (k_+)  \right]
    +  
    \mathrm{Tr}\left[\Sigma_2 (k_-) \right]
    \\
        =  \mathrm{Tr}\left[
         \int_q^\Lambda  \gamma_\nu \frac{\lambda^a}{2} \left[   
         S_1 (q_+)
        + 
         S_2 (q_-)
        \right] \right.
        \\
        \left. \qquad \quad \times \  
        {\cal G} (q,k;P) D^{\rm f}_{\mu\nu} (k-q) 
          \gamma_\mu \frac{\lambda^a}{2}\ \right] \, ,
\end{multline}
and
\begin{multline}
\mathrm{Tr}\left[\gamma\cdot k \Sigma_1 (k_+) \right]
- \mathrm{Tr}\left[ \gamma\cdot k
\Sigma_2 (k_-) \right]
\\
=  \mathrm{Tr}\left[ \gamma\cdot k
 \int_q^\Lambda  \gamma_\nu \frac{\lambda^a}{2} \left[   
 S_1 (q_+)  
-
 S_2 (q_-)
\right] \right.
\\
\times \Biggl.
{\cal G} (q,k;P) D^{\rm f}_{\mu\nu} (k-q) 
  \gamma_\mu \frac{\lambda^a}{2} \Bigg] \, .
\end{multline}
\end{subequations}
Comparing Eqs.~\eqref{eq:DSEprojected} and \eqref{eq:AVWTIprojected} for $P = 0$, Eq.~\eqref{eq:AVWTI} becomes
\begin{subequations}
\begin{align}
    \bar B - \bar B_0 
    = & \;\Sigma_\mathrm{S}\left[ {\cal G},\sigma_{\mathrm{S},1} + \sigma_{\mathrm{S},2}\right] \, ,
    \\
    \bar A - \bar A_0
    = & \;\Sigma_\mathrm{V}\left[ {\cal G},\sigma_{\mathrm{V},1} - \sigma_{\mathrm{V},2}\right] \, ,
\end{align}
\label{eq:AVWTIDSE}
\end{subequations}
where we have defined
\begin{subequations}
\begin{align}
\bar A \equiv & \;A_1 - A_2 \, ,
\\
\bar B \equiv & \;B_1 + B_2 \, .
\end{align}
\end{subequations}
Constructing the effective interaction which enters the \gls{BSE} amounts to finding a function $\mathcal{G}$ which, e.\,g., for $P = 0$, maps $\sigma_{\mathrm{S},1} + \sigma_{\mathrm{S},2}$ and $\sigma_{\mathrm{V},1} - \sigma_{\mathrm{V},2}$ to $\bar B$ and $\bar A$ simultaneously.

If $\bar A, \bar B$ ought to be proper quark-propagator dressing functions associated by virtue of the usual $(A, \sigma_\mathrm{V})$--$(B, \sigma_\mathrm{S})$ interrelation, i.\,e.,
\begin{equation}
    \bar B^2 = \frac{\bar A}{\bar\sigma_\mathrm{V}} - k^2 \bar A^2
    \, , \quad
    \bar A^2 = \frac{1}{k^2} \left(\frac{\bar B}{\bar \sigma_\mathrm{S}} - \bar B^2 \right)
    \, ,
\label{Bpropagator}
\end{equation}
then
\begin{equation}
\begin{split}
    k^2  A_1 A_2 - B_1 B_2
        &=
        \frac{1}{2} \frac{ A_1 {\Delta_2}^2 - A_2 {\Delta_1}^2}{A_1 \Delta_2 - A_2 \Delta_1 }
        \\ &=
        \frac{1}{2} \frac{ B_1 {\Delta_2}^2 +B_2 {\Delta_1}^2}{B_1 \Delta_2 + B_2 \Delta_1 } \, ,
\end{split}
\label{eq:AVWTIpropRel}
\end{equation}
must be satisfied, where we have defined $\Delta_i \equiv k^2 A_i^2 + B_i^2$.
From the first equality in Eq.~\eqref{eq:AVWTIpropRel} one concludes that $A_1 = A_2$ requires $B_1 = B_2$.
From the latter equality one then concludes that if the propagators are equal, $S_1 = S_2$, Eq.~\eqref{eq:AVWTIpropRel} demands that $B_1 = B_2 = 0$, which is fulfilled in the chiral \gls{WW} phase.

\section{Exotic pseudoscalar decay constants}
\label{sec:f0--}
In this appendix, we briefly demonstrate the result $f_{0^{--}}=r_{0^{--}}=0$.
For $f_{0^{--}}$, the relevant part of the integral in Eq.~(\ref{eq:f0-}) is
\begin{equation}
     \int^\Lambda_q \mathrm{Tr}[\gamma_5\; \gamma\!\cdot\! P \; \chi^{0^-}\!\!(q;P) ]
\end{equation}
and, by going back to the \gls{BSA} via Eq.~(\ref{eq:bsa-bsw}), one obtains
\begin{equation}
     \int^\Lambda_q \mathrm{Tr}[\gamma_5\; \gamma\!\cdot\! P \; S_1(q_+) \, \Gamma^{0^-}(q;P) \, S_2(q_-) ]\;.
\end{equation}
The relevant part of the four-dimensional Euclidean integration is \cite{Blank:2010bp}
\begin{equation}
\int_{-1}^{1}dz \sqrt{1-z^2}\;,
\end{equation}
which is carried out in $z$ with symmetric integration weights over an interval symmetric around the origin. 
In the following, we will show that the integrand, consisting of Dirac traces and amplitude functions, is antisymmetric in $z$ and thus the integral vanishes for a $\mathcal{C}=-$ \gls{BSA}.

First of all, we note that the Dirac traces are the same for $\mathcal{C}=\pm$ and depend only on the total momentum squared $P^2$, the momentum partitioning $\eta$, the relative momentum squared $q^2$, the scalar product $q\cdot P$, and the quark mass functions $M_1(q_+^2)$ and $M_2(q_-^2)$. 
One obtains, using the basis given for pseudoscalars in Appendix A of \cite{Hilger:2015ora}, a vector $A$ of four functions to be multiplied by the corresponding scalar coefficients $\Gamma_i$ of each covariant tensor $T_i$, as given by Eq.~(\ref{eq:bsadecomp}),
\begin{eqnarray}\nonumber
A_1&=&2 P^2[M_1 M_2-P^2(\eta-\eta^2)+q\cdot P(2\eta-1)+q^2]\;,\\\nonumber
A_2&=&2\sqrt{P^2}\{P^2[M_1(\eta-1)-M_2\eta]+q\cdot P(M_1-M_2)\}\;,\\\nonumber
A_3&=&2P^2(M_1-M_2)\sqrt{q^2-\frac{(q\cdot P)^2}{P^2}}\;,\\ \label{eq:traces1}
A_4&=&-2i(P^2)^{3/2}\sqrt{q^2-\frac{(q\cdot P)^2}{P^2}}\;.
\end{eqnarray}
We now need to investigate the (anti)symmetry of these terms with respect to the hyper-angle cosine $z$.
Concretely, we consider the equal-mass case, only in which $\mathcal{C}$ is actually a well-defined quantum number. Thus we can set $\eta=1/2$ and the two mass functions are identical, $M_1(q^2)=M_2(q^2)=M(q^2)$.
The $z$-dependent terms in the expressions (\ref{eq:traces1}) are $q\cdot P$, $M_1(q_+^2)$, and $M_2(q_-^2)$ via
\begin{eqnarray}
q\cdot P&=&\sqrt{q^2} \sqrt{P^2} z\;,\\\label{eq:qps}
q_+^2&=&q^2-\frac{M^2}{4}+i\mu\sqrt{q^2}z\;,\\\label{eq:qms}
q_-^2&=&q^2-\frac{M^2}{4}-i\mu\sqrt{q^2}z\;,
\end{eqnarray} 
where $\mu=\sqrt{-P^2}$ and it is apparent that $q_+^2$ and $q_-^2$ are complex conjugate, i.\,e., $q_+^2=(q_-^2)^\ast$, because $q^2$, $\mu$, and $z$ are all real variables. 
Writing
\begin{eqnarray}
M_1(q_+^2)+M_2(q_-^2)&=&M(q_+^2)+M(q_-^2)\;,\\
M_1(q_+^2)-M_2(q_-^2)&=&M(q_+^2)-M(q_-^2)\;,\\
M_1(q_+^2) M_2(q_-^2)&=&M(q_+^2) M(q_-^2)\;,
\end{eqnarray} 
it is obvious by inserting Eqs.~(\ref{eq:qps}) and (\ref{eq:qms}) in these expressions (\ref{eq:traces1}) that $M_1-M_2$ is antisymmetric in $z$, while $M_1+M_2$ and $M_1 M_2$ are symmetric.
Writing in addition the expressions for $A$ again and setting $\eta=1/2$, one obtains, retaining only $z$-dependent terms,
\begin{eqnarray}\nonumber
A_1&\sim &2 M_1 M_2-\frac{P^2}{4}+q^2\;,\\\nonumber
A_2&\sim &\frac{P^2}{2}(M_1+M_2)-q\cdot P(M_1-M_2)\;,\\\nonumber
A_3&\sim &(M_1-M_2)\sqrt{q^2-\frac{(q\cdot P)^2}{P^2}}\;,\\\label{eq:traces2}
A_4&\sim &\sqrt{q^2-\frac{(q\cdot P)^2}{P^2}}\;.
\end{eqnarray}
Now one can see that $A_3$ is antisymmetric in $z$, while the others are symmetric. 
Since the covariants $T_i$ given as the basis for pseudoscalars in Appendix A of \cite{Hilger:2015ora} have $\mathcal{C}={+,+,-,+}$, the corresponding $\Gamma_i$ must have the symmetry properties ${-,-,+,-}$ with respect to $z$ in order to result in a \gls{BSA} with $\mathcal{C}=-$, as required for an exotic pseudoscalar.
As a consequence, the sum
\begin{equation}
\sum_{i=1}^4 A_i(z) \Gamma_i(z)
\end{equation}
is a function antisymmetric in $z$, and the integral over $z$ in $f_{0^{--}}$ vanishes.

The argument for $r_{0^{--}}=0$ is analogous.

\section{Tabulated results}
\label{sec:numbers}
In this appendix, we collect all results, plotted above in Sec.~\ref{sec:results}, in tabulated form.

\begin{center}
\tablecaption{The quark-bilinear meson spectrum, and, where applicable, leptonic decay constants and in-hadron condensates in the \gls{AWW} model \eqref{eq:AWW}.
All numbers are in GeV.
The level of excitation can be inferred from the mass ordering.
$n$ refers to the eigenvalue which generates the on-shell solution.}
\label{tbl:AWW}
\tablefirsthead{
\toprule
$(m_q, m_{{\bar q}^\prime}) $&$  J^\mathcal{PC} $&$  n $&$       M $&$  f $&$  \sqrt[3]{\left|\Braket{\bar qq}\right|} $\\
\midrule}
\tablehead{
\multicolumn{6}{c}{$\ldots$ continuation of Tab.~\ref{tbl:AWW} $\ldots$}\\
\toprule
$(m_q, m_{{\bar q}^\prime}) $&$  J^\mathcal{PC} $&$  n $&$       M $&$  f $&$  \sqrt[3]{\left|\Braket{\bar qq}\right|} $\\
\midrule}
\begin{mpxtabular}{cccccc}
$
              0 $&$  0^{++} $&$         0 $&$   0.613 $&$  0.000 $&$    0.000 $\\$
              0 $&$  0^{-+} $&$         0 $&$   0.000 $&$  0.130 $&$    0.251 $\\$
              0 $&$  0^{--} $&$         0 $&$   0.967 $&$  0.000 $&$    0.000 $\\$
              0 $&$  1^{++} $&$         0 $&$   0.912 $&$  0.192 $&$    0.000 $\\$
              0 $&$  1^{+-} $&$         0 $&$   0.896 $&$  0.000 $&$    0.000 $\\$
              0 $&$  1^{--} $&$         0 $&$   0.739 $&$  0.215 $&$    0.275 $\\$
              0 $&$  1^{--} $&$         1 $&$   1.016 $&$  0.063 $&$    0.142 $\\$
          0.005 $&$  0^{++} $&$         0 $&$   0.645 $&$  0.000 $&$    0.000 $\\$
          0.005 $&$  0^{-+} $&$         0 $&$   0.137 $&$  0.133 $&$    0.256 $\\$
          0.005 $&$  0^{--} $&$         0 $&$   0.985 $&$  0.000 $&$    0.000 $\\$
          0.005 $&$  1^{++} $&$         0 $&$   0.935 $&$  0.193 $&$    0.000 $\\$
          0.005 $&$  1^{+-} $&$         0 $&$   0.915 $&$  0.000 $&$    0.000 $\\$
          0.005 $&$  1^{-+} $&$         0 $&$   1.062 $&$  0.000 $&$    0.000 $\\$
          0.005 $&$  1^{--} $&$         0 $&$   0.758 $&$  0.219 $&$    0.277 $\\$
          0.005 $&$  1^{--} $&$         1 $&$   1.041 $&$  0.064 $&$    0.142 $\\$
           0.09 $&$  0^{++} $&$         0 $&$   1.015 $&$  0.000 $&$    0.000 $\\$
           0.09 $&$  0^{-+} $&$         0 $&$   0.606 $&$  0.172 $&$    0.311 $\\$
           0.09 $&$  0^{--} $&$         0 $&$   1.219 $&$  0.000 $&$    0.000 $\\$
           0.09 $&$  1^{++} $&$         0 $&$   1.221 $&$  0.209 $&$    0.000 $\\$
           0.09 $&$  1^{+-} $&$         0 $&$   1.169 $&$  0.000 $&$    0.000 $\\$
           0.09 $&$  1^{-+} $&$         0 $&$   1.337 $&$  0.000 $&$    0.000 $\\$
           0.09 $&$  1^{--} $&$         0 $&$   1.014 $&$  0.256 $&$    0.290 $\\$
           0.09 $&$  1^{--} $&$         1 $&$   1.354 $&$  0.070 $&$    0.130 $\\$
          0.115 $&$  0^{++} $&$         0 $&$   1.096 $&$  0.000 $&$    0.000 $\\$
          0.115 $&$  0^{-+} $&$         0 $&$   0.693 $&$  0.181 $&$    0.324 $\\$
          0.115 $&$  0^{--} $&$         0 $&$   1.278 $&$  0.000 $&$    0.000 $\\$
          0.115 $&$  1^{++} $&$         0 $&$   1.289 $&$  0.212 $&$    0.000 $\\$
          0.115 $&$  1^{+-} $&$         0 $&$   1.233 $&$  0.000 $&$    0.000 $\\$
          0.115 $&$  1^{-+} $&$         0 $&$   1.400 $&$  0.000 $&$    0.000 $\\$
          0.115 $&$  1^{--} $&$         0 $&$   1.078 $&$  0.265 $&$    0.292 $\\$
          0.115 $&$  1^{--} $&$         1 $&$   1.428 $&$  0.070 $&$    0.128 $\\$
          0.975 $&$  0^{++} $&$         0 $&$   3.021 $&$  0.000 $&$    0.000 $\\$
          0.975 $&$  0^{+-} $&$         0 $&$   3.242 $&$  0.000 $&$    0.000 $\\$
          0.975 $&$  0^{-+} $&$         0 $&$   2.672 $&$  0.322 $&$    0.574 $\\$
          0.975 $&$  0^{-+} $&$         1 $&$   3.256 $&$  0.137 $&$    0.372 $\\$
          0.975 $&$  0^{--} $&$         0 $&$   2.970 $&$  0.000 $&$    0.000 $\\$
          0.975 $&$  1^{++} $&$         0 $&$   3.087 $&$  0.215 $&$    0.000 $\\$
          0.975 $&$  1^{++} $&$         1 $&$   3.275 $&$  0.004 $&$    0.000 $\\$
          0.975 $&$  1^{+-} $&$         0 $&$   3.023 $&$  0.000 $&$    0.000 $\\$
          0.975 $&$  1^{+-} $&$         1 $&$   3.293 $&$  0.000 $&$    0.000 $\\$
          0.975 $&$  1^{-+} $&$         0 $&$   3.060 $&$  0.000 $&$    0.000 $\\$
          0.975 $&$  1^{--} $&$         0 $&$   2.840 $&$  0.383 $&$    0.286 $\\$
          0.975 $&$  1^{--} $&$         1 $&$   3.294 $&$  0.031 $&$    0.081 $\\$
          0.975 $&$  1^{--} $&$         2 $&$   3.307 $&$  0.193 $&$    0.152 $\\$
          0.975 $&$  2^{++} $&$         0 $&$   3.191 $&$    - $&$      - $\\$
          0.975 $&$  2^{+-} $&$         0 $&$   3.360 $&$    - $&$      - $\\$
          0.975 $&$  2^{-+} $&$         0 $&$   3.363 $&$    - $&$      - $\\$
           1.11 $&$  0^{++} $&$         0 $&$   3.288 $&$  0.000 $&$    0.000 $\\$
           1.11 $&$  0^{+-} $&$         0 $&$   3.497 $&$  0.000 $&$    0.000 $\\$
           1.11 $&$  0^{-+} $&$         0 $&$   2.944 $&$  0.332 $&$    0.599 $\\$
           1.11 $&$  0^{-+} $&$         1 $&$   3.508 $&$  0.147 $&$    0.392 $\\$
           1.11 $&$  0^{--} $&$         0 $&$   3.225 $&$  0.000 $&$    0.000 $\\$
           1.11 $&$  1^{++} $&$         0 $&$   3.347 $&$  0.211 $&$    0.000 $\\$
           1.11 $&$  1^{++} $&$         1 $&$   3.523 $&$  0.004 $&$    0.000 $\\$
           1.11 $&$  1^{+-} $&$         0 $&$   3.286 $&$  0.000 $&$    0.000 $\\$
           1.11 $&$  1^{+-} $&$         1 $&$   3.543 $&$  0.000 $&$    0.000 $\\$
           1.11 $&$  1^{-+} $&$         0 $&$   3.309 $&$  0.000 $&$    0.000 $\\$
           1.11 $&$  1^{--} $&$         0 $&$   3.098 $&$  0.389 $&$    0.282 $\\$
           1.11 $&$  1^{--} $&$         1 $&$   3.553 $&$  0.136 $&$    0.148 $\\$
           1.11 $&$  1^{--} $&$         2 $&$   3.563 $&$  0.152 $&$    0.103 $\\$
           1.11 $&$  2^{++} $&$         0 $&$   3.441 $&$    - $&$      - $\\$
           1.11 $&$  2^{+-} $&$         0 $&$   3.604 $&$    - $&$      - $\\$
           1.11 $&$  2^{-+} $&$         0 $&$   3.617 $&$    - $&$      - $\\$
           4.43 $&$  0^{++} $&$         0 $&$   9.681 $&$  0.000 $&$    0.000 $\\$
           4.43 $&$  0^{++} $&$         1 $&$  10.004 $&$  0.000 $&$    0.000 $\\$
           4.43 $&$  0^{+-} $&$         0 $&$   9.792 $&$  0.000 $&$    0.000 $\\$
           4.43 $&$  0^{-+} $&$         0 $&$   9.424 $&$  0.378 $&$    0.894 $\\$
           4.43 $&$  0^{-+} $&$         1 $&$   9.820 $&$  0.263 $&$    0.723 $\\$
           4.43 $&$  0^{-+} $&$         2 $&$  10.018 $&$  0.235 $&$    0.679 $\\$
           4.43 $&$  0^{--} $&$         0 $&$   9.574 $&$  0.000 $&$    0.000 $\\$
           4.43 $&$  0^{--} $&$         1 $&$   9.956 $&$  0.000 $&$    0.000 $\\$
           4.43 $&$  1^{++} $&$         0 $&$   9.692 $&$  0.118 $&$    0.000 $\\$
           4.43 $&$  1^{++} $&$         1 $&$   9.790 $&$  0.001 $&$    0.000 $\\$
           4.43 $&$  1^{++} $&$         2 $&$  10.012 $&$  0.103 $&$    0.000 $\\$
           4.43 $&$  1^{+-} $&$         0 $&$   9.671 $&$  0.000 $&$    0.000 $\\$
           4.43 $&$  1^{+-} $&$         1 $&$   9.800 $&$  0.000 $&$    0.000 $\\$
           4.43 $&$  1^{+-} $&$         2 $&$  10.004 $&$  0.000 $&$    0.000 $\\$
           4.43 $&$  1^{-+} $&$         0 $&$   9.597 $&$  0.000 $&$    0.000 $\\$
           4.43 $&$  1^{-+} $&$         1 $&$   9.967 $&$  0.000 $&$    0.000 $\\$
           4.43 $&$  1^{-+} $&$         2 $&$   9.986 $&$  0.000 $&$    0.000 $\\$
           4.43 $&$  1^{--} $&$         0 $&$   9.463 $&$  0.393 $&$    0.200 $\\$
           4.43 $&$  1^{--} $&$         1 $&$   9.838 $&$  0.283 $&$    0.158 $\\$
           4.43 $&$  1^{--} $&$         2 $&$   9.905 $&$  0.018 $&$    0.023 $\\$
           4.43 $&$  1^{--} $&$         3 $&$  10.039 $&$  0.226 $&$    0.135 $\\$
           4.43 $&$  2^{++} $&$         0 $&$   9.712 $&$    - $&$      - $\\$
           4.43 $&$  2^{++} $&$         1 $&$  10.024 $&$    - $&$      - $\\$
           4.43 $&$  2^{+-} $&$         0 $&$   9.814 $&$    - $&$      - $\\$
           4.43 $&$  2^{-+} $&$         0 $&$   9.909 $&$    - $&$      - $\\$
           4.43 $&$  2^{-+} $&$         1 $&$   9.999 $&$    - $&$      - $\\$
           4.43 $&$  2^{--} $&$         0 $&$   9.924 $&$    - $&$      - $\\$
           4.43 $&$  2^{--} $&$         1 $&$   9.993 $&$    - $&$      - $\\$
           4.43 $&$  3^{-+} $&$         0 $&$  10.017 $&$    - $&$      - $\\$
           4.43 $&$  3^{--} $&$         0 $&$   9.950 $&$    - $&$      - $\\$
   (0.0, 0.005) $&$  0^{+} $&$         0 $&$   0.629 $&$  0.003 $&$    0.076 $\\$
   (0.0, 0.005) $&$  0^{-} $&$         0 $&$   0.097 $&$  0.131 $&$    0.253 $\\$
   (0.0, 0.005) $&$  0^{-} $&$         1 $&$   0.977 $&$  0.000 $&$    0.005 $\\$
   (0.0, 0.005) $&$  1^{+} $&$         0 $&$   0.906 $&$  0.024 $&$    0.123 $\\$
   (0.0, 0.005) $&$  1^{+} $&$         1 $&$   0.924 $&$  0.191 $&$    0.113 $\\$
   (0.0, 0.005) $&$  1^{-} $&$         0 $&$   0.749 $&$  0.217 $&$    0.276 $\\$
   (0.0, 0.005) $&$  1^{-} $&$         1 $&$   1.029 $&$  0.065 $&$    0.143 $\\$
    (0.0, 0.09) $&$  0^{+} $&$         0 $&$   0.832 $&$  0.042 $&$    0.191 $\\$
    (0.0, 0.09) $&$  0^{-} $&$         0 $&$   0.423 $&$  0.149 $&$    0.281 $\\$
    (0.0, 0.09) $&$  0^{-} $&$         1 $&$   1.120 $&$  0.009 $&$    0.080 $\\$
    (0.0, 0.09) $&$  1^{+} $&$         0 $&$   1.040 $&$  0.114 $&$    0.199 $\\$
    (0.0, 0.09) $&$  1^{+} $&$         1 $&$   1.103 $&$  0.179 $&$    0.146 $\\$
    (0.0, 0.09) $&$  1^{-} $&$         0 $&$   0.899 $&$  0.240 $&$    0.284 $\\$
   (0.0, 0.115) $&$  0^{+} $&$         0 $&$   0.881 $&$  0.051 $&$    0.206 $\\$
   (0.0, 0.115) $&$  0^{-} $&$         0 $&$   0.482 $&$  0.153 $&$    0.287 $\\$
   (0.0, 0.115) $&$  0^{-} $&$         1 $&$   1.158 $&$  0.013 $&$    0.099 $\\$
   (0.0, 0.115) $&$  1^{+} $&$         0 $&$   1.078 $&$  0.124 $&$    0.204 $\\$
   (0.0, 0.115) $&$  1^{+} $&$         1 $&$   1.149 $&$  0.181 $&$    0.142 $\\$
   (0.0, 0.115) $&$  1^{-} $&$         0 $&$   0.941 $&$  0.247 $&$    0.286 $\\$
   (0.0, 0.975) $&$  0^{-} $&$         0 $&$   2.034 $&$  0.323 $&$    0.605 $\\$
   (0.0, 0.975) $&$  0^{-} $&$         0 $&$   1.877 $&$  0.343 $&$    0.597 $\\$
  (0.005, 0.09) $&$  0^{+} $&$         0 $&$   0.846 $&$  0.039 $&$    0.186 $\\$
  (0.005, 0.09) $&$  0^{-} $&$         0 $&$   0.434 $&$  0.151 $&$    0.283 $\\$
  (0.005, 0.09) $&$  0^{-} $&$         1 $&$   1.124 $&$  0.008 $&$    0.076 $\\$
  (0.005, 0.09) $&$  1^{+} $&$         0 $&$   1.049 $&$  0.106 $&$    0.194 $\\$
  (0.005, 0.09) $&$  1^{+} $&$         1 $&$   1.108 $&$  0.183 $&$    0.143 $\\$
  (0.005, 0.09) $&$  1^{-} $&$         0 $&$   0.905 $&$  0.241 $&$    0.285 $\\$
 (0.005, 0.115) $&$  0^{+} $&$         0 $&$   0.893 $&$  0.048 $&$    0.202 $\\$
 (0.005, 0.115) $&$  0^{-} $&$         0 $&$   0.492 $&$  0.155 $&$    0.289 $\\$
 (0.005, 0.115) $&$  0^{-} $&$         1 $&$   1.162 $&$  0.012 $&$    0.094 $\\$
 (0.005, 0.115) $&$  1^{+} $&$         0 $&$   1.085 $&$  0.117 $&$    0.200 $\\$
 (0.005, 0.115) $&$  1^{+} $&$         1 $&$   1.154 $&$  0.183 $&$    0.139 $\\$
 (0.005, 0.115) $&$  1^{-} $&$         0 $&$   0.946 $&$  0.247 $&$    0.287 $\\$
 (0.005, 0.975) $&$  0^{-} $&$         0 $&$   2.046 $&$  0.299 $&$    0.575 $\\$
 (0.005, 0.975) $&$  0^{-} $&$         0 $&$   1.868 $&$  0.323 $&$    0.571 $\\$
   (0.09, 1.11) $&$  0^{-} $&$         0 $&$   2.267 $&$  0.241 $&$    0.500 $\\$
   (0.09, 1.11) $&$  0^{-} $&$         0 $&$   2.041 $&$  0.303 $&$    0.543 $\\$
 (0.115, 0.975) $&$  0^{+} $&$         0 $&$   2.416 $&$  0.211 $&$    0.533 $\\$
 (0.115, 0.975) $&$  0^{+} $&$         0 $&$   2.265 $&$  0.216 $&$    0.519 $\\$
 (0.115, 0.975) $&$  0^{-} $&$         0 $&$   1.872 $&$  0.269 $&$    0.488 $\\$
 (0.115, 0.975) $&$  0^{-} $&$         1 $&$   2.175 $&$  0.178 $&$    0.410 $\\$
 (0.115, 0.975) $&$  1^{+} $&$         0 $&$   2.354 $&$  0.327 $&$    0.296 $\\$
   (1.11, 4.43) $&$  0^{-} $&$         2 $&$   6.779 $&$  0.211 $&$    0.569 $\\$
   (1.11, 4.43) $&$  1^{-} $&$         2 $&$   6.815 $&$  0.254 $&$    0.164 $\\
\bottomrule
\end{mpxtabular}

\bigskip

\tablecaption{Same as Tab.~\ref{tbl:AWW}, but for the \gls{MT} model \eqref{eq:MT}.}
\label{tbl:MT}
\tablefirsthead{
\toprule
$(m_q, m_{{\bar q}^\prime}) $&$  J^\mathcal{PC} $&$  n $&$       M $&$  f $&$  \sqrt[3]{\left|\Braket{\bar qq}\right|} $\\
\midrule}
\tablehead{
\multicolumn{6}{c}{$\ldots$ continuation of Tab.~\ref{tbl:MT} $\ldots$}\\
\toprule
$(m_q, m_{{\bar q}^\prime}) $&$  J^\mathcal{PC} $&$  n $&$       M $&$  f $&$  \sqrt[3]{\left|\Braket{\bar qq}\right|} $\\
\midrule}
\begin{xtabular}{cccccc}
$
               0 $&$  0^{++} $&$         0 $&$  0.627 $&$  0.000 $&$    0.000 $\\$
               0 $&$  0^{+-} $&$         0 $&$  1.003 $&$  0.000 $&$    0.000 $\\$
               0 $&$  0^{-+} $&$         0 $&$  0.001 $&$  0.124 $&$    0.266 $\\$
               0 $&$  0^{--} $&$         0 $&$  0.834 $&$  0.000 $&$    0.000 $\\$
               0 $&$  1^{++} $&$         0 $&$  0.864 $&$  0.174 $&$    0.000 $\\$
               0 $&$  1^{+-} $&$         0 $&$  0.798 $&$  0.000 $&$    0.000 $\\$
               0 $&$  1^{-+} $&$         0 $&$  0.967 $&$  0.000 $&$    0.000 $\\$
               0 $&$  1^{--} $&$         0 $&$  0.705 $&$  0.200 $&$    0.233 $\\$
               0 $&$  1^{--} $&$         1 $&$  0.954 $&$  0.075 $&$    0.122 $\\$
          0.0038 $&$  0^{++} $&$         0 $&$  0.654 $&$  0.000 $&$    0.000 $\\$
          0.0038 $&$  0^{+-} $&$         0 $&$  1.023 $&$  0.000 $&$    0.000 $\\$
          0.0038 $&$  0^{-+} $&$         0 $&$  0.137 $&$  0.128 $&$    0.272 $\\$
          0.0038 $&$  0^{--} $&$         0 $&$  0.851 $&$  0.000 $&$    0.000 $\\$
          0.0038 $&$  1^{++} $&$         0 $&$  0.885 $&$  0.176 $&$    0.000 $\\$
          0.0038 $&$  1^{+-} $&$         0 $&$  0.818 $&$  0.000 $&$    0.000 $\\$
          0.0038 $&$  1^{-+} $&$         0 $&$  0.987 $&$  0.000 $&$    0.000 $\\$
          0.0038 $&$  1^{--} $&$         0 $&$  0.725 $&$  0.203 $&$    0.235 $\\$
          0.0038 $&$  1^{--} $&$         1 $&$  0.977 $&$  0.075 $&$    0.121 $\\$
           0.075 $&$  0^{++} $&$         0 $&$  1.020 $&$  0.000 $&$    0.000 $\\$
           0.075 $&$  0^{+-} $&$         0 $&$  1.330 $&$  0.000 $&$    0.000 $\\$
           0.075 $&$  0^{-+} $&$         0 $&$  0.642 $&$  0.173 $&$    0.346 $\\$
           0.075 $&$  0^{-+} $&$         1 $&$  1.386 $&$  0.028 $&$    0.172 $\\$
           0.075 $&$  0^{--} $&$         0 $&$  1.122 $&$  0.000 $&$    0.000 $\\$
           0.075 $&$  1^{++} $&$         0 $&$  1.194 $&$  0.195 $&$    0.000 $\\$
           0.075 $&$  1^{+-} $&$         0 $&$  1.114 $&$  0.000 $&$    0.000 $\\$
           0.075 $&$  1^{-+} $&$         0 $&$  1.265 $&$  0.000 $&$    0.000 $\\$
           0.075 $&$  1^{--} $&$         0 $&$  1.018 $&$  0.249 $&$    0.248 $\\$
           0.075 $&$  1^{--} $&$         1 $&$  1.312 $&$  0.072 $&$    0.108 $\\$
           0.075 $&$  2^{++} $&$         0 $&$  1.397 $&$    - $&$      - $\\$
           0.085 $&$  0^{++} $&$         0 $&$  1.062 $&$  0.000 $&$    0.000 $\\$
           0.085 $&$  0^{+-} $&$         0 $&$  1.367 $&$  0.000 $&$    0.000 $\\$
           0.085 $&$  0^{-+} $&$         0 $&$  0.687 $&$  0.179 $&$    0.354 $\\$
           0.085 $&$  0^{-+} $&$         1 $&$  1.418 $&$  0.031 $&$    0.177 $\\$
           0.085 $&$  0^{--} $&$         0 $&$  1.155 $&$  0.000 $&$    0.000 $\\$
           0.085 $&$  1^{++} $&$         0 $&$  1.231 $&$  0.196 $&$    0.000 $\\$
           0.085 $&$  1^{+-} $&$         0 $&$  1.150 $&$  0.000 $&$    0.000 $\\$
           0.085 $&$  1^{-+} $&$         0 $&$  1.298 $&$  0.000 $&$    0.000 $\\$
           0.085 $&$  1^{--} $&$         0 $&$  1.053 $&$  0.254 $&$    0.250 $\\$
           0.085 $&$  1^{--} $&$         1 $&$  1.351 $&$  0.072 $&$    0.107 $\\$
           0.085 $&$  2^{++} $&$         0 $&$  1.428 $&$    - $&$      - $\\$
           0.695 $&$  0^{++} $&$         0 $&$  2.855 $&$  0.000 $&$    0.000 $\\$
           0.695 $&$  0^{++} $&$         1 $&$  3.198 $&$  0.000 $&$    0.000 $\\$
           0.695 $&$  0^{+-} $&$         0 $&$  3.034 $&$  0.000 $&$    0.000 $\\$
           0.695 $&$  0^{-+} $&$         0 $&$  2.520 $&$  0.347 $&$    0.650 $\\$
           0.695 $&$  0^{-+} $&$         1 $&$  3.009 $&$  0.114 $&$    0.348 $\\$
           0.695 $&$  0^{--} $&$         0 $&$  2.785 $&$  0.000 $&$    0.000 $\\$
           0.695 $&$  0^{--} $&$         1 $&$  3.189 $&$  0.000 $&$    0.000 $\\$
           0.695 $&$  1^{++} $&$         0 $&$  2.936 $&$  0.210 $&$    0.000 $\\$
           0.695 $&$  1^{++} $&$         1 $&$  3.044 $&$  0.004 $&$    0.000 $\\$
           0.695 $&$  1^{+-} $&$         0 $&$  2.870 $&$  0.000 $&$    0.000 $\\$
           0.695 $&$  1^{+-} $&$         1 $&$  3.074 $&$  0.000 $&$    0.000 $\\$
           0.695 $&$  1^{-+} $&$         0 $&$  2.877 $&$  0.000 $&$    0.000 $\\$
           0.695 $&$  1^{--} $&$         0 $&$  2.715 $&$  0.394 $&$    0.258 $\\$
           0.695 $&$  1^{--} $&$         1 $&$  3.067 $&$  0.156 $&$    0.134 $\\$
           0.695 $&$  1^{--} $&$         2 $&$  3.115 $&$  0.064 $&$    0.064 $\\$
           0.695 $&$  2^{++} $&$         0 $&$  3.026 $&$    - $&$      - $\\$
           0.695 $&$  2^{+-} $&$         0 $&$  3.129 $&$    - $&$      - $\\$
           0.695 $&$  2^{-+} $&$         0 $&$  3.156 $&$    - $&$      - $\\$
           0.695 $&$  2^{--} $&$         0 $&$  3.198 $&$    - $&$      - $\\$
           0.855 $&$  0^{++} $&$         0 $&$  3.252 $&$  0.000 $&$    0.000 $\\$
           0.855 $&$  0^{++} $&$         1 $&$  3.575 $&$  0.000 $&$    0.000 $\\$
           0.855 $&$  0^{+-} $&$         0 $&$  3.414 $&$  0.000 $&$    0.000 $\\$
           0.855 $&$  0^{-+} $&$         0 $&$  2.918 $&$  0.373 $&$    0.703 $\\$
           0.855 $&$  0^{-+} $&$         1 $&$  3.384 $&$  0.126 $&$    0.377 $\\$
           0.855 $&$  0^{--} $&$         0 $&$  3.167 $&$  0.000 $&$    0.000 $\\$
           0.855 $&$  0^{--} $&$         1 $&$  3.563 $&$  0.000 $&$    0.000 $\\$
           0.855 $&$  1^{++} $&$         0 $&$  3.325 $&$  0.207 $&$    0.000 $\\$
           0.855 $&$  1^{++} $&$         1 $&$  3.420 $&$  0.005 $&$    0.000 $\\$
           0.855 $&$  1^{+-} $&$         0 $&$  3.264 $&$  0.000 $&$    0.000 $\\$
           0.855 $&$  1^{+-} $&$         1 $&$  3.449 $&$  0.000 $&$    0.000 $\\$
           0.855 $&$  1^{+-} $&$         2 $&$  3.592 $&$  0.000 $&$    0.000 $\\$
           0.855 $&$  1^{-+} $&$         0 $&$  3.250 $&$  0.000 $&$    0.000 $\\$
           0.855 $&$  1^{-+} $&$         1 $&$  3.591 $&$  0.000 $&$    0.000 $\\$
           0.855 $&$  1^{--} $&$         0 $&$  3.096 $&$  0.412 $&$    0.256 $\\$
           0.855 $&$  1^{--} $&$         1 $&$  3.438 $&$  0.171 $&$    0.137 $\\$
           0.855 $&$  1^{--} $&$         2 $&$  3.509 $&$  0.053 $&$    0.059 $\\$
           0.855 $&$  2^{++} $&$         0 $&$  3.405 $&$    - $&$      - $\\$
           0.855 $&$  2^{+-} $&$         0 $&$  3.496 $&$    - $&$      - $\\$
           0.855 $&$  2^{-+} $&$         0 $&$  3.542 $&$    - $&$      - $\\$
           0.855 $&$  2^{--} $&$         0 $&$  3.581 $&$    - $&$      - $\\$
            3.77 $&$  0^{++} $&$         0 $&$  9.710 $&$  0.000 $&$    0.000 $\\$
            3.77 $&$  0^{++} $&$         1 $&$  9.917 $&$  0.000 $&$    0.000 $\\$
            3.77 $&$  0^{+-} $&$         0 $&$  9.764 $&$  0.000 $&$    0.000 $\\$
            3.77 $&$  0^{-+} $&$         0 $&$  9.385 $&$  0.632 $&$    1.327 $\\$
            3.77 $&$  0^{-+} $&$         1 $&$  9.728 $&$  0.293 $&$    0.814 $\\$
            3.77 $&$  0^{-+} $&$         2 $&$  9.945 $&$  0.126 $&$    0.471 $\\$
            3.77 $&$  0^{--} $&$         0 $&$  9.561 $&$  0.000 $&$    0.000 $\\$
            3.77 $&$  0^{--} $&$         1 $&$  9.869 $&$  0.000 $&$    0.000 $\\$
            3.77 $&$  1^{++} $&$         0 $&$  9.733 $&$  0.137 $&$    0.000 $\\$
            3.77 $&$  1^{++} $&$         1 $&$  9.761 $&$  0.004 $&$    0.000 $\\$
            3.77 $&$  1^{++} $&$         2 $&$  9.935 $&$  0.141 $&$    0.000 $\\$
            3.77 $&$  1^{+-} $&$         0 $&$  9.718 $&$  0.000 $&$    0.000 $\\$
            3.77 $&$  1^{+-} $&$         1 $&$  9.771 $&$  0.000 $&$    0.000 $\\$
            3.77 $&$  1^{+-} $&$         2 $&$  9.925 $&$  0.000 $&$    0.000 $\\$
            3.77 $&$  1^{-+} $&$         0 $&$  9.585 $&$  0.000 $&$    0.000 $\\$
            3.77 $&$  1^{-+} $&$         1 $&$  9.882 $&$  0.000 $&$    0.000 $\\$
            3.77 $&$  1^{-+} $&$         2 $&$  9.927 $&$  0.000 $&$    0.000 $\\$
            3.77 $&$  1^{--} $&$         0 $&$  9.457 $&$  0.599 $&$    0.235 $\\$
            3.77 $&$  1^{--} $&$         1 $&$  9.754 $&$  0.325 $&$    0.154 $\\$
            3.77 $&$  1^{--} $&$         2 $&$  9.913 $&$  0.010 $&$    0.017 $\\$
            3.77 $&$  1^{--} $&$         3 $&$  9.961 $&$  0.061 $&$    0.050 $\\$
            3.77 $&$  2^{++} $&$         0 $&$  9.755 $&$    - $&$      - $\\$
            3.77 $&$  2^{++} $&$         1 $&$  9.954 $&$    - $&$      - $\\$
            3.77 $&$  2^{+-} $&$         0 $&$  9.783 $&$    - $&$      - $\\$
            3.77 $&$  2^{-+} $&$         0 $&$  9.917 $&$    - $&$      - $\\$
            3.77 $&$  2^{-+} $&$         1 $&$  9.937 $&$    - $&$      - $\\$
            3.77 $&$  2^{--} $&$         0 $&$  9.926 $&$    - $&$      - $\\$
            3.77 $&$  2^{--} $&$         1 $&$  9.930 $&$    - $&$      - $\\$
            3.77 $&$  3^{-+} $&$         0 $&$  9.949 $&$    - $&$      - $\\$
            3.77 $&$  3^{--} $&$         0 $&$  9.942 $&$    - $&$      - $\\$
   (0.0, 0.0038) $&$  0^{+} $&$         0 $&$  0.641 $&$  0.003 $&$    0.085 $\\$
   (0.0, 0.0038) $&$  0^{+} $&$         1 $&$  1.013 $&$  0.000 $&$    0.010 $\\$
   (0.0, 0.0038) $&$  0^{-} $&$         0 $&$  0.097 $&$  0.126 $&$    0.269 $\\$
   (0.0, 0.0038) $&$  0^{-} $&$         1 $&$  0.843 $&$  0.000 $&$    0.006 $\\$
   (0.0, 0.0038) $&$  1^{+} $&$         0 $&$  0.808 $&$  0.006 $&$    0.067 $\\$
   (0.0, 0.0038) $&$  1^{+} $&$         1 $&$  0.875 $&$  0.175 $&$    0.038 $\\$
   (0.0, 0.0038) $&$  1^{-} $&$         0 $&$  0.715 $&$  0.201 $&$    0.234 $\\$
   (0.0, 0.0038) $&$  1^{-} $&$         1 $&$  0.965 $&$  0.077 $&$    0.122 $\\$
   (0.0, 0.0038) $&$  1^{-} $&$         2 $&$  0.977 $&$  0.016 $&$    0.028 $\\$
    (0.0, 0.075) $&$  0^{+} $&$         0 $&$  0.842 $&$  0.048 $&$    0.220 $\\$
    (0.0, 0.075) $&$  0^{+} $&$         1 $&$  1.204 $&$  0.014 $&$    0.124 $\\$
    (0.0, 0.075) $&$  0^{-} $&$         0 $&$  0.448 $&$  0.146 $&$    0.304 $\\$
    (0.0, 0.075) $&$  0^{-} $&$         1 $&$  1.000 $&$  0.011 $&$    0.093 $\\$
    (0.0, 0.075) $&$  1^{+} $&$         0 $&$  0.975 $&$  0.076 $&$    0.150 $\\$
    (0.0, 0.075) $&$  1^{+} $&$         1 $&$  1.063 $&$  0.186 $&$    0.042 $\\$
    (0.0, 0.075) $&$  1^{-} $&$         0 $&$  0.890 $&$  0.232 $&$    0.243 $\\$
    (0.0, 0.085) $&$  0^{+} $&$         0 $&$  0.866 $&$  0.053 $&$    0.229 $\\$
    (0.0, 0.085) $&$  0^{+} $&$         1 $&$  1.228 $&$  0.017 $&$    0.138 $\\$
    (0.0, 0.085) $&$  0^{-} $&$         0 $&$  0.479 $&$  0.148 $&$    0.308 $\\$
    (0.0, 0.085) $&$  0^{-} $&$         1 $&$  1.021 $&$  0.014 $&$    0.104 $\\$
    (0.0, 0.085) $&$  1^{+} $&$         0 $&$  0.997 $&$  0.083 $&$    0.153 $\\$
    (0.0, 0.085) $&$  1^{+} $&$         1 $&$  1.088 $&$  0.188 $&$    0.029 $\\$
    (0.0, 0.085) $&$  1^{-} $&$         0 $&$  0.914 $&$  0.237 $&$    0.245 $\\$
    (0.0, 0.695) $&$  0^{+} $&$         0 $&$  2.096 $&$  0.376 $&$    0.759 $\\$
 (0.0038, 0.075) $&$  0^{+} $&$         0 $&$  0.853 $&$  0.045 $&$    0.216 $\\$
 (0.0038, 0.075) $&$  0^{+} $&$         1 $&$  1.208 $&$  0.012 $&$    0.113 $\\$
 (0.0038, 0.075) $&$  0^{-} $&$         0 $&$  0.459 $&$  0.148 $&$    0.307 $\\$
 (0.0038, 0.075) $&$  0^{-} $&$         1 $&$  1.006 $&$  0.011 $&$    0.090 $\\$
 (0.0038, 0.075) $&$  1^{+} $&$         0 $&$  0.982 $&$  0.070 $&$    0.146 $\\$
 (0.0038, 0.075) $&$  1^{+} $&$         1 $&$  1.068 $&$  0.187 $&$    0.034 $\\$
 (0.0038, 0.075) $&$  1^{-} $&$         0 $&$  0.896 $&$  0.233 $&$    0.244 $\\$
 (0.0038, 0.085) $&$  0^{+} $&$         0 $&$  0.876 $&$  0.049 $&$    0.225 $\\$
 (0.0038, 0.085) $&$  0^{+} $&$         1 $&$  1.232 $&$  0.015 $&$    0.127 $\\$
 (0.0038, 0.085) $&$  0^{-} $&$         0 $&$  0.489 $&$  0.150 $&$    0.311 $\\$
 (0.0038, 0.085) $&$  0^{-} $&$         1 $&$  1.026 $&$  0.013 $&$    0.101 $\\$
 (0.0038, 0.085) $&$  1^{+} $&$         0 $&$  1.004 $&$  0.077 $&$    0.150 $\\$
 (0.0038, 0.085) $&$  1^{+} $&$         1 $&$  1.093 $&$  0.188 $&$    0.019 $\\$
 (0.0038, 0.085) $&$  1^{-} $&$         0 $&$  0.919 $&$  0.237 $&$    0.246 $\\$
 (0.0038, 0.695) $&$  0^{+} $&$         0 $&$  2.100 $&$  0.372 $&$    0.758 $\\$
 (0.0038, 0.695) $&$  0^{-} $&$         0 $&$  1.895 $&$  0.975 $&$    1.341 $\\$
 (0.0038, 0.695) $&$  0^{-} $&$         0 $&$  1.869 $&$  0.960 $&$    1.315 $\\$
  (0.075, 0.855) $&$  0^{-} $&$         0 $&$  2.496 $&$  0.251 $&$    0.595 $\\$
  (0.075, 0.855) $&$  0^{-} $&$         0 $&$  2.211 $&$  0.441 $&$    0.799 $\\$
  (0.075, 0.855) $&$  0^{-} $&$         0 $&$  2.095 $&$  0.453 $&$    0.785 $\\$
  (0.085, 0.695) $&$  0^{+} $&$         0 $&$  2.161 $&$  0.238 $&$    0.601 $\\$
  (0.085, 0.695) $&$  0^{-} $&$         0 $&$  1.802 $&$  0.295 $&$    0.566 $\\$
  (0.085, 0.695) $&$  0^{-} $&$         1 $&$  2.052 $&$  0.231 $&$    0.524 $\\$
  (0.085, 0.695) $&$  0^{-} $&$         2 $&$  2.327 $&$  0.211 $&$    0.536 $\\$
   (0.855, 3.77) $&$  0^{-} $&$         0 $&$  6.608 $&$  0.433 $&$    0.958 $\\$
   (0.855, 3.77) $&$  1^{-} $&$         0 $&$  6.690 $&$  0.387 $&$    0.193 $\\
\bottomrule
\end{xtabular}
\end{center}

\input{AWW-OpenFlavor-EtaShift.bbl}

\end{document}

%% file: glossaries.tex
\newacronym{OAMD}{OAMD}{orbital-angular momentum decomposition}
\newacronym{DS}{DS}{Dyson-Schwinger}
\newacronym{BS}{BS}{Bethe-Salpeter}
\newacronym{MT}{MT}{Maris-Tandy}
\newacronym{AWW}{AWW}{Alkofer-Watson-Weigel}
\newacronym{NG}{NG}{Nambu-Goldstone}
\newacronym{WW}{WW}{Wigner-Weyl}
\newacronym{SM}{SM}{Standard Model}
\newacronym{UV}{UV}{ultraviolet}
\newacronym{DSE}{DSE}{Dyson-Schwinger equation}
\newacronym{BSE}{BSE}{Bethe-Salpeter equation}
\newacronym{DSBS}{DSBS}{Dyson-Schwinger--Bethe-Salpeter}
\newacronym{DSBSE}{DSBSE}{Dyson-Schwinger-Bethe-Salpeter-equation}
\newacronym{BSA}{BSA}{Bethe-Salpeter amplitude}
\newacronym{BSM}{BSM}{Bethe-Salpeter matrix}
\newacronym{BSV}{BSV}{Bethe-Salpeter vertex}
\newacronym{BSW}{BSW}{Bethe-Salpeter wave function}
\newacronym{QED}{QED}{quantum electrodynamics}
\newacronym{QCD}{QCD}{quantum chromodynamics}
\newacronym{QCDg}{QCD}{Quantenchromodynamik}
\newacronym{DCSB}{D$\upchi$SB}{dynamical chiral symmetry breaking}
\newacronym{VMD}{VMD}{vector meson dominance}
\newacronym{FAIR}{FAIR}{Facility for Antiproton and Ion Research}
\newacronym{HIC_for_FAIR}{HIC for FAIR}{Helmholtz International Center for FAIR}
\newacronym{WASAatCOSY}{WASA at COSY}{Wide Angle Shower Apparatus at Cooler Synchrotron}
\newacronym{WASA}{WASA}{Wide Angle Shower Apparatus}
\newacronym{COSY}{COSY}{Cooler Synchrotron}
\newacronym{MAMI}{MAMI}{\emph{Mainzer Mikrotron}}
\newacronym{LHC}{LHC}{Large Hadron Collider}
\newacronym{LHCb}{LHC{\scshape b}}{Large Hadron Collider beauty experiment}
\newacronym{KLOE}{KLOE}{K Long Experiment}
\newacronym{DAFNE}{DA$\Phi$NE}{Double Annular $\Phi$ Factory for Nice Experiments}
\newacronym{DESY}{DESY}{\emph{Deutsches Elektronen-Synchrotron}}
\newacronym{JLab}{JLab}{Thomas Jefferson National Accelerator Facility}
\newacronym{QSR}{QSR}{QCD sum rule}
\newacronym{OPE}{OPE}{operator product expansion}
\newacronym{JINR}{JINR}{Joint Institute for Nuclear Research}
\newacronym{HZDR}{HZDR}{Helmholtz-Zentrum Dresden-Rossendorf}
\newacronym{ANL}{ANL}{Argonne National Laboratory}
\newacronym{PANDA}{PANDA}{facility for anti-Proton ANnihilation at DArmstadt}
\newacronym{EFT}{EFT}{effective field theory}
\newacronym{VEV}{VEV}{vacuum expectation value}
\newacronym{KFU}{KFU}{Karl-Franzens University}
\newacronym{lQCD}{{\scshape l}QCD}{lattice-regularized QCD}
\newacronym{WTI}{WTI}{Ward-Takahashi identity}
\newacronym{AVWTI}{{\scshape av}WTI}{axialvector Ward-Takahashi identity}
\newacronym{RL}{RL}{rainbow-ladder}
\newacronym{TUD}{TUD}{Technical University of Dresden}
\newacronym{UU}{UU}{Uppsala University}
\newacronym{NJL}{NJL}{Nambu--Jona-Lasinio model}
\newacronym{QFT}{QFT}{quantum field theory}
\newacronym{BRST}{BRST}{Becchi-Rouet-Stora-Tyutin}
\newacronym{FAKT}{FAKT}{\emph{``Fachausschuss Kern- und Teilchenphysik''}}
\newacronym{OPG}{{\"O}PG}{\emph{``{\"O}sterreichische Physikalische Gesellschaft''}}
\newacronym{APS}{{\"O}PG}{Austrian Physical Society}
\newacronym{SLAC}{SLAC}{Stanford Linear Accelerator Center}
\newacronym{DPG}{DPG}{\emph{``Deutsche Physikalische Gesellschaft''}}
\newacronym{GPS}{DPG}{German Physical Society}
\newacronym{FWF}{FWF}{\emph{``Fonds zur F{\"o}rderung der wissenschaftlichen Forschung''}}
\newacronym{KSU}{KSU}{Kent State University}
\newacronym{GMOR}{GMOR}{Gell-Mann--Oakes--Renner}
\newacronym{QGV}{QGV}{quark-gluon interaction vertex}
\newacronym{STI}{STI}{Slavnov-Taylor identity}
\newacronym{GU}{GU}{Giessen University}
\newacronym{GSI}{GSI}{\emph{Gesellschaft f{\"u}r Schwerionenforschung}}
\newacronym{ERA}{ERA}{European Research Area}
\newacronym{BRL}{BRL}{beyond-rainbow-ladder}
\newacronym{WP}{WP}{work package}
\newacronym{D}{D}{major deliverable}
\newacronym{BEPC}{BEPC}{Beijing Electron Positron Collider}
\newacronym{BES}{BES-III}{Bejing Spectrometer}
\newacronym{KEK}{KEK}{High Energy Accelerator Research Organization}

%% file: AWW-OpenFlavor-EtaShift.bbl
\begin{thebibliography}{100}
\providecommand{\url}[1]{{#1}}
\providecommand{\urlprefix}{URL }
\expandafter\ifx\csname urlstyle\endcsname\relax
  \providecommand{\doi}[1]{DOI \discretionary{}{}{}#1}\else
  \providecommand{\doi}{DOI \discretionary{}{}{}\begingroup
  \urlstyle{rm}\Url}\fi

\bibitem{Hilger:2008jg}
T.~Hilger, R.~Thomas, B.~K\"ampfer, Phys. Rev. C \textbf{79}, 025202 (2009)

\bibitem{Hilger:2011cq}
T.~Hilger, B.~K{\"a}mpfer, S.~Leupold, Phys. Rev. C \textbf{84}, 045202 (2011)

\bibitem{Eichten:1979ms}
E.~Eichten, K.~Gottfried, T.~Kinoshita, K.D. Lane, T.M. Yan, Phys. Rev. D
  \textbf{21}, 203 (1980)

\bibitem{Godfrey:1985xj}
S.~Godfrey, N.~Isgur, Phys. Rev. D \textbf{32}, 189 (1985)

\bibitem{Roberts:2007ni}
W.~Roberts, M.~Pervin, Int. J. Mod. Phys. A \textbf{23}, 2817 (2008)

\bibitem{Melde:2008yr}
T.~Melde, W.~Plessas, B.~Sengl, Phys. Rev. D \textbf{77}, 114002 (2008)

\bibitem{GomezRocha:2012zd}
M.~Gomez-Rocha, W.~Schweiger, Phys. Rev. D \textbf{86}, 053010 (2012)

\bibitem{Gomez-Rocha:2014aoa}
M.~Gomez-Rocha, Phys. Rev. D \textbf{90}(7), 076003 (2014)

\bibitem{Li:2015zda}
Y.~Li, P.~Maris, X.~Zhao, J.P. Vary, Phys. Lett. B \textbf{758}, 118 (2016)

\bibitem{Segovia:2016xqb}
J.~Segovia, P.G. Ortega, D.R. Entem, F.~Fernandez, Phys. Rev. D \textbf{93}(7),
  074027 (2016)

\bibitem{Vary:2016emi}
J.P. Vary, L.~Adhikari, G.~Chen, Y.~Li, P.~Maris, X.~Zhao, Few Body Syst.
  \textbf{57}(8), 695 (2016)

\bibitem{Berwein:2015vca}
M.~Berwein, N.~Brambilla, J.T. Castella, A.~Vairo, Phys. Rev. D
  \textbf{92}(11), 114019 (2015)

\bibitem{Pelaez:2015zoa}
J.R. Pel{\'a}ez, T.~Cohen, F.J. Llanes-Estrada, J.~Ruiz~de Elvira, Acta Phys.
  Polon. Supp. \textbf{8}, 465 (2015)

\bibitem{Terschlusen:2016cfw}
C.~Terschl{\"u}sen, S.~Leupold, arXiv:1604.01682  (2016)

\bibitem{Parganlija:2016yxq}
D.~Parganlija, F.~Giacosa, arXiv:1612.09218  (2016)

\bibitem{Shifman:1978bx}
M.A. Shifman, A.I. Vainshtein, V.I. Zakharov, Nucl. Phys. B \textbf{147}, 385
  (1979)

\bibitem{Nielsen:2009uh}
M.~Nielsen, F.S. Navarra, S.H. Lee, Phys. Rept. \textbf{497}, 41 (2010)

\bibitem{Narison:2014wqa}
S.~Narison, Nucl. Part. Phys. Proc. \textbf{258--259}, 189 (2015)

\bibitem{Ho:2016owu}
J.~Ho, D.~Harnett, T.G. Steele, arXiv:1609.06750  (2016)

\bibitem{Dudek:2010wm}
J.J. Dudek, R.G. Edwards, M.J. Peardon, D.G. Richards, C.E. Thomas, Phys. Rev.
  D \textbf{82}, 034508 (2010)

\bibitem{Bouchard:2014eea}
C.M. Bouchard, E.~Freeland, C.W. Bernard, C.C. Chang, A.X. El-Khadra, M.E.
  G{\'a}miz, A.S. Kronfeld, J.~Laiho, R.S. Van~de Water, PoS
  \textbf{LATTICE2014}, 378 (2014)

\bibitem{Lang:2015hza}
C.B. Lang, D.~Mohler, S.~Prelovsek, R.M. Woloshyn, Phys. Lett. B \textbf{750},
  17 (2015)

\bibitem{Hutauruk:2016sug}
P.T.P. Hutauruk, I.C. Clo{\"e}t, A.W. Thomas, Phys. Rev. C \textbf{94}(3),
  035201 (2016)

\bibitem{Lang:2015ljt}
C.B. Lang, AIP Conf. Proc. \textbf{1735}, 020002 (2016)

\bibitem{Eichmann:2016yit}
G.~Eichmann, H.~Sanchis-Alepuz, R.~Williams, R.~Alkofer, C.S. Fischer, Prog.
  Part. Nucl. Phys. \textbf{91}, 1 (2016)

\bibitem{Blaschke:2000gd}
D.~Blaschke, G.~Burau, Y.L. Kalinovsky, P.~Maris, P.C. Tandy, Int. J. Mod.
  Phys. A \textbf{16}, 2267 (2001)

\bibitem{Krassnigg:2003wy}
A.~Krassnigg, C.D. Roberts, Fizika B \textbf{13}, 143 (2004)

\bibitem{Alkofer:2005ug}
R.~Alkofer, M.~Kloker, A.~Krassnigg, R.F. Wagenbrunn, Phys. Rev. Lett.
  \textbf{96}, 022001 (2006)

\bibitem{Eichmann:2007nn}
G.~Eichmann, A.~Krassnigg, M.~Schwinzerl, R.~Alkofer, Ann. Phys. \textbf{323},
  2505 (2008)

\bibitem{Eichmann:2009qa}
G.~Eichmann, R.~Alkofer, A.~Krassnigg, D.~Nicmorus, Phys. Rev. Lett.
  \textbf{104}, 201601 (2010)

\bibitem{Blank:2010bz}
M.~Blank, A.~Krassnigg, Phys. Rev. D \textbf{82}, 034006 (2010)

\bibitem{Carbonell:2013kwa}
J.~Carbonell, V.~Karmanov, Phys. Lett. B \textbf{727}, 319 (2013)

\bibitem{Sauli:2014uxa}
V.~{\v S}auli, Int. J. Theor. Phys. \textbf{54}(11), 4131 (2015)

\bibitem{Biernat:2014xaa}
E.P. Biernat, M.T. Pe{\~n}a, J.E. Ribeiro, A.~Stadler, F.~Gross, Phys. Rev. D
  \textbf{90}(9), 096008 (2014)

\bibitem{Blum:2014gna}
A.~Blum, M.Q. Huber, M.~Mitter, L.~von Smekal, Phys. Rev. D \textbf{89}(6),
  061703 (2014)

\bibitem{Biernat:2015xya}
E.P. Biernat, F.~Gross, M.T. Pe{\~n}a, A.~Stadler, Phys. Rev. D \textbf{92}(7),
  076011 (2015)

\bibitem{Bedolla:2015mpa}
M.A. Bedolla, J.~Cobos-Martinez, A.~Bashir, Phys. Rev. D \textbf{92}(5), 054031
  (2015)

\bibitem{Raya:2015gva}
K.~Raya, L.~Chang, A.~Bashir, J.J. Cobos-Martinez, L.X. Gutierrez-Guerrero,
  C.D. Roberts, P.C. Tandy, Phys. Rev. D \textbf{93}(7), 074017 (2016)

\bibitem{Binosi:2016rxz}
D.~Binosi, L.~Chang, J.~Papavassiliou, S.X. Qin, C.D. Roberts, Phys. Rev. D
  \textbf{93}(9), 096010 (2016)

\bibitem{Serna:2016ifh}
F.E. Serna, G.~Krein, EPJ Web Conf. \textbf{137}, 13015 (2017)

\bibitem{Maas:2016tci}
A.~Maas, W.A. Mian, Eur. Phys. J. A \textbf{53}(2), 22 (2017)

\bibitem{Bettoni:2014vka}
D.~Bettoni, EPJ Web Conf. \textbf{73}, 01006 (2014)

\bibitem{Geesaman:2015fha}
A.~Aprahamian, et~al., {Reaching for the horizon: The 2015 long range plan for
  nuclear science}  (2015)

\bibitem{Olive:2016xmw}
C.~Patrignani, et~al., Chin. Phys. C \textbf{40}(10), 100001 (2016)

\bibitem{Maris:1997tm}
P.~Maris, C.D. Roberts, Phys. Rev. C \textbf{56}, 3369 (1997)

\bibitem{Maris:2000sk}
P.~Maris, P.C. Tandy, Phys. Rev. C \textbf{62}, 055204 (2000)

\bibitem{Burden:2002ps}
C.J. Burden, M.A. Pichowsky, Few-Body Syst. \textbf{32}, 119 (2002)

\bibitem{Maris:2005tt}
P.~Maris, P.C. Tandy, Nucl. Phys. B, Proc. Suppl. \textbf{161}, 136 (2006)

\bibitem{Maris:2006ea}
P.~Maris, AIP Conf. Proc. \textbf{892}, 65 (2007)

\bibitem{Bhagwat:2007rj}
M.S. Bhagwat, A.~H{\"o}ll, A.~Krassnigg, C.D. Roberts, S.V. Wright, Few-Body
  Syst. \textbf{40}, 209 (2007)

\bibitem{Eichmann:2008ef}
G.~Eichmann, I.C. Clo{\"e}t, R.~Alkofer, A.~Krassnigg, C.D. Roberts, Phys. Rev.
  C \textbf{79}, 012202(R) (2009)

\bibitem{Blank:2010sn}
M.~Blank, A.~Krassnigg, AIP Conf. Proc. \textbf{1343}, 349 (2011)

\bibitem{Mader:2011zf}
V.~Mader, G.~Eichmann, M.~Blank, A.~Krassnigg, Phys. Rev. D \textbf{84}, 034012
  (2011)

\bibitem{Popovici:2014pha}
C.~Popovici, T.~Hilger, M.~Gomez-Rocha, A.~Krassnigg, Few Body Syst.
  \textbf{56}(6), 481 (2015)

\bibitem{Sanchis-Alepuz:2014sca}
H.~Sanchis-Alepuz, C.S. Fischer, Phys. Rev. D \textbf{90}(9), 096001 (2014)

\bibitem{Eichmann:2015cra}
G.~Eichmann, C.S. Fischer, W.~Heupel, Phys. Lett. B \textbf{753}, 282 (2016)

\bibitem{Krassnigg:2009zh}
A.~Krassnigg, Phys. Rev. D \textbf{80}, 114010 (2009)

\bibitem{Blank:2011ha}
M.~Blank, A.~Krassnigg, Phys. Rev. D \textbf{84}, 096014 (2011)

\bibitem{Hilger:2014nma}
T.~Hilger, C.~Popovici, M.~Gomez-Rocha, A.~Krassnigg, Phys. Rev. D \textbf{91},
  034013 (2015)

\bibitem{Hilger:2015hka}
T.~Hilger, M.~Gomez-Rocha, A.~Krassnigg, Phys. Rev. D \textbf{91}, 114004
  (2015)

\bibitem{Hilger:2015ora}
T.~Hilger, M.~Gomez-Rocha, A.~Krassnigg, arXiv:1508.07183  (2015)

\bibitem{Hilger:2016efh}
T.~Hilger, A.~Krassnigg, Eur. Phys. J. A \textbf{53}, 142 (2017)

\bibitem{Hilger:2016drj}
T.~Hilger, A.~Krassnigg, EPJ Web Conf. \textbf{137}, 01010 (2017)

\bibitem{Friman:2011zz}
B.~Friman, C.~Hohne, J.~Knoll, S.~Leupold, J.~Randrup, R.~Rapp, P.~Senger,
  Lect. Notes Phys. \textbf{814}, 1 (2011)

\bibitem{Rapp:2011zz}
R.~Rapp, et~al., Lect. Notes Phys. \textbf{814}, 335 (2011)

\bibitem{Holl:2003dq}
A.~H{\"o}ll, A.~Krassnigg, C.D. Roberts, arXiv:nucl-th/0311033  (2003)

\bibitem{Williams:2007ey}
R.~Williams, C.S. Fischer, M.R. Pennington, arXiv:0704.2296 [hep-ph]  (2007)

\bibitem{Wang:2012me}
K.l. Wang, S.x. Qin, Y.x. Liu, L.~Chang, C.D. Roberts, S.M. Schmidt, Phys. Rev.
  D \textbf{86}, 114001 (2012)

\bibitem{Hilger:2015zva}
T.~Hilger, Phys. Rev. D \textbf{93}, 054020 (2016)

\bibitem{Bloch:1995dd}
J.C.R. Bloch, {Numerical investigation of fermion mass generation in QED}.
\newblock Ph.D. thesis, University of Durham (1995), arXiv:hep-ph/0208074

\bibitem{Krassnigg:2008gd}
A.~Krassnigg, {PoS} \textbf{{CONFINEMENT8}}, 075 (2008)

\bibitem{Dorkin:2013rsa}
S.M. Dorkin, L.P. Kaptari, T.~Hilger, B.~K{\"a}mpfer, Phys. Rev. C \textbf{89},
  034005 (2014)

\bibitem{Krassnigg:2016hml}
A.~Krassnigg, M.~Gomez-Rocha, T.~Hilger, J. Phys. Conf. Ser. \textbf{742}(1),
  012032 (2016)

\bibitem{Maris:1999nt}
P.~Maris, P.C. Tandy, Phys. Rev. C \textbf{60}, 055214 (1999)

\bibitem{Blank:2010pa}
M.~Blank, A.~Krassnigg, A.~Maas, Phys. Rev. D \textbf{83}, 034020 (2011)

\bibitem{Alkofer:2002bp}
R.~Alkofer, P.~Watson, H.~Weigel, Phys. Rev. D \textbf{65}, 094026 (2002)

\bibitem{Jain:1993qh}
P.~Jain, H.J. Munczek, Phys. Rev. D \textbf{48}, 5403 (1993)

\bibitem{Munczek:1994zz}
H.J. Munczek, Phys. Rev. D \textbf{52}, 4736 (1995)

\bibitem{Eichmann:2008ae}
G.~Eichmann, R.~Alkofer, I.C. Clo{\"e}t, A.~Krassnigg, C.D. Roberts, Phys. Rev.
  C \textbf{77}, 042202(R) (2008)

\bibitem{Bender:2002as}
A.~Bender, W.~Detmold, C.D. Roberts, A.W. Thomas, Phys. Rev. C \textbf{65},
  065203 (2002)

\bibitem{Bhagwat:2004hn}
M.S. Bhagwat, A.~H{\"o}ll, A.~Krassnigg, C.D. Roberts, P.C. Tandy, Phys. Rev. C
  \textbf{70}, 035205 (2004)

\bibitem{Holl:2004qn}
A.~H{\"o}ll, A.~Krassnigg, C.D. Roberts, Nucl. Phys. Proc. Suppl. \textbf{141},
  47 (2005)

\bibitem{Matevosyan:2006bk}
H.H. Matevosyan, A.W. Thomas, P.C. Tandy, Phys. Rev. C \textbf{75}, 045201
  (2007)

\bibitem{Matevosyan:2007cx}
H.H. Matevosyan, A.W. Thomas, P.C. Tandy, J. Phys. G \textbf{34}, 2153 (2007)

\bibitem{Gomez-Rocha:2014vsa}
M.~Gomez-Rocha, T.~Hilger, A.~Krassnigg, Few Body Syst. \textbf{56}(6), 475
  (2015)

\bibitem{Gomez-Rocha:2015qga}
M.~Gomez-Rocha, T.~Hilger, A.~Krassnigg, Phys. Rev. D \textbf{92}(5), 054030
  (2015)

\bibitem{Jinno:2015sea}
R.~Jinno, T.~Kitahara, G.~Mishima, Phys. Rev. D \textbf{91}(7), 076011 (2015)

\bibitem{Gomez-Rocha:2016cji}
M.~Gomez-Rocha, T.~Hilger, A.~Krassnigg, Phys. Rev. D \textbf{93}, 074010
  (2016)

\bibitem{Fischer:2009jm}
C.S. Fischer, R.~Williams, Phys. Rev. Lett. \textbf{103}, 122001 (2009)

\bibitem{Chang:2009zb}
L.~Chang, C.D. Roberts, Phys. Rev. Lett. \textbf{103}, 081601 (2009)

\bibitem{Williams:2015cvx}
R.~Williams, C.S. Fischer, W.~Heupel, Phys. Rev. D \textbf{93}(3), 034026
  (2016)

\bibitem{Dorkin:2010ut}
S.M. Dorkin, T.~Hilger, L.P. Kaptari, B.~K{\"a}mpfer, Few Body Syst.
  \textbf{49}, 247 (2011)

\bibitem{Blank:2010bp}
M.~Blank, A.~Krassnigg, Comput. Phys. Commun. \textbf{182}, 1391 (2011)

\bibitem{UweHilger:2012uua}
T.U. Hilger, {Medium Modifications of Mesons}.
\newblock Ph.D. thesis, TU Dresden (2012)

\bibitem{Krassnigg:2010mh}
A.~Krassnigg, M.~Blank, Phys. Rev. D \textbf{83}, 096006 (2011)

\bibitem{Fischer:2014xha}
C.S. Fischer, S.~Kubrak, R.~Williams, Eur. Phys. J. A \textbf{50}, 126 (2014)

\bibitem{Krassnigg:2003dr}
A.~Krassnigg, C.D. Roberts, Nucl. Phys. A \textbf{737}, 7 (2004)

\bibitem{Bhagwat:2006pu}
M.S. Bhagwat, P.~Maris, Phys. Rev. C \textbf{77}, 025203 (2008)

\bibitem{Blank:2011qk}
M.~Blank, {Properties of quarks and mesons in the
  Dyson-Schwinger/Bethe-Salpeter approach}.
\newblock Ph.D. thesis, University of Graz (2011), arXiv:1106.4843

\bibitem{Maris:1997hd}
P.~Maris, C.D. Roberts, P.C. Tandy, Phys. Lett. B \textbf{420}, 267 (1998)

\bibitem{Holl:2004fr}
A.~H{\"o}ll, A.~Krassnigg, C.D. Roberts, Phys. Rev. C \textbf{70}, 042203(R)
  (2004)

\bibitem{Roberts:2012sv}
C.D. Roberts, arXiv:1203.5341  (2012)

\bibitem{Qin:2011xq}
S.X. Qin, L.~Chang, Y.X. Liu, C.D. Roberts, D.J. Wilson, Phys. Rev. C
  \textbf{85}, 035202 (2012)

\bibitem{Maris:2000ig}
P.~Maris, C.D. Roberts, S.M. Schmidt, P.C. Tandy, Phys. Rev. C \textbf{63},
  025202 (2001)

\bibitem{Bhagwat:2006py}
M.S. Bhagwat, A.~H{\"o}ll, A.~Krassnigg, C.D. Roberts, Nucl. Phys. A
  \textbf{790}, 10 (2007)

\bibitem{Cloet:2013jya}
I.C. Clo{\"e}t, C.D. Roberts, Prog. Part. Nucl. Phys. \textbf{77}, 1 (2014)

\bibitem{Horn:2016rip}
T.~Horn, C.D. Roberts, J. Phys. G \textbf{43}(7), 073001 (2016)

\bibitem{Brodsky:2010xf}
S.J. Brodsky, C.D. Roberts, R.~Shrock, P.C. Tandy, Phys. Rev. C \textbf{82},
  022201 (2010)

\bibitem{Chang:2011mu}
L.~Chang, C.D. Roberts, P.C. Tandy, Phys. Rev. C \textbf{85}, 012201(R) (2012)

\bibitem{Brodsky:2012ku}
S.J. Brodsky, C.D. Roberts, R.~Shrock, P.C. Tandy, Phys. Rev. C \textbf{85},
  065202 (2012)

\bibitem{Williams:2006vva}
R.~Williams, C.S. Fischer, M.R. Pennington, Phys. Lett. B \textbf{645}, 167
  (2007)

\bibitem{Williams:2007ef}
R.~Williams, C.S. Fischer, M.R. Pennington, Acta Phys. Polon. B \textbf{38},
  2803 (2007)

\bibitem{Fischer:2014ata}
C.S. Fischer, J.~Luecker, C.A. Welzbacher, Phys. Rev. D \textbf{90}(3), 034022
  (2014)

\bibitem{Brodsky:2008xu}
S.J. Brodsky, R.~Shrock, Science \textbf{108}, 45 (2011)

\bibitem{Brodsky:2009zd}
S.J. Brodsky, R.~Shrock, Proc. Nat. Acad. Sci. \textbf{108}, 45 (2011)

\bibitem{Shifman:1978by}
M.A. Shifman, A.I. Vainshtein, V.I. Zakharov, Nucl. Phys. B \textbf{147}, 448
  (1979)

\bibitem{Buchheim:2015yyc}
T.~Buchheim, B.~K{\"a}mpfer, T.~Hilger, J. Phys. G \textbf{43}(5), 055105
  (2015)

\bibitem{Buchheim:2014uda}
T.~Buchheim, T.~Hilger, B.~K{\"a}mpfer, EPJ Web Conf. \textbf{81}, 05007 (2014)

\bibitem{Morita:2007pt}
K.~Morita, S.H. Lee, Phys. Rev. Lett. \textbf{100}, 022301 (2008)

\bibitem{Hilger:2010zb}
T.~Hilger, R.~Schulze, B.~K{\"a}mpfer, J. Phys. G \textbf{37}, 094054 (2010)

\bibitem{Hilger:2009kn}
T.~Hilger, B.~K{\"a}mpfer, arXiv:0904.3491  (2009)

\bibitem{Hilger:2012db}
T.~Hilger, T.~Buchheim, B.~K{\"a}mpfer, S.~Leupold, Prog. Part. Nucl. Phys.
  \textbf{67}, 188 (2012)

\bibitem{Zschocke:2011aa}
S.~Zschocke, T.~Hilger, B.~K\"ampfer, Eur. Phys. J. A \textbf{47}, 151 (2011)

\bibitem{Buchheim:2015xka}
T.~Buchheim, T.~Hilger, B.~K\"ampfer, J. Phys. Conf. Ser. \textbf{668}(1),
  012047 (2016)

\bibitem{Buchheim:2014rpa}
T.~Buchheim, T.~Hilger, B.~K{\"a}mpfer, Phys. Rev. C \textbf{91}, 015205 (2015)

\bibitem{Thomas:2006nk}
R.~Thomas, T.~Hilger, S.~Zschocke, B.~K{\"a}mpfer, AIP Conf. Proc.
  \textbf{892}, 274 (2007)

\bibitem{Thomas:2007gx}
R.~Thomas, T.~Hilger, B.~K\"ampfer, Nucl. Phys. A \textbf{795}, 19 (2007)

\bibitem{Thomas:2007es}
R.~Thomas, T.~Hilger, B.~K{\"a}mpfer, Prog. Part. Nucl. Phys. \textbf{61}, 297
  (2008)

\bibitem{Hilger:2010cn}
T.~Hilger, R.~Thomas, B.~K{\"a}mpfer, S.~Leupold, Phys. Lett. B \textbf{709},
  200 (2012)

\bibitem{Jarecke:2002xd}
D.~Jarecke, P.~Maris, P.C. Tandy, Phys. Rev. C \textbf{67}, 035202 (2003)

\bibitem{Horvatic:2007qs}
D.~Horvatic, D.~Klabucar, A.E. Radzhabov, Phys. Rev. D \textbf{76}, 096009
  (2007)

\bibitem{Bashir:2012fs}
A.~Bashir, L.~Chang, I.C. Clo{\"e}t, B.~El-Bennich, Y.X. Liu, C.D. Roberts,
  P.C. Tandy, Commun. Theor. Phys. \textbf{58}, 79 (2012)

\bibitem{Fischer:2014cfa}
C.S. Fischer, S.~Kubrak, R.~Williams, Eur. Phys. J. A \textbf{51}, 10 (2015)

\bibitem{Rojas:2014aka}
E.~Rojas, B.~El-Bennich, J.~de~Melo, Phys. Rev. D \textbf{90}, 074025 (2014)

\bibitem{Serna:2016kdb}
F.E. Serna, M.A. Brito, G.~Krein, AIP Conf. Proc. \textbf{1701}, 100018 (2016)

\bibitem{El-Bennich:2016qmb}
B.~El-Bennich, G.~Krein, E.~Rojas, F.E. Serna, Few Body Syst. \textbf{57}(10),
  955 (2016)

\bibitem{Bender:1996bb}
A.~Bender, C.D. Roberts, L.~von Smekal, Phys. Lett. B \textbf{380}, 7 (1996)

\bibitem{Holl:2005vu}
A.~H{\"o}ll, A.~Krassnigg, P.~Maris, C.D. Roberts, S.V. Wright, Phys. Rev. C
  \textbf{71}, 065204 (2005)

\bibitem{Smith:1969zk}
C.H.L. Smith, Nuovo Cim. A \textbf{60}, 348 (1969)

\bibitem{Eichmann:2016nsu}
G.~Eichmann, Few Body Syst. \textbf{58}(2), 81 (2017)

\bibitem{Gross:1982nz}
F.~Gross, Phys. Rev. C \textbf{26}, 2203 (1982)

\bibitem{Gross:1993}
F.~Gross, \emph{Relativistic quantum mechanics and field theory} (John Wiley
  {\&} Sons, New York, 1993)

\bibitem{Eides:2000xc}
M.I. Eides, H.~Grotch, V.A. Shelyuto, Phys. Rept. \textbf{342}, 63 (2001)

\bibitem{Mian:2016eel}
W.A. Mian, A.~Maas, EPJ Web Conf. \textbf{137}, 05015 (2017)

\bibitem{Eichmann:2009zx}
G.~Eichmann, {Hadron properties from QCD bound-state equations}.
\newblock Ph.D. thesis, University of Graz (2009), arXiv:0909.0703

\bibitem{Windisch:2016iud}
A.~Windisch, Phys. Rev. C \textbf{95}(4), 045204 (2017)

\bibitem{Dorkin:2014lxa}
S.~Dorkin, L.~Kaptari, B.~K{\"a}mpfer, Phys. Rev. C \textbf{91}(5), 055201
  (2015)

\bibitem{Souchlas:2010boa}
N.~Souchlas, J. Phys. G \textbf{37}(11), 115001 (2010)

\bibitem{Knoll2004357}
D.~Knoll, D.~Keyes, J. Comp. Phys. \textbf{193}(2), 357  (2004)

\bibitem{Wales:1997}
D.J. Wales, J.P.K. Doye, J. of Phys. Chem. A \textbf{101}(28), 5111 (1997)

\bibitem{Fischer:2005en}
C.S. Fischer, P.~Watson, W.~Cassing, Phys. Rev. D \textbf{72}, 094025 (2005)

\bibitem{Maris:2004pc}
P.~Maris,
\newblock private communication

\end{thebibliography}
